\documentclass[12pt,a4paper]{book}

\usepackage{amsmath}
\usepackage{amsthm}
\usepackage{amssymb}
\usepackage{dcolumn} 
\usepackage{bm} 
\usepackage[pdftex]{hyperref} 
\usepackage{ulem}

\usepackage{graphicx}
\usepackage{curves}
\usepackage{epic}
\usepackage{wasysym}
\usepackage{epsf,times}

\newcommand{\bs}[1]{{\boldsymbol{#1}}}

\providecommand{\myopenone}{\leavevmode\hbox{\small1\kern-3.8pt\normalsize1}}

\def\:={\,\raisebox{0.85pt}{.}\hspace{-2.78pt}\raisebox{2.85pt}{.}\!\!=\,}
\def\=:{\,=\!\!\raisebox{0.85pt}{.}\hspace{-2.78pt}\raisebox{2.85pt}{.}\,}

\title{\textbf{Fractional Abelian topological phases of matter for fermions 
in two-dimensional space}}

\author{Christopher Mudry\\
Condensed Matter Theory Group\\
Paul Scherrer Institute, CH-5232 Villigen PSI, Switzerland}

\begin{document}
\maketitle
\tableofcontents


\setcounter{chapter}{7}

\begin{figure}[t]
\begin{center}
\includegraphics[width=0.35\textwidth]{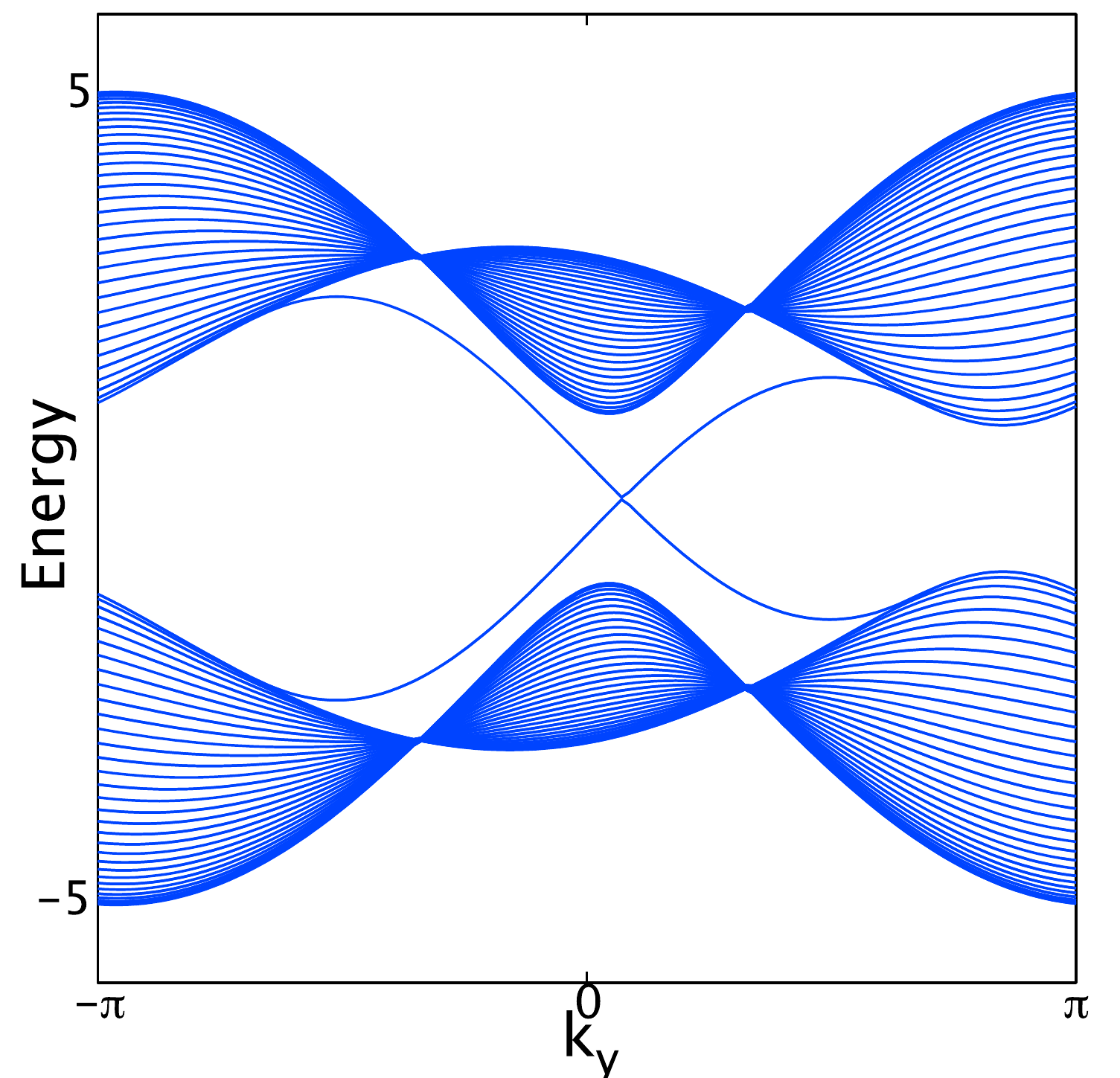}
\end{center}
\caption{
{[Colour online]} 
Single-particle spectrum of a Bogoliubov-de-Gennes  
superconductor in a cylindrical geometry which is the
direct sum of a $p^{\,}_{x}+\mathrm{i}p^{\,}_{y}$
and of a $p^{\,}_{x}-\mathrm{i}p^{\,}_{y}$
superconductor.
A two-fold degenerate dispersion of two chiral edge
states is seen to cross the mean-field superconducting gap.
There is a single pair of Kramers degenerate edge state
that disperses along one edge of the cylinder.
(Adapted from  Ref.~{}\protect\cite{Chang14}.
\textcopyright
IOP Publishing. Reproduced with Permission. All rights reserved.)
\label{fig: example edge states}
        }
\end{figure}

\section{Introduction}

In these lectures, I will focus exclusively on 
two-dimensional realizations of fractional topological
insulators. However, before doing so, I need to revisit the definition
of non-interacting topological phases of matter for fermions and, 
for this matter, I would like to attempt to place some of the recurrent
concepts that have been used during this school on a time line that starts 
in 1931.

Topology in physics enters the scene in 1931 when Dirac 
\cite{Dirac31} showed that
the existence of magnetic monopoles in quantum mechanics implies the 
quantization of the electric and magnetic charge.

In the same decade, Tamm \cite{Tamm32}
and Shockley \cite{Shockley39} surmised from the band theory of Bloch
that surface states can appear at the boundaries of band insulators
(see Fig.\ \ref{fig: example edge states}).

\begin{figure}
\begin{center}
(a)
\includegraphics[width=0.45\textwidth]{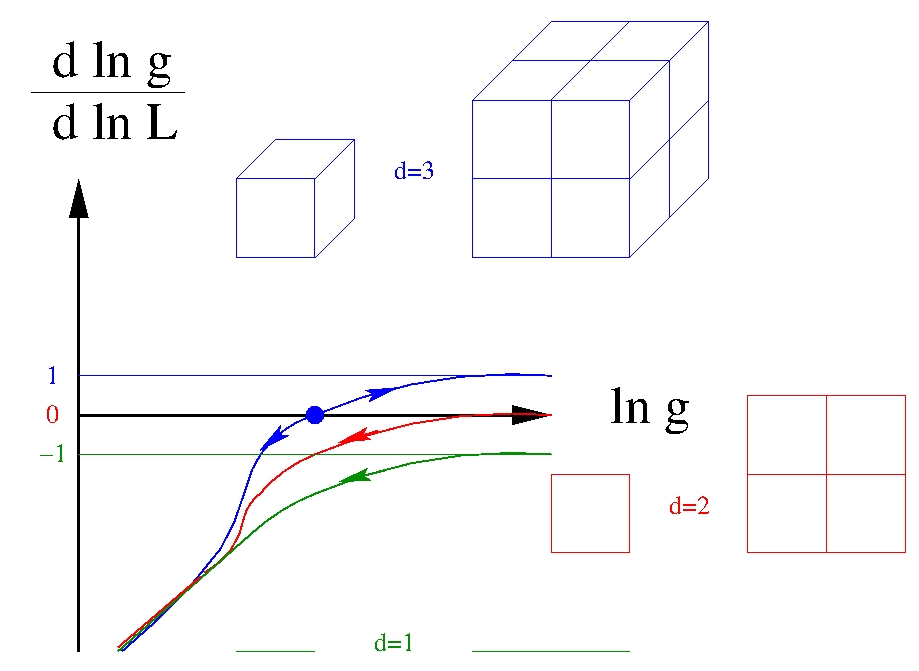}
\hfill
(b)
\includegraphics[width=0.45\textwidth]{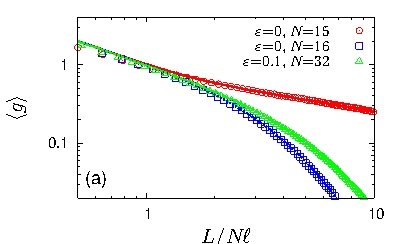}
\end{center}
\caption{
{[Colour online]}
(a) The beta function of the dimensionless conductance $g$ 
is plotted (qualitatively) as a function of the linear system size $L$ 
in the orthogonal symmetry class
($\beta=1$) for space dimensions $d=1$, $d=2$, and $d=3$.
(b) The dependence of the mean Landauer conductance $\langle g\rangle$ 
for a quasi-one-dimensional wire
as a function of the length of the wire $L$ in the symmetry class BD1.
The number $N$ of channel is varied as well as the chemical potential
$\varepsilon$ of the leads. 
(Part (b) taken from \protect\cite{Brouwer05}.)
\label{fig: gang of four and g for Q1d BD1}
       }
\end{figure}

The dramatic importance of static and local disorder 
for electronic quantum transport had
been overlooked until 1957 when Anderson \cite{Anderson58}
showed that sufficiently strong 
disorder ``generically'' localizes a bulk electron.
That there can be exceptions to this rule follows from reinterpreting
the demonstration in 1953 by Dyson \cite{Dyson53} 
that disordered phonons in a linear
chain can acquire a diverging density of states at zero energy
with the help of bosonization tools in one-dimensional space
(see Fig.\ \ref{fig: gang of four and g for Q1d BD1}).

Following the proposal by Wigner to model nuclear interactions with the
help of random matrix theory, Dyson \cite{Dyson62} in 1962
introduced the threefold way,
i.e., the study of the joint probability distribution 
\begin{equation}
P(\theta^{\,}_{1},\cdots,\theta^{\,}_{N})
\propto
\prod_{1\leq j<k\leq N}
\left|
e^{\mathrm{i}\theta^{\,}_{j}}
-
e^{\mathrm{i}\theta^{\,}_{k}}
\right|^{\beta},
\qquad
\beta=1,2,4,
\end{equation}
for the eigenvalues of unitary matrices of rank $N$
generated by random Hamiltonians without any symmetry $(\beta=2)$,
by random Hamiltonians with time-reversal symmetry corresponding to spin-$0$ 
particles $(\beta=1)$, and 
by random Hamiltonians with time-reversal symmetry corresponding to spin-$1/2$ 
particles $(\beta=4)$.

\begin{figure}[t]
\begin{center}
(a)
\includegraphics[width=0.35\textwidth,angle=0]{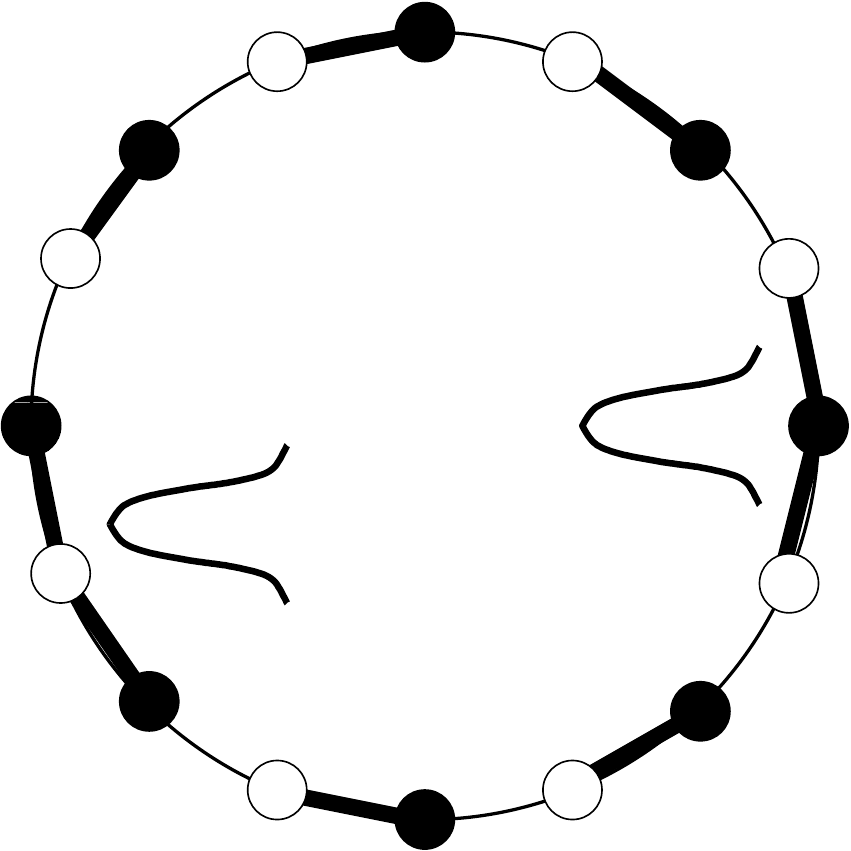}
\hfill
(b)
\includegraphics[width=0.35\textwidth,angle=0]{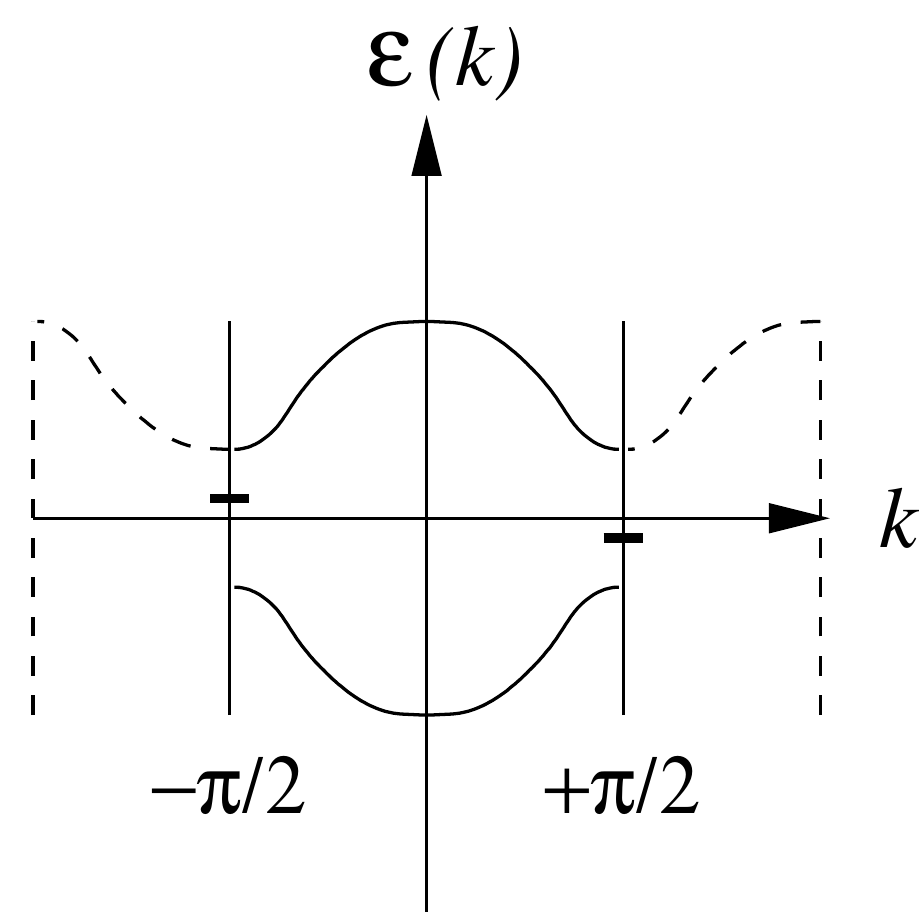}
\end{center}
\caption{
{(Taken from \protect\cite{Ryu12}.)}
(a)
Nearest-neighbor hopping of a spinless fermion along a ring with a 
real-valued hopping amplitude that is larger on the thick bonds than 
on the thin bonds. There are two defective sites, 
each of which are shared by two thick bonds.
(b) The single-particle spectrum is gapped at half-filling.
There are two bound states within this gap, 
each exponentially localized around one of the defective sites, 
whose energy is split from the band center by an
energy that decreases exponentially fast with the separation of the two
defects.
\label{fig: Jackiw+Rebbi zero mode}
        }
\end{figure}

Topology acquired a mainstream status in physics in 1973
with the disovery of Berezinskii 
\cite{Berezinski71}
and of Kosterlitz and Thousless 
\cite{Kosterlitz73,Kosterlitz74}
that topological defects 
in magnetic classical textures can drive a phase transition.
In turn, there is an intimate connection between topological defects
of classical background fields in the presence of which electrons
propagate and fermionic zero modes, as was demonstrated by Jackiw and Rebbi
\cite{Jackiw76}
in 1976 (see Fig.\ \ref{fig: Jackiw+Rebbi zero mode});
{see also the work by Su, Schrieffer, and Heeger \cite{Schrieffer80}}.

The 1970's witnessed the birth of lattice gauge theory as a mean to
regularize quantum chromodynamics (QCD${}^{\,}_{4}$). Regularizing
the standard model on the lattice proved to be more difficult
because of the Nielsen-Ninomiya no-go theorem
that prohibits defining a theory of chiral fermions on a lattice 
in odd-dimensional space without violating locality or time-reversal
symmetry \cite{Nielsen81a,Nielsen81b,Nielsen81c}.
This is known as the fermion-doubling problem when regularizing the
Dirac equation in $d$-dimensional space on a $d$-dimensional lattice.

\begin{figure}[t]
\begin{center}
(a)
\includegraphics[width=0.25\textwidth,angle=0]{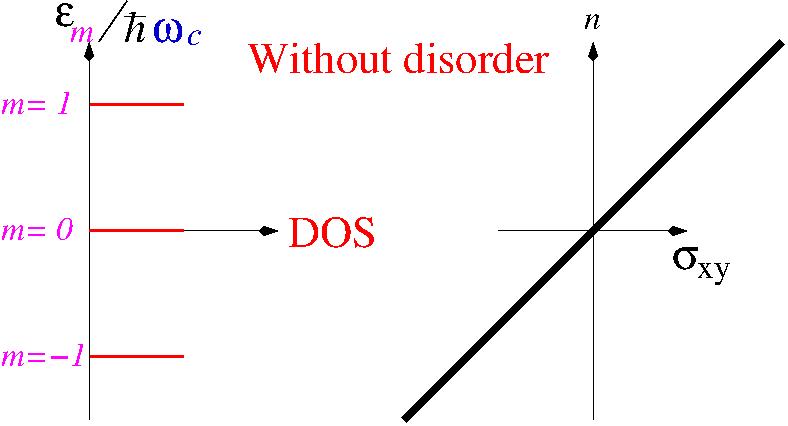}
\hfill
(b)
\includegraphics[width=0.25\textwidth,angle=0]{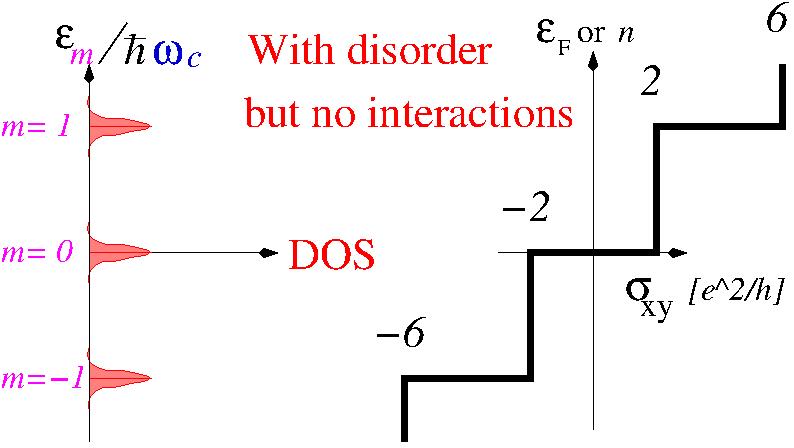}
\hfill
(c)
\includegraphics[width=0.3\textwidth]{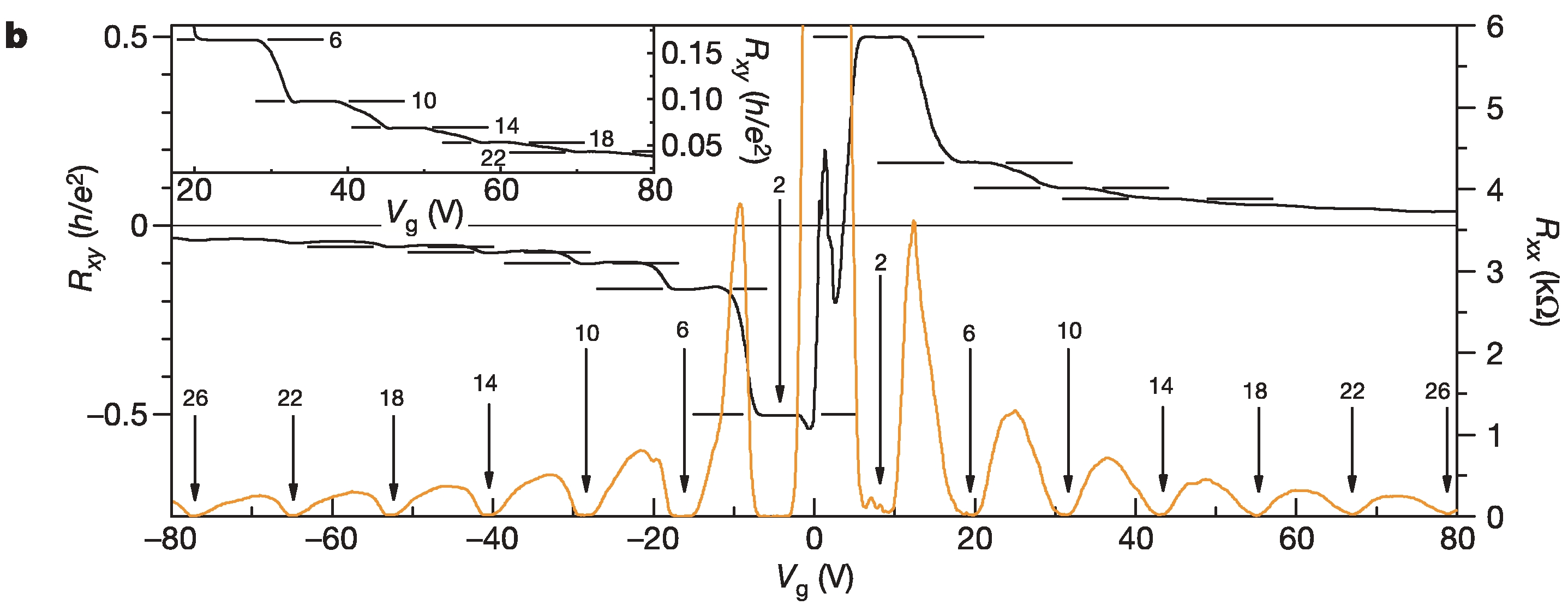}
\end{center}
\caption{
{[Colour online]}
(a)
The Hall conductivity is a linear function of the electron density
if Galilean invariance holds.
(b)
Galilean invariance is broken in the presence of disorder,
so that plateaus become evident
at integral filling fractions of the Landau levels.
(c)
Graphene deposited on SiO${}_2$/Si, 
$T$=1.6 K and $B$=9 T (inset $T$=30 mK)
support the integer quantum Hall effect
at the filling fractions
$\nu=\pm2,\pm6,\pm10,\cdots=\pm2(2n+1)$, $n\in\mathbb{N}$.
(Part (c) taken from \cite{Zhang05}. Reprinted by permission from
Macmillan Publishers Ltd: Nature, copyright 2005.)
\label{eq: IQHE in graphene 2005}
        }
\end{figure}

The 1980's opened with a big bang. 
The integer quantum Hall effect (IQHE) was discovered in 1980 by
von Klitzing, Dorda, and Pepper \cite{Klitzing80}
(see Fig.\ \ref{eq: IQHE in graphene 2005}),
while the fractional quantum Hall effect (FQHE) was discovered
in 1982 by Tsui, Stormer, and Gossard \cite{Tsui82}.
At integer fillings of the Landau levels, the non-interacting
ground state is unique and the screened Coulomb interaction
$V^{\,}_{\mathrm{int}}$
can be treated perturbatively,
as long as transitions between Landau levels
or outside the confining potential $V^{\,}_{\mathrm{conf}}$
along the magnetic field
are suppressed by the single-particle gaps 
$\hbar\,\omega^{\,}_{\mathrm{c}}$ and 
$V^{\,}_{\mathrm{conf}}$, respectively,
\begin{equation}
V^{\,}_{\mathrm{int}}\ll 
\hbar\,\omega^{\,}_{\mathrm{c}}\ll
V^{\,}_{\mathrm{conf}},
\qquad
\omega^{\,}_{\mathrm{c}}=
{e\,B}/{(m\,c)}.
\end{equation}
When Galilean invariance is not broken,
the conductivity tensor is then given by the classical Drude formula 
\begin{equation}
\lim_{\tau\to\infty}\bs{j}=
\begin{pmatrix}
0
&
+
\left(B\,R^{\,}_{\text{H}}\right)^{-1}
\\
-
\left(B\,R^{\,}_{\text{H}}\right)^{-1}
&
0
\end{pmatrix}
\bs{E},
\qquad
R^{-1}_{\text{H}}\:=
-n\,e\,c,
\end{equation}
that relates the (expectation value of the)
electronic current density $\bm{j}\in\mathbb{R}^{2}$ 
to an applied electric field $\bm{E}\in\mathbb{R}^{2}$
within the plane perpendicular to the applied static and uniform magnetic
field $\bm{B}$ 
in the ballistic regime ($\tau\to\infty$ is the scattering time).
The electronic density,
the electronic charge,
and the speed of light are denoted $n$, $e$, and $c$, respectively.
Moderate disorder is an essential ingredient to observe the
IQHE, for it allows the Hall conductivity to
develop plateaus at sufficiently low temperatures that are readily
visible experimentally (see Fig.\ \ref{eq: IQHE in graphene 2005}). 
These plateaus are a consequence of the fact that
most single-particle states in a Landau level are localized by disorder,
according to Anderson's insight 
that any quantum interference induced by a static and local disorder
almost always lead to localization in one- and two-dimensional space.
The caveat ``almost'' is crucial here, for the very observation of transitions
between Landau plateaus implies that not all single-particle Landau levels
are localized. 

\begin{figure}[t]
\begin{center}
\includegraphics[scale=0.25,angle=0]{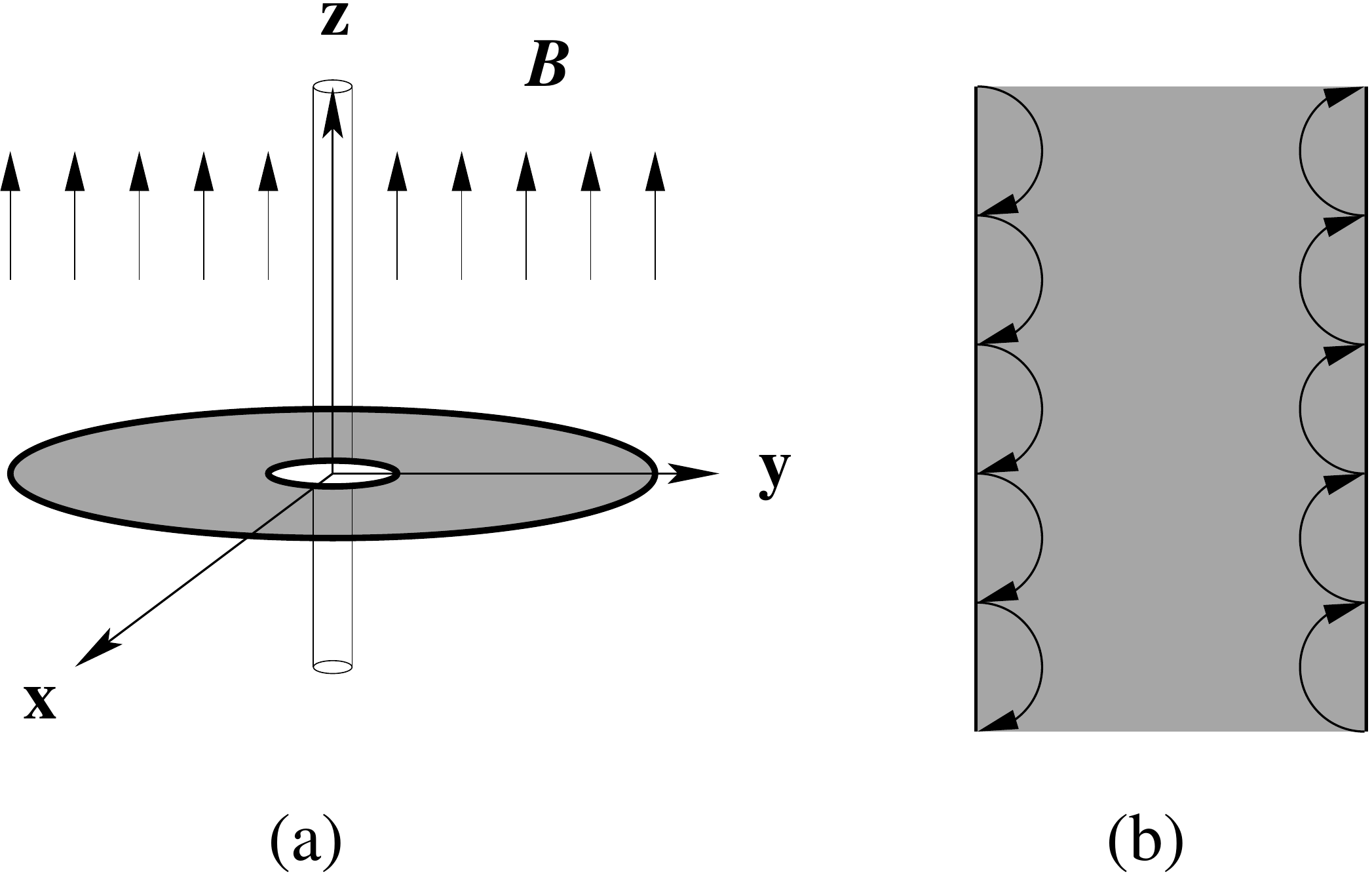}
\hfill
(c)
\includegraphics[scale=0.25,angle=0]{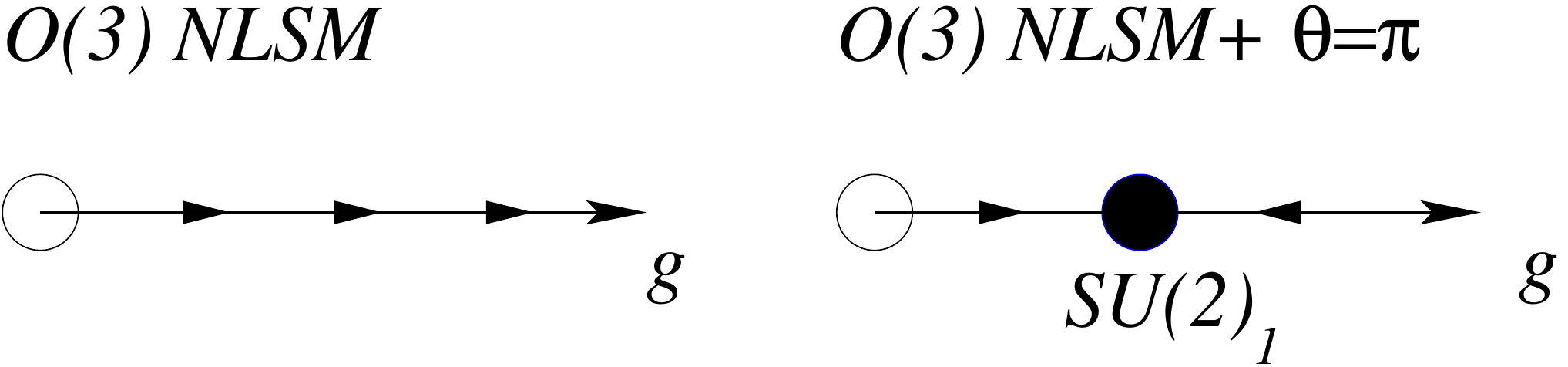}
\end{center}
\caption{
{[Colour online]}
{
(a) A gas of electrons confined to the geometry of a punctured disk
is subjected to a uniform magnetic field $\boldsymbol{B}$ 
perpendicular to the disk.
(b) The classical skipping orbits of electrons confined to 
the geometry of a Hall bar. The Hall-bar geometry of panel (b)
is topologically equivalent to the Corbino geometry of panel (a). 
(c) }
Chiral edges are immune to backscattering within each traffic lane.
\label{Chiral edges states}
        }
\end{figure}

The explanation for the integer quantum Hall effect followed quickly 
its discovery owing to a very general argument of Laughlin \cite{Laughlin81}
based on gauge invariance that implies that the Hall conductivity must take a
fractional value if the longitudinal conductivity vanishes
(mobility gap).
This argument was complemented by an argument of Halperin \cite{Halperin82}
stressing the crucial role played by edge states when electrons
in the quantum Hall effect are confined to a strip geometry
(see Fig.\ \ref{Chiral edges states}),
while works from Thouless, Kohmoto, Nightingale, den Nijs and Niu
\cite{Thouless82,Simon83,Niu84,Niu85} 
demonstrated that the Hall response is, 
within linear response theory,
proportional to the topological invariant 
\begin{equation}
C\:=
-
\frac{\mathrm{i}}{2\pi}
\int\limits_{0}^{2\pi}\mathrm{d}\phi
\int\limits_{0}^{2\pi}\mathrm{d}\varphi
\left[
\left\langle
\frac{\partial\Psi}{\partial\phi}
\right|
\left.
\frac{\partial\Psi}{\partial\varphi}
\right\rangle
-
\left\langle
\frac{\partial\Psi}{\partial\varphi}
\right|
\left.
\frac{\partial\Psi}{\partial\phi}
\right\rangle
\right]
\end{equation}
that characterizes 
the many-body ground state $|\Psi\rangle$ obeying twisted boundary conditions 
in the quantum Hall effect.
Together, these arguments constitute the first example of 
the bulk-edge correspondence with observable consequences, namely the
distinctive signatures of both the
integer and the fractional quantum Hall effect.

\begin{figure}[t]
\begin{center}
(a)
\includegraphics[width=0.4\textwidth]{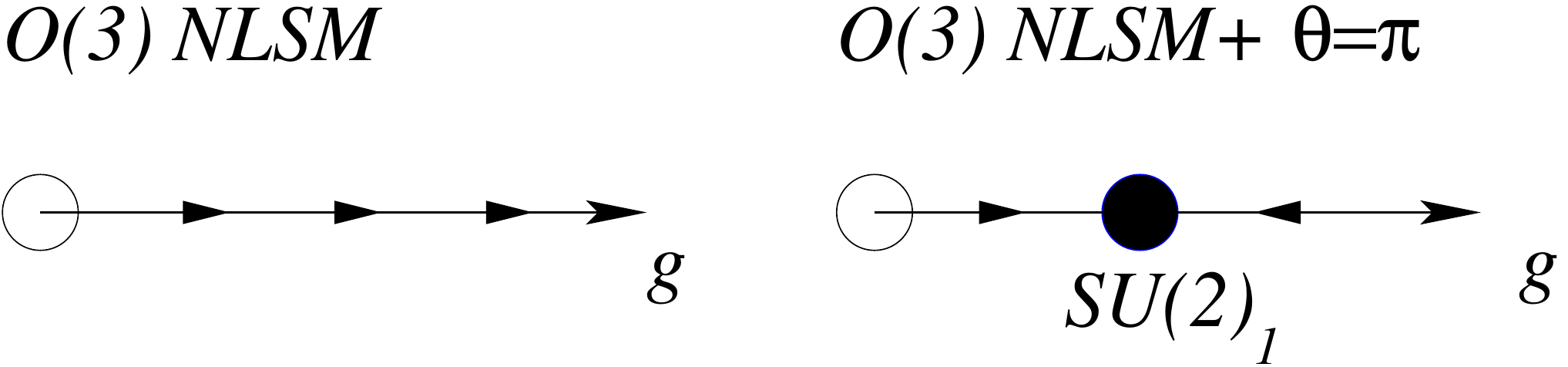}
\hfill
(b)
\includegraphics[scale=0.25,angle=0]{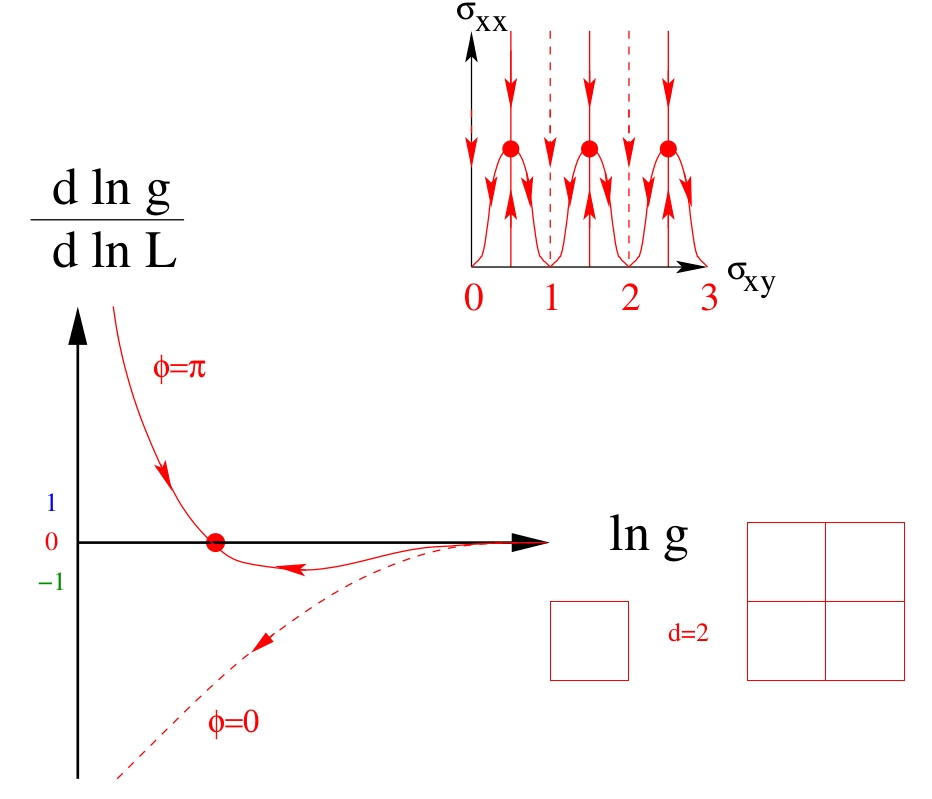}
\end{center}
\caption{
{[Colour online]}
(a) A topological $\theta=\pi$ term modifies the RG flow to strong coupling
in the two-dimensional $O(3)$ non-linear-sigma model.
There exists a stable critical point at intermediary coupling that
realizes the conformal field theory $SU(2)^{\,}_{1}$.
(b) Pruisken argued that the phenomenological two-parameter flow diagram of
Khmelnitskii is a consequence of augmenting the NLSM in the unitary
symmetry class by a topological term.
\label{fig: Haldane conjecture and Pruisken term}
        }
\end{figure}

The transitions between plateaus in the quantum Hall effect are
the manifestations at finite temperature and for a system of finite size
of a continuous quantum phase transition, i.e., of a singular dependence
of the conductivity tensor on the magnetic field (filling fraction)
that is rounded by a non-vanishing temperature or by the finite linear size
of a sample. In the non-interacting limit,
as was the case for the Dyson singularity at the band center, an isolated
bulk single-particle state must become critical in the presence 
of not-too strong disorder. The one-parameter
scaling theory of Anderson localization that
had been initiated by Wegner
and was encoded by a class on non-linear-sigma models (NLSMs)
has to be incomplete \cite{Wegner76,Abrahams79,Hikami80}.
Khmelnitskii 
\cite{Khmelnitskii83}, on the one hand, and 
Levine, Libby, and Pruisken
\cite{Levine83,Pruisken84}, on the other hand,
introduced in 1983 a two-parameter scaling theory for the IQHE
on phenomenological grounds.
They also argued that the NLSM for the IQHE, 
when augmented by a topological $\theta$ term, would reproduce
the two-parameter flow diagram
(see Fig.\ \ref{fig: Haldane conjecture and Pruisken term}).
This remarkable development took place simultaneously with the
works on Haldane \cite{Haldane83a,Haldane83b} 
on encoding the difference between half-integer and integer
spin chains (Haldane's conjecture) by the presence of a $\theta=\pi$ 
topological term in the $O(3)$ NLSM 
and by the work of Witten \cite{Witten84} on principal chiral
models augmented by a Wess-Zumino-Novikow-Witten (WZNW) term.

\begin{figure}[t]
\begin{center}
(a)
\includegraphics[angle=0,width=0.25\textwidth]{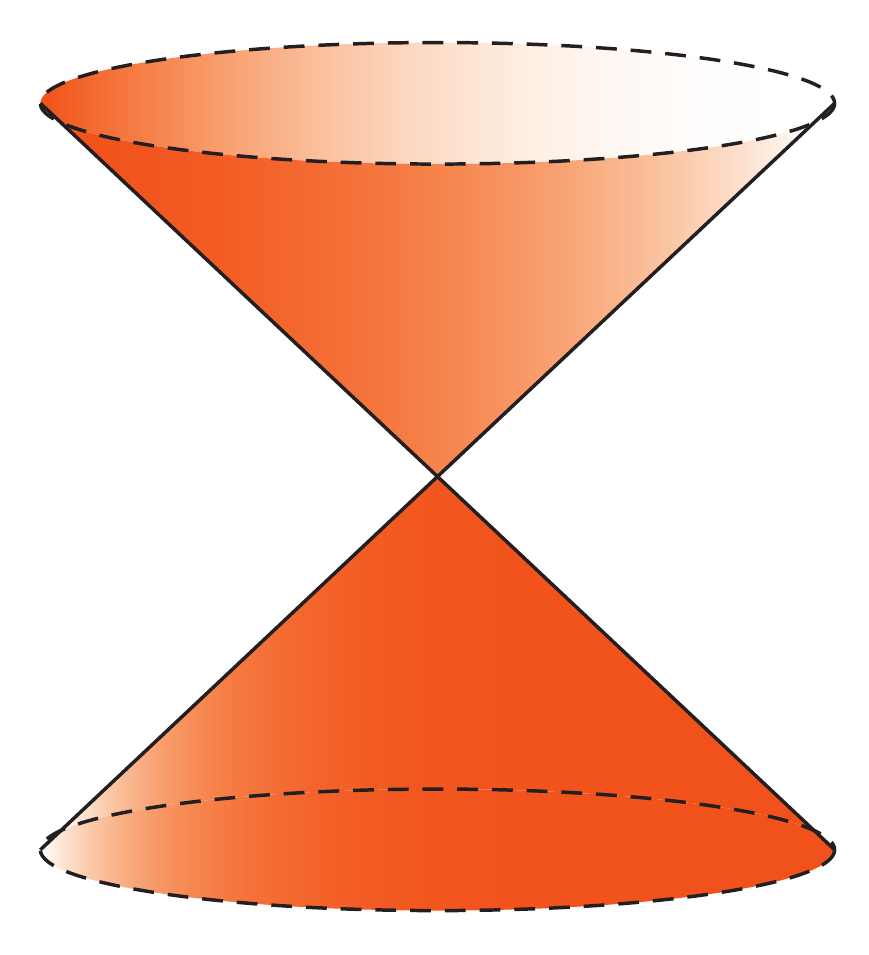}
\hfill
(b)
\includegraphics[angle=0,width=0.35\textwidth]{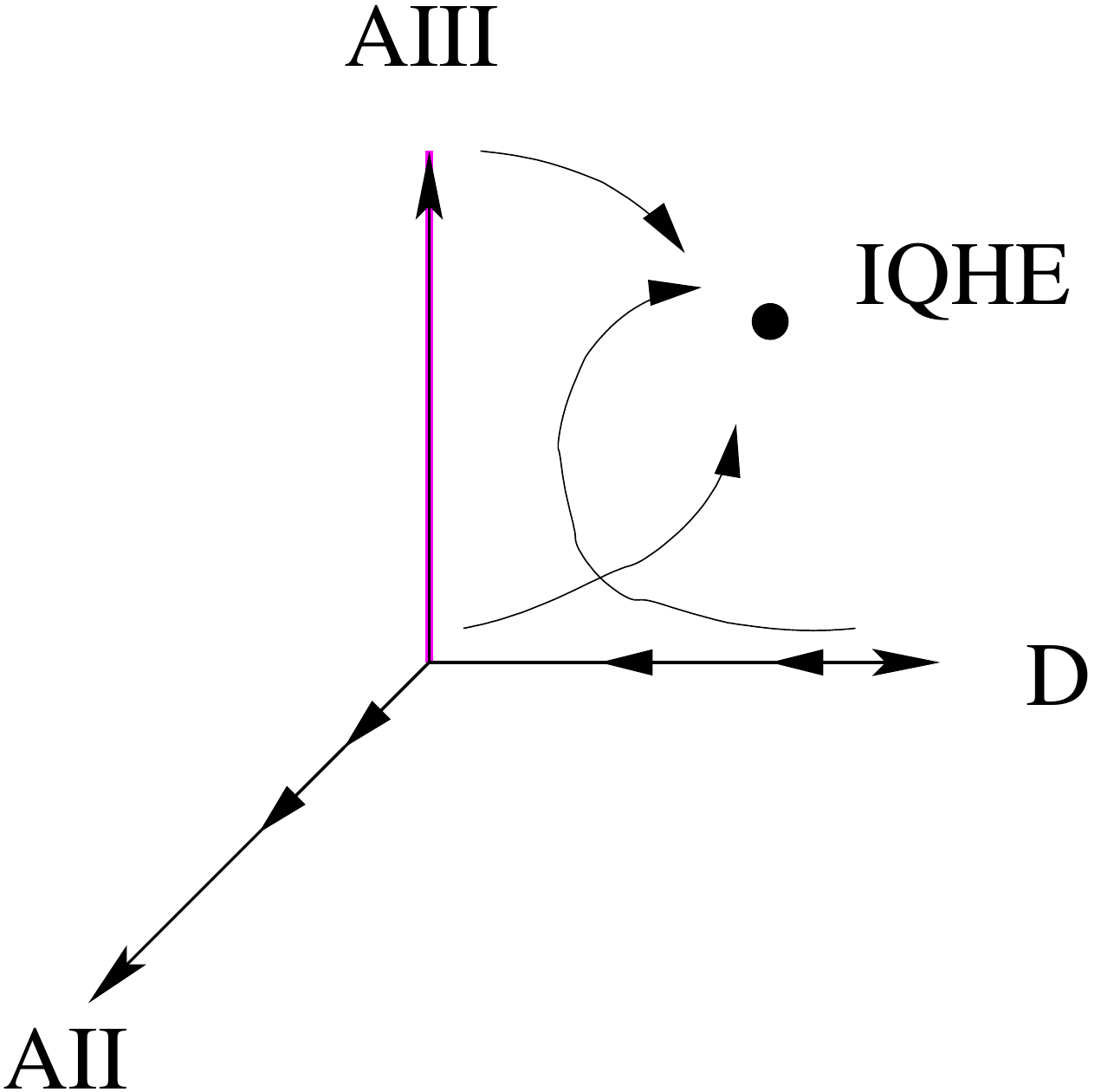}
\\
(c)
\includegraphics[angle=0,width=0.25\textwidth]{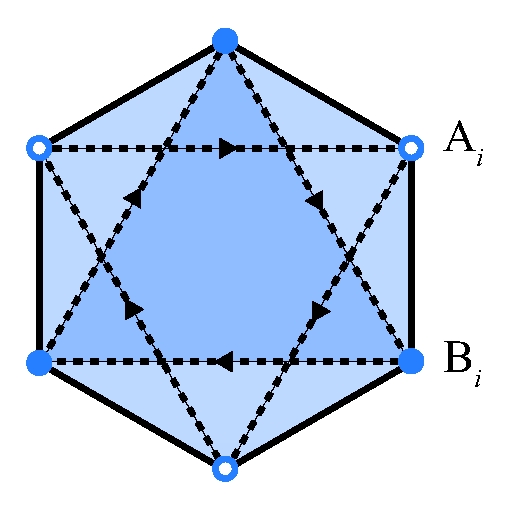}
\hfill
(d)
\includegraphics[angle=0,width=0.4\textwidth]{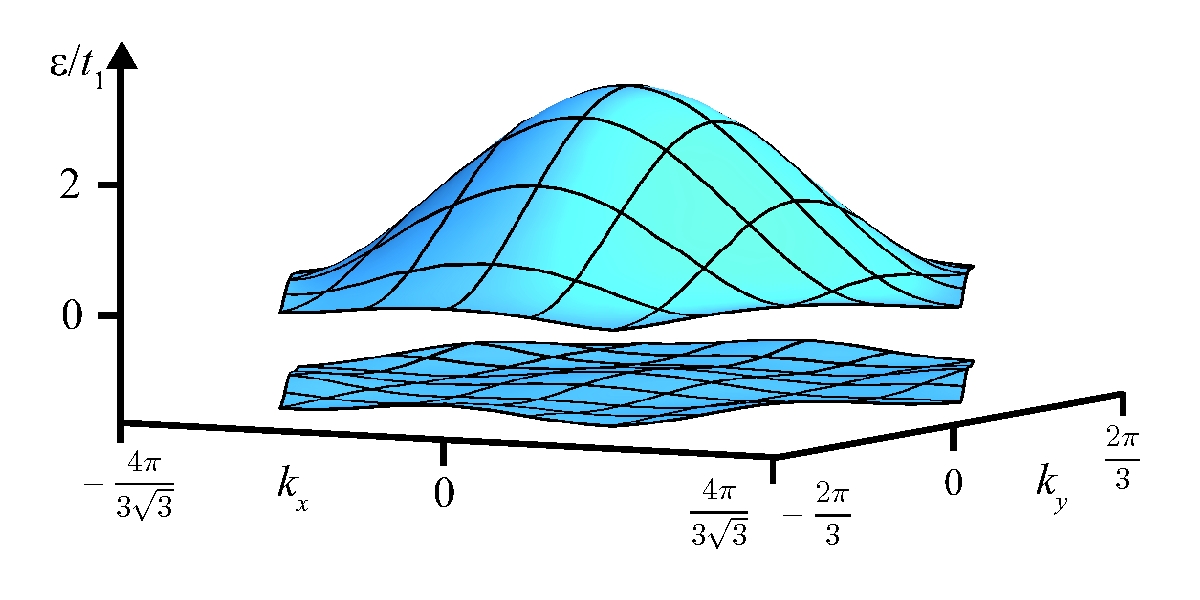}
\end{center}
\caption{
(a)
A single (non-degenerate) cone of Dirac fermions in two-dimensional space
realizes a critical point between two massive phases of Dirac fermions,
each of which carries the Hall conductance 
$\sigma^{\mathrm{Dirac}}_{\mathrm{H}}=\pm(1/2)$ in units of $e^{2}/h$.
(b)
A generic static and local random perturbation
of a single Dirac cone is encoded by three channels. 
There is a random vector potential that realizes the
symmetry class AIII if it is the only one present.
There is a random scalar potential that realizes the
symmetry class AII if it is the only one present.
There is a random mass that realizes the 
symmetry class D if it is the only one present.
It is conjectured in \protect\cite{Ludwig94} that
in the presence of all three channels, the renormalization group
flow to strong coupling (the variance of the disorder in each channel) 
is to the plateau transition in the universality class of the IQHE.
(c)
Unit cell of the honeycomb lattice with the pattern of 
nearest- and next-nearest-neighbor hopping amplitude that realizes
a Chen insulators with two bands, shown in (d),
each of which carries the Chern number $\pm1$.
\label{Fig: random Dirac fermions}
        }
\end{figure}

Deciphering the critical theory for the plateau transition is
perhaps the most tantalizing challenge in the theory of Anderson
localization. Among the many interesting avenues that have been proposed
to reach this goal (that remains elusive so far), 
Ludwig, Fisher, Shankar, and Grinstein \cite{Ludwig94}
studied random Dirac fermions in two-dimensional space
in 1994 (see Fig.\ \ref{Fig: random Dirac fermions}),
motivated as they were
by the fact that a massive Dirac fermion in two-dimensional space carries 
the fractional value 
\begin{equation}
\sigma^{\mathrm{Dirac}}_{\mathrm{H}}=
\pm\frac{1}{2}\,\frac{e^{2}}{h}
\end{equation}
according to Deser, Jackiw, and Templeton \cite{Deser82}
and that it is possible to regularize two such massive Dirac fermions on
a two-band lattice model realizing a Chern insulator according to Haldane
\cite{Haldane88}.

The early 90's were also the golden age of mesoscopic physics, the application
of random matrix theory to condensed matter physics. The threefold way had
been applied successfully to quantum dots and quantum transport in
quasi-one-dimensional geometries. Zirnbauer \cite{Zirnbauer96}
in 1996 and Altland and Zirnbauer \cite{Altland97}
in 1997 extended the threefold way of Dyson
to the tenfold way by including three symmetry classes of relevance
to quantum chromodynamics called the chiral classes, and four
symmetry classes of relevance to superconducting quantum dots
(see Table \ref{TableSymmetryClassesTwo}) \cite{Heinzner05}.
Quantum transport in quasi-one-dimensional wires belonging to the 
chiral and superconducting classes
was studied by Brouwer, Mudry, Simons, and Altland
and by Brouwer, Furusaki, Gruzberg, Mudry, respectively
\cite{Brouwer98,Brouwer00a,Brouwer00c,Brouwer05}
(see Fig.\ \ref{fig: Brown walk on symmetric spaces}).
Unlike in the threefold way, the three chiral symmetry classes
and two of the four superconducting classes 
(the symmetry classes D and DIII)
were shown to realize quantum critical point separating localized phases
in quasi-one-dimensional arrays of wires.
The diverging nature 
of the density of states at the band center 
(the disorder is of vanishing mean) 
for five of the ten symmetry classes
in Table \ref{Table: Butsuri}
is a signature of topologically protected zero modes 
bound to point defects. These point defects are
vanishing values of an order parameter (domain walls). 
{The order parameter, if translation symmetry holds, 
is here also responsible for a spectral gap.}

Lattice realizations of $\mathbb{Z}^{\,}_{2}$ topological band 
insulators in two-dimensional space were proposed by Kane and Mele
\cite{Kane05a,Kane05b}
and in three-dimensional space by Moore and Balents 
\cite{Moore07},
Roy
\cite{Roy09},
and {Fu, Kane, and Mele}
\cite{Fu07a}.
This theoretical discovery initiated in \cite{Ryu07b,Ryu12}
the search of Dirac Hamiltonians
belonging to the two-dimensional symmetry classes AII and CII
from Table \ref{TableSymmetryClassesTwo}
for which the corresponding NLSM 
encoding the effects of static and local disorder 
were augmented by a topological term so as to evade
Anderson localization on the boundary of a $d=3$-dimensional topological
insulators. Following this route for all symmetry classes
and for all dimensions, Ryu, Schnyder, Furusaki, and Ludwig
\cite{Schnyder08,Schnyder09,Ryu10}  
arrived at the periodic Table \ref{periodic table}.
The same table was derived independently by Kitaev \cite{Kitaev09}
using a mathematical construction
known as K theory that he applied to gapped Hamiltonians 
in the bulk (upon the imposition of 
periodic boundary conditions, say) in the clean limit.
This table specifies in any given dimension $d$ of space,
for which symmetry classes it is possible to realize a many-body
ground state for non-interacting fermions subject to a static and local
disorder such that all bulk states are localized but there exist
a certain (topological) number of boundary states,  
that remain delocalized.

The goals of these lectures are the following:
first, to rederive the tenfold way 
for non-interacting fermions in the presence
of local interactions and static local disorder, and,
second, to decide if interactions between fermions
can produce topological phases of matter with protected boundary states
that are not captured by the tenfold way.
This program will be applied in two-dimensional space.

Section \ref{sec: The tenfold way in quasi-one-dimensional space}
motivates the tenfold way by deriving it explicitly 
in quasi-one-dimensional space.
Section \ref{sec: Fractionalization from Abelian bosonization}
reviews Abelian chiral bosonization, a technical tool
allowing one to go beyond the tenfold way
to incorporate the effects of many-body interactions.
Abelian chiral bosonization is applied in Section
\ref{sec: Stability analysis for the edge theory in the symmetry class AII}
to demonstrate the stability of the gapless helical edge states in
the symmetry class AII in the presence of disorder and many-body interactions
and then in Section
\ref{sec: Construction of two-dimensional topological phases from coupled wires}
to construct microscopically long-ranged entangled phases of 
two-dimensional quantum matter.
{
Sections 
\ref{sec: Stability analysis for the edge theory in the symmetry class AII}
and
\ref{sec: Construction of two-dimensional topological phases from coupled wires}
follow the presentations made in
\cite{Neupert11b}
and
\cite{Neupert14},
respectively.}

\begin{table}[t]
\caption{
\label{TableSymmetryClassesTwo}
The ten Altland-Zirnbauer (AZ) symmetry classes of single-particle
Hamiltonians $\mathcal{H}$, 
classified according to their behavior under time-reversal symmetry 
$\mathcal{T}$,
charge-conjugation (particle-hole) symmetry 
$\mathcal{C}$, 
and `sublattice'
(`chiral') symmetry $\mathcal{S}$.
The entries in the columns headed T, C, and S indicate 
the presence/absence of these respective symmetries as well as their types.
These operations square to $+$ or $-$times the unit operator
when they are symmetries. An entry $0$ indicates that the operation
is not a symmetry. The column `Hamiltonian' lists, 
for each AZ symmetry classes, 
the symmetric space of which the quantum mechanical time-evolution operator 
$\exp(\mathrm{i}t\,\mathcal{H})$
is an element.
The column `Cartan label' shows the name given to the corresponding symmetric 
space listed in the column ‘Hamiltonian’
in \'Elie Cartan’s classification scheme (dating back to the year 1926). 
The column `$G/H$ (fermionic NLSM)' lists the
(compact sectors of the) target space of the NLSM describing 
{the physics of} Anderson localization
at long wavelength in the given symmetry class.
       }
\begin{scriptsize}
\begin{tabular}{c||c|c|c||c||c}
{\small  Cartan label} & 
$\mathrm{T}$&$\mathrm{C}$ &
$\mathrm{S}$ & 
{\small Hamiltonian}    & 
$G/H$ (fermionic NLSM)
\\\hline\hline
A {\small(unitary)} &
$0$ &
$0$ &
$0$ &
$U(N)$ &
$U(2n)/U(n)\times U(n)$ 
\\ \hline
AI {\small(orthogonal)} &
$+1$ &
$0$ &
$0$ &
$U(N)/O(N)$&
$Sp(2n)/Sp(n)\times Sp(n)$ 
\\ \hline
AII {\small(symplectic)} &
$-1$ &
$0$ &
$0$ &
$U(2N)/Sp(2N)$ &
$O(2n)/O(n)\times O(n)$
\\\hline\hline
AIII {\small(ch. unit.)} &
$0$ &
$0$ &
$+1$ &
${U(N+M)/U(N)\times U(M)}$ &
$U(n)$
\\\hline 
 BDI {\small(ch. orth.)} &
$+1$ &
$+1$ &
$+1$ &
$O(N+M)/O(N)\times O(M)$ & 
$U(2n)/Sp(2n)$
\\\hline 
CII {\small(ch. sympl.)} &
$-1$ &
$-1$ &
$+1$ &
$Sp(N+M)/Sp(N)\times Sp(M)$ & 
$U(2n)/O(2n)$ 
\\\hline\hline
D ({\small BdG}) &
$0$ &
$+1$ &
$0$ &
$\mathrm{SO}(2N)$ &
$O(2n)/U(n)$
\\\hline
C ({\small BdG}) &
$0$ &
$-1$ &
$0$ &
$Sp(2N)$ & 
$Sp(2n)/U(n)$ 
\\\hline
DIII ({\small BdG}) &
$-1$ &
$+1$ &
$-1$ & 
$\mathrm{SO}(2N)/U(N)$ & 
$O(2n)$ 
\\\hline
CI ({\small BdG}) &
$+1$ &
$-1$ &
$-1$ &
$Sp(2N)/U(N)$ & 
$Sp(2n)$
\end{tabular}
\end{scriptsize}
\end{table}

\newpage
\begin{table}[t]
\caption{
\label{Table: Butsuri}
AZ symmetry classes for disordered quantum wires. Symmetry classes are
defined by the presence or absence of time-reversal symmetry and
spin-rotation symmetry (SRS) and by the single-particle spectral
symmetries of sublattice symmetry (SLS) (random hopping model at the
band centre) also known as chiral symmetries, and particle-hole
symmetry (zero-energy quasiparticles in superconductors). For
historical reasons, the first three rows are referred to as the
orthogonal (O), unitary (U), and symplectic (S) symmetry classes when
the disorder is generic. A prefix ‘ch’ (‘chiral’) is added when the
disorder respects a SLS, as in the next three rows. The last four rows
correspond to dirty superconductors and are named after the symmetric
spaces associated with their Hamiltonians. 
$m^{\,}_{o\pm}$ and $m^{\,}_{l}$
are the multiplicities of the ordinary and long roots of the symmetric spaces
associated with the transfer matrix. For the three chiral classes,
$m^{\,}_{o+}=0$, $m^{\,}_{o-}=m^{\,}_{o}$;
otherwise, $m^{\,}_{o+} =m^{\,}_{o-}=m^{\,}_{o}$.
$D$ is the degeneracy of the transfer matrix eigenvalues. 
The next two columns show the symbols for the symmetric spaces associated 
with the transfer matrix 
$\mathcal{M}$ andthe Hamiltonian $\mathcal{H}$.
With $g$ denoting the dimensionless Landauer conductance and 
$\rho(\varepsilon)$ the (self-averaging) density of states (DOS)
per unit energy and per unit length,
the last three columns list theoretical results for 
the weak-localization correction $\delta g$ for $\ell\ll L\ll N\ell$,
the disorder average $\overline{\ln g}$ of $\ln g$ for $L\gg N\ell$, 
and the DOS near $\varepsilon=0$. 
The results for $\overline{\ln g}$ and $\rho(\varepsilon)$
in the chiral classes are for even $N$. For odd $N$,
$\overline{\ln g}$ and $\rho(\varepsilon)$ are the same as in class D. 
       }
\begin{footnotesize}
\begin{tabular}{ccccccccc}
Symmetry class
&
$m^{\,}_{o}$
& 
$m^{\,}_l$ 
& 
$D$ 
& 
$\mathcal{M}$ 
& 
$\mathcal{H}$
&
$\delta g$
&
$-\overline{\ln g}$
&
$\rho(\varepsilon)$ for $0<\varepsilon \tau_c \ll 1$ 
\\ 
\hline
AI
&
1 
& 
1 
& 
2 
& 
CI
& 
AI
&
$-2/3$
&
$2L/(\gamma\ell)$
&
$\rho^{\,}_{0}$
\\
A
& 
2 
& 
1 
& 
2(1)
& 
AIII
& 
A 
&
0
&
$2L/(\gamma\ell)$
&
$\rho^{\,}_{0}$
\\
AII
& 
4 
& 
1 
& 
2 
& 
DIII
& 
AII
&
$+1/3$
&
$2L/(\gamma\ell)$
&
$\rho^{\,}_{0}$
\\ 
\hline 
BDI
& 
1 
& 
0 
& 
2 
& 
AI
& 
BDI
&
0
&
$2m^{\,}_{o}L/(\gamma\ell)$
& 
$\rho_0 |\ln|\varepsilon\tau_c||$
\\
AIII
& 
2 
& 
0 
& 
2(1) 
& 
A
& 
AIII
&
0
&
$2m^{\,}_{o}L/(\gamma\ell)$
&
$\pi \rho_0 |\varepsilon \tau_c \ln|\varepsilon\tau_c||$ 
\\
CII
& 
4 
& 
0 
& 
2 
& 
AII
& 
CII
&
0
&
$2m^{\,}_{o}L/(\gamma\ell)$
&
$(\pi \rho_0/3) |(\varepsilon \tau_c)^3\ln|\varepsilon\tau_c||$
\\ 
\hline  
CI
& 
2 
& 
2 
& 
4 
& 
C
&
CI 
&
$-{4}/{3}$
&
$2m^{\,}_{l}L/(\gamma\ell)$
&
$(\pi\rho_0/2) |\varepsilon \tau_c|$  
\\
C 
& 
4 
& 
3 
& 
4 
& 
CII
& 
C
&
$-{2}/{3}$
&
$2m^{\,}_{l}L/(\gamma\ell)$
&
$\rho_0 |\varepsilon \tau_c|^2$  
\\
DIII
& 
2 
& 
0 
& 
2 
& 
D& 
DIII
&
$+{2}/{3}$
&
$4\sqrt{L/(2\pi\gamma\ell)}$
&
$\pi \rho_0/|\varepsilon \tau_c \ln^3|\varepsilon\tau_c||$
\\
D
& 
1 
& 
0 
& 
1 
& 
BDI
& 
D
&
$+{1}/{3}$
&
$4\sqrt{L/(2\pi\gamma\ell)}$
&
$\pi \rho_0/|\varepsilon \tau_c \ln^3|\varepsilon\tau_c||$
\end{tabular}
\end{footnotesize}
\end{table} 

\newpage 

\begin{table}[t!]
\caption{\label{periodic table}
Classification of topological insulators and 
superconductors as a function of spatial dimension $d$
and AZ symmetry class, indicated by the ``Cartan label'' (first column).
The definition of the ten AZ symmetry classes of single particle Hamiltonians
is given in Table \protect\ref{TableSymmetryClassesTwo}.
The symmetry classes are grouped in two separate lists
of complex and real cases, respectively, depending on whether
the Hamiltonian is complex or whether one
(or more) reality conditions (arising from
time-reversal or charge conjugation symmetries)
are imposed on it; the AZ symmetry classes are ordered in such a way
that a periodic pattern in dimensionality
becomes visible \protect\cite{Kitaev09}.
{Entries $\mathbb{Z}$, $2\mathbb{Z}$, and $\mathbb{Z}^{\,}_{2}$} 
indicate that the topologically distinct phases within a given 
symmetry class of topological insulators (superconductors) are
{characterized by an integer invariant, 
an even-integer invariant, 
or a $\mathbb{Z}^{\,}_{2}$ quantity, respectively}. 
An entry ‘0’ indicates that there exists
no topological insulator (superconductor), 
i.e., a case where all quantum ground states are topologically
equivalent to the trivial state.
}
\vskip 10 true pt
\textbf{Complex case}
\vskip 10 true pt
\begin{tabular}{cccccccccccccc}\hline
Cartan$\backslash d$ & 0  & 1 & 2 & 3 & 4 & 5 & 6 & 7 & 8 & 9 & 10 & 11 & $\cdots$ \\ \hline\hline
A   & $\mathbb{Z}$ & 0 & $\mathbb{Z}$ & 0 
    & $\mathbb{Z}$ & 0 & $\mathbb{Z}$ & 0
    & $\mathbb{Z}$ & 0 & $\mathbb{Z}$ & 0 
    & $\cdots$
\\\hline
AIII  & 0 & $\mathbb{Z}$ & 0 & $\mathbb{Z}$
      & 0 & $\mathbb{Z}$ & 0 & $\mathbb{Z}$
      & 0 & $\mathbb{Z}$ & 0 & $\mathbb{Z}$
      & $\cdots$
\\\hline
\end{tabular}
\vskip 10 true pt
\textbf{Real case}
\vskip 10 true pt
\begin{tabular}{cccccccccccccc}\hline
Cartan$\backslash d$ & 0  & 1 & 2 & 3 & 4 & 5 & 6 & 7 & 8 & 9 & 10 & 11 & $\cdots$ \\ \hline\hline
AI  & $\mathbb{Z}$ & 0 & 0 
    & 0 & $2\mathbb{Z}$ & 0 
    & $\mathbb{Z}_2$ & $\mathbb{Z}_2$ & $\mathbb{Z}$ 
    & 0 & 0 & 0 & $\cdots$ \\ \hline
BDI & $\mathbb{Z}_2$ & $\mathbb{Z}$ & 0 & 0 
    & 0 & $2\mathbb{Z}$ & 0 
    & $\mathbb{Z}_2$ & $\mathbb{Z}_2$ & $\mathbb{Z}$ 
    & 0 & 0 &  $\cdots$ \\ \hline
D   & $\mathbb{Z}_2$ & $\mathbb{Z}_2$ & $\mathbb{Z}$ 
    & 0 & 0 & 0 
    & $2\mathbb{Z}$  & 0 & $\mathbb{Z}_2$ 
    & $\mathbb{Z}_2$ & $\mathbb{Z}$  & 0 & $\cdots$ \\ \hline
DIII& 0 & $\mathbb{Z}_2$ & $\mathbb{Z}_2$ & $\mathbb{Z}$ 
    & 0 & 0 & 0 
    & $2\mathbb{Z}$  & 0 & $\mathbb{Z}_2$ 
    & $\mathbb{Z}_2$ & $\mathbb{Z}$  &  $\cdots$ \\ \hline
AII & $2\mathbb{Z}$  & 0 & $\mathbb{Z}_2$ 
    & $\mathbb{Z}_2$ & $\mathbb{Z}$ & 0 
    & 0 & 0 & $2\mathbb{Z}$ 
    & 0 & $\mathbb{Z}_2$ & $\mathbb{Z}_2$&  $\cdots$\\ \hline
CII & 0 & $2\mathbb{Z}$  & 0 & $\mathbb{Z}_2$ 
    & $\mathbb{Z}_2$ & $\mathbb{Z}$ & 0 
    & 0 & 0 & $2\mathbb{Z}$ 
    & 0 & $\mathbb{Z}_2$ &  $\cdots$\\ \hline
C   & 0  & 0 & $2\mathbb{Z}$  
    & 0 & $\mathbb{Z}_2$  & $\mathbb{Z}_2$ 
    & $\mathbb{Z}$ & 0 & 0 
    & 0 & $2\mathbb{Z}$ & 0 & $\cdots$ \\ \hline
CI  & 0 & 0  & 0 & $2\mathbb{Z}$  
    & 0 & $\mathbb{Z}_2$  & $\mathbb{Z}_2$ 
    & $\mathbb{Z}$ & 0 & 0 
    & 0 & $2\mathbb{Z}$  & $\cdots$ \\ \hline
\end{tabular}
\end{table}

\section{The tenfold way in quasi-one-dimensional space}
\label{sec: The tenfold way in quasi-one-dimensional space}

This section is dedicated to a non-vanishing density
of non-interacting fermions hopping between the sites
of quasi-one-dimensional lattices or between the sites
defining the one-dimensional boundary of a two-dimensional lattice.
According to the Pauli exclusion principle, the non-interacting ground state
is obtained by filling all the single-particle energy eigenstates
up to the Fermi energy fixed by the fermion density. The fate of this
single-particle energy eigenstate when a static and local random
potential is present is known as the problem of Anderson localization. 
The effect of disorder on a single-particle extended energy eigenstate state 
can be threefold:
\begin{itemize}
\item
The extended nature of the single-particle energy eigenstate 
is robust to disorder.
\item
The extended single-particle energy eigenstate is turned into a critical state.
\item
The extended single-particle energy eigenstate is turned into a localized state.
\end{itemize}
There are several methods allowing to decide which one of these three
outcomes takes place. Irrespectively of the dimensionality $d$ of space, 
the symmetries obeyed by the static and local random potential matter 
for the outcome in a dramatic fashion. 
To illustrate this point, let us consider the problem
of Anderson localization in quasi-one-dimensional space.

\subsection{Symmetries for the case of one one-dimensional channel}
\label{subsec: Symmetries for the case of one one-dimensional channel}

For simplicity, consider first the case of an infinitely long
one-dimensional chain with the lattice spacing $\mathfrak{a}\equiv1$
along which a non-vanishing but finite density of
spinless fermions hop with the uniform nearest-neighbor hopping amplitude $t$.
If periodic boundary conditions are imposed, the single-particle Hamiltonian 
is the direct sum over all momenta $-\pi\leq k\leq+\pi$
within the first Brillouin zone of
\begin{equation}
\mathcal{H}(k)\:= -2t\cos k.
\label{eq: dispersion if uniform nn hopping along chain}
\end{equation}
The Fermi energy $\varepsilon^{\,}_{\mathrm{F}}$
intersects the dispersion 
(\ref{eq: dispersion if uniform nn hopping along chain})
at the two Fermi points
$\pm k^{\,}_{\mathrm{F}}$. Linearization of the dispersion
(\ref{eq: dispersion if uniform nn hopping along chain})
about these two Fermi points delivers the Dirac Hamiltonian
\begin{subequations}
\label{eq: def massless 2x2 Dirac Hamiltonian in 1d}
\begin{equation}
\mathcal{H}^{\,}_{\mathrm{D}}\:=
-
\tau^{\,}_{3}\,
\mathrm{i}\frac{\partial}{\partial x}
\label{eq: def massless 2x2 Dirac Hamiltonian in 1d a}
\end{equation}
in the units defined by 
\begin{equation}
\hbar\equiv1,
\qquad
v^{\,}_{\mathrm{F}}=
2t\,|\sin k^{\,}_{\mathrm{F}}|\equiv
1.
\label{eq: def massless 2x2 Dirac Hamiltonian in 1d b}
\end{equation} 
Here, $\tau^{\,}_{3}$ is the third Pauli matrices
among the four $2\times 2$ matrices
\begin{equation}
\tau^{\,}_{0}\:=
\begin{pmatrix}
1
&
0
\\
0
&
1
\end{pmatrix},
\qquad
\tau^{\,}_{1}\:=
\begin{pmatrix}
0
&
1
\\
1
&
0
\end{pmatrix},
\qquad
\tau^{\,}_{2}\:=
\begin{pmatrix}
0
&
-\mathrm{i}
\\
+\mathrm{i}
&
0
\end{pmatrix},
\qquad
\tau^{\,}_{3}\:=
\begin{pmatrix}
+1
&
0
\\
0
&
-1
\end{pmatrix}.
\label{eq: def massless 2x2 Dirac Hamiltonian in 1d c}
\end{equation}
\end{subequations}
The momentum eigenstate
\begin{subequations}
\begin{equation}
\Psi^{\,}_{\mathrm{R},p}(x)\:=
e^{+\mathrm{i}p\,x}\,
\begin{pmatrix}
1\\
0
\end{pmatrix}
\end{equation}
is an eigenstate with the single-particle energy
$\varepsilon^{\,}_{\mathrm{R}}(p)=+p$.
The momentum eigenstate
\begin{equation}
\Psi^{\,}_{\mathrm{L},p}(x)\:=
e^{+\mathrm{i}p\,x}\,
\begin{pmatrix}
0\\
1
\end{pmatrix}
\end{equation}
\end{subequations}
is an eigenstate with the single-particle energy
$\varepsilon^{\,}_{\mathrm{L}}(p)=-p$.
The plane waves
\begin{subequations}
\label{eq: plane wave solutions to 2x2 massless Dirac equation}
\begin{equation}
\Psi^{\,}_{\mathrm{R},p}(x,t)\:=
e^{+\mathrm{i}p\,(x-t)}\,
\begin{pmatrix}
1\\
0
\end{pmatrix}
\end{equation}
and
\begin{equation}
\Psi^{\,}_{\mathrm{L},p}(x,t)\:=
e^{+\mathrm{i}p\,(x+t)}\,
\begin{pmatrix}
0
\\
1
\end{pmatrix}
\end{equation}
are respectively right-moving and left-moving solutions to the 
massless Dirac equation
\begin{equation}
\mathrm{i}\frac{\partial}{\partial t}\Psi=
\mathcal{H}^{\,}_{\mathrm{D}}\,\Psi.
\end{equation}
\end{subequations}

We perturb the massless Dirac Hamiltonian 
(\ref{eq: def massless 2x2 Dirac Hamiltonian in 1d})
with the most generic static and local one-body potential
\begin{equation}
\mathcal{V}(x)\:=
a^{\,}_{0}(x)\,
\tau^{\,}_{0}
+
m^{\,}_{1}(x)\,
\tau^{\,}_{1}
+
m^{\,}_{2}(x)\,
\tau^{\,}_{2}
+
a^{\,}_{1}(x)\,
\tau^{\,}_{3}.
\label{eq: def 2x2 generic mathcal V in class A}
\end{equation}
The real-valued function $a^{\,}_{0}$
is a space-dependent chemical potential.
It couples to the spinless fermions as 
the scalar part of the electromagnetic gauge potential does.
The real-valued function $a^{\,}_{1}$
is a space-dependent modulation of the Fermi point.
It couples to the spinless fermions as 
the vector part of the electromagnetic gauge potential does.
Both $a^{\,}_{0}$ and $a^{\,}_{1}$ multiply Pauli matrices such that
each commutes with the massless Dirac Hamiltonian 
(\ref{eq: def massless 2x2 Dirac Hamiltonian in 1d}). 
Neither channels are confining (localizing). 
The real-valued functions $m^{\,}_{1}$ and $m^{\,}_{2}$
are space-dependent mass terms, for they multiply Pauli matrices such
that each anticommutes with the massless Dirac Hamiltonian
and with each other. Either channels are confining (localizing).

\medskip\noindent
\textbf{\textit{7.2.1.1\hskip 10 true pt Symmetry class A}}

\medskip\noindent
The only symmetry preserved by
\begin{equation}
\mathcal{H}\:=
\mathcal{H}^{\,}_{\mathrm{D}}
+
\mathcal{V}(x)
\label{eq: def 2x2 generic Hamiltonian in class A}
\end{equation}
with $\mathcal{V}$ defined in Eq.\ 
(\ref{eq: def 2x2 generic mathcal V in class A})
is the global symmetry under multiplication of all
states in the single-particle Hilbert space over which
$\mathcal{H}$
acts by the same $U(1)$ phase. Correspondingly, the local two-current
\begin{equation}
J^{\mu}(x,t)\:=
\Big(
\Psi^{\dag}\,\Psi,
\Psi^{\dag}\,\tau^{\,}_{3}\,\Psi
\Big)(x,t),
\end{equation}
obeys the continuity equation
\begin{equation}
\partial^{\,}_{\mu}\,J^{\mu}=0,
\qquad
\partial^{\,}_{0}\:=
\frac{\partial}{\partial t},
\qquad
\partial^{\,}_{1}\:=
\frac{\partial}{\partial x}.
\label{eq: U(1) conservation law}
\end{equation}
The family of Hamiltonian (\ref{eq: def 2x2 generic Hamiltonian in class A}) 
labelled by the potential $\mathcal{V}$ of the form%
 (\ref{eq: def 2x2 generic mathcal V in class A})
is said to belong to the symmetry class A because of the conservation law
(\ref{eq: U(1) conservation law}). 

One would like to reverse time in the Dirac equation
\begin{equation}
\left(\mathrm{i}\frac{\partial}{\partial t}\Psi\right)(x,t)=
\left(\mathcal{H}\,\Psi\right)(x,t)
\label{eq: Dirac equation for class A}
\end{equation}
where $\mathcal{H}$ is defined by
Eq.\ (\ref{eq: def 2x2 generic Hamiltonian in class A}).
Under reversal of time
\begin{equation}
t=-t',
\end{equation} 
the Dirac equation%
 (\ref{eq: Dirac equation for class A})
becomes
\begin{equation}
\left(-\mathrm{i}\frac{\partial}{\partial t'}\Psi\right)(x,-t')=
\left(\mathcal{H}\,\Psi\right)(x,-t').
\end{equation}
Complex conjugation removes the minus sign on the left-hand side,
\begin{equation}
\left(\mathrm{i}\frac{\partial}{\partial t'}\Psi^{*}\right)(x,-t')=
\left(\mathcal{H}\,\Psi\right)^{*}(x,-t').
\end{equation}
Form invariance of the Dirac equation under reversal of time
then follows if one postulates the existence of an unitary
$2\times2$ matrix
$\mathcal{U}^{\,}_{\mathcal{T}}$ 
and of a phase $0\leq\phi^{\,}_{\mathcal{T}}<2\pi$
such that (complex conjugation will be denoted by $\mathsf{K}$)
\begin{subequations}
\begin{equation}
\left(\mathcal{U}^{\,}_{\mathcal{T}}\,\mathsf{K}\right)^{2}=
e^{\mathrm{i}\phi^{\,}_{\mathcal{T}}}\,\tau^{\,}_{0},
\qquad
\Psi^{*}(x,-t)\=:
\mathcal{U}^{\,}_{\mathcal{T}}\,\Psi^{\,}_{\mathcal{T}}(x,t),
\qquad
\mathcal{H}^{\,}_{\mathcal{T}}\:=
\mathcal{U}^{-1}_{\mathcal{T}}\,
\mathcal{H}^{*}\,
\mathcal{U}^{\,}_{\mathcal{T}},
\end{equation}
in which case
\begin{equation}
\left(\mathrm{i}\frac{\partial}{\partial t}\Psi^{\,}_{\mathcal{T}}\right)(x,t)=
\left(\mathcal{H}^{\,}_{\mathcal{T}}\,\Psi^{\,}_{\mathcal{T}}\right)(x,t).
\end{equation}
\end{subequations}
Time-reversal symmetry then holds if and only if
\begin{equation}
\mathcal{U}^{-1}_{\mathcal{T}}\,
\mathcal{H}^{*}\,
\mathcal{U}^{\,}_{\mathcal{T}}=
\mathcal{H}.
\label{eq: def TRS for generic 2x2 Hamiltonian}
\end{equation}

Time-reversal symmetry must hold for the massless Dirac equation.
By inspection of the right- and left-moving solutions
(\ref{eq: plane wave solutions to 2x2 massless Dirac equation}), 
one deduces that $\mathcal{U}^{\,}_{\mathcal{T}}$ must interchange right
and left movers. There are two possibilities, either
\begin{equation}
\mathcal{U}^{\,}_{\mathcal{T}}=\tau^{\,}_{2},
\qquad
\phi^{\,}_{\mathcal{T}}=\pi,
\label{eq: def reversal time AII}
\end{equation}
or
\begin{equation}
\mathcal{U}^{\,}_{\mathcal{T}}=\tau^{\,}_{1},
\qquad
\phi^{\,}_{\mathcal{T}}=0.
\label{eq: def reversal time AI}
\end{equation}

\medskip\noindent
\textbf{\textit{7.2.1.2\hskip 10 true pt Symmetry class AII}}

\medskip\noindent
Imposing time-reversal symmetry
using the definition (\ref{eq: def reversal time AII})
restricts the family of Dirac Hamiltonians 
(\ref{eq: def 2x2 generic Hamiltonian in class A})
to
\begin{equation}
\mathcal{H}(x)\:=
-
\mathrm{i}\tau^{\,}_{3}\,\frac{\partial}{\partial t}
+
a^{\,}_{0}(x)\,
\tau^{\,}_{0}.
\label{eq: def 2x2 generic mathcal V in class AII}
\end{equation}
The family of Hamiltonian (\ref{eq: def 2x2 generic Hamiltonian in class A}) 
labelled by the potential $\mathcal{V}$ of the form%
 (\ref{eq: def 2x2 generic mathcal V in class AII})
is said to belong to the symmetry class AII because of the conservation law
(\ref{eq: U(1) conservation law})
and of the time-reversal symmetry
 (\ref{eq: def TRS for generic 2x2 Hamiltonian})
with the representation (\ref{eq: def reversal time AII}).

\medskip\noindent
\textbf{\textit{7.2.1.3\hskip 10 true pt Symmetry class AI}}

\medskip\noindent
Imposing time-reversal symmetry
using the definition (\ref{eq: def reversal time AI})
restricts the family of Dirac Hamiltonians 
(\ref{eq: def 2x2 generic Hamiltonian in class A})
to
\begin{equation}
\mathcal{H}(x)\:=
-
\mathrm{i}\tau^{\,}_{3}\,\frac{\partial}{\partial t}
+
a^{\,}_{0}(x)\,
\tau^{\,}_{0}
+
m^{\,}_{1}(x)\,
\tau^{\,}_{1}
+
m^{\,}_{2}(x)\,
\tau^{\,}_{2}.
\label{eq: def 2x2 generic mathcal V in class AI}
\end{equation}
The family of Hamiltonian (\ref{eq: def 2x2 generic Hamiltonian in class A}) 
labelled by the potential $\mathcal{V}$ of the form%
 (\ref{eq: def 2x2 generic mathcal V in class AI})
is said to belong to the symmetry class AI because of the conservation law
(\ref{eq: U(1) conservation law})
and of the time-reversal symmetry
(\ref{eq: def TRS for generic 2x2 Hamiltonian})
with the representation (\ref{eq: def reversal time AI}).

Take advantage of the fact that the dispersion relation
(\ref{eq: dispersion if uniform nn hopping along chain})
obeys the symmetry
\begin{equation}
\mathcal{H}(k)=
-\mathcal{H}(k+\pi).
\label{eq: spectral symmetry cos k band}
\end{equation}
This spectral symmetry is a consequence of the fact that
the lattice Hamiltonian anticommutes with the local gauge transformation
that maps the basis of single-particle localized wave functions
\begin{equation}
\psi^{\,}_{i}:\mathbb{Z}\to\mathbb{C},
j\mapsto\psi^{\,}_{i}(j)\:=\delta^{\,}_{ij}
\end{equation}
into the basis
\begin{equation}
\psi^{\prime}_{i}:\mathbb{Z}\to\mathbb{C},
j\mapsto\psi^{\prime}_{i}(j)\:= (-1)^{j}\,\delta^{\,}_{ij}.
\end{equation}
Such a spectral symmetry is an example of a sublattice symmetry
in condensed matter physics. So far, the chemical potential 
\begin{equation}
\varepsilon^{\,}_{\mathrm{F}}\equiv
-2\,t\,\cos k^{\,}_{\mathrm{F}}
\end{equation}
defined in Eq.\
(\ref{eq: dispersion if uniform nn hopping along chain})
has been arbitrary. However, in view of the spectral symmetry
(\ref{eq: spectral symmetry cos k band}),
the single-particle energy eigenvalue
\begin{equation}
0=
\varepsilon^{\,}_{\mathrm{F}}\equiv
-2\,t\,\cos k^{\,}_{\mathrm{F}},
\qquad
k^{\,}_{\mathrm{F}}=
\frac{\pi}{2},
\end{equation}
is special. It is the center of symmetry of the single-particle
spectrum (\ref{eq: dispersion if uniform nn hopping along chain}).
The spectral symmetry (\ref{eq: spectral symmetry cos k band})
is also known as a chiral symmetry
of the Dirac equation (\ref{eq: def massless 2x2 Dirac Hamiltonian in 1d a})
by which
\begin{equation}
\mathcal{H}^{\,}_{\mathrm{D}}=
-
\tau^{\,}_{1}\,\mathcal{H}^{\,}_{\mathrm{D}}\,\tau^{\,}_{1},
\end{equation} 
after an expansion to leading order in powers of the deviation
of the momenta away from the two Fermi points $\pm\pi/2$.

\medskip\noindent
\textbf{\textit{7.2.1.4\hskip 10 true pt Symmetry class AIII}}

\medskip\noindent
If charge conservation holds together with the chiral symmetry
\begin{subequations}
\label{eq: AIII 2x2}
\begin{equation}
\mathcal{H}=
-
\tau^{\,}_{1}\,
\mathcal{H}\,
\tau^{\,}_{1},
\label{eq: AIII 2x2 a}
\end{equation}
then
\begin{equation}
\mathcal{H}=
-
\tau^{\,}_{3}\,\mathrm{i}\partial^{\,}_{x}
+
a^{\,}_{1}(x)\,
\tau^{\,}_{3}
+
m^{\,}_{2}(x)\,
\tau^{\,}_{2}
\label{eq: AIII 2x2 b}
\end{equation}
\end{subequations}
is said to belong to the symmetry class AIII.

\medskip\noindent
\textbf{\textit{7.2.1.5\hskip 10 true pt Symmetry class CII}}

\medskip\noindent
It is not possible to write down a $2\times2$ Dirac equation
in the symmetry class CII. For example, 
if charge conservation holds together 
with the chiral and time-reversal symmetries
\begin{subequations}
\begin{equation}
\mathcal{H}=
-
\tau^{\,}_{1}\,
\mathcal{H}\,
\tau^{\,}_{1},
\qquad
\mathcal{H}=
+
\tau^{\,}_{2}\,
\mathcal{H}^{*}\,
\tau^{\,}_{2},
\end{equation}
respectively, then
\begin{equation}
\mathcal{H}=
-
\tau^{\,}_{3}\,\mathrm{i}\partial^{\,}_{x}
\end{equation}
\end{subequations}
does not belong to the symmetry class CII,
as the composition of the chiral transformation with reversal of time
squares to unity instead of minus times unity.

\medskip\noindent
\textbf{\textit{7.2.1.6\hskip 10 true pt Symmetry class BDI}}

\medskip\noindent
If charge conservation holds together 
with the chiral and time-reversal symmetries
\begin{subequations}
\label{eq: BDI 2x2}
\begin{equation}
\mathcal{H}=
-
\tau^{\,}_{1}\,
\mathcal{H}\,
\tau^{\,}_{1},
\qquad
\mathcal{H}=
+
\tau^{\,}_{1}\,
\mathcal{H}^{*}\,
\tau^{\,}_{1},
\label{eq: BDI 2x2 a}
\end{equation}
then
\begin{equation}
\mathcal{H}=
-
\tau^{\,}_{3}\,\mathrm{i}\partial^{\,}_{x}
+
m^{\,}_{2}(x)\,
\tau^{\,}_{2}
\label{eq: BDI 2x2 b}
\end{equation}
\end{subequations}
is said to belong to the symmetry class BDI.

The global $U(1)$ gauge symmetry responsible for the continuity equation
(\ref{eq: U(1) conservation law})
demands that one treats the two components of the Dirac spinors as independent.
This is not desirable if the global $U(1)$ gauge symmetry is to be restricted
to a global $\mathbb{Z}^{\,}_{2}$ gauge symmetry, as occurs in a 
mean-field treatment of superconductivity. If the possibility
of restricting the global $U(1)$ to a global $\mathbb{Z}^{\,}_{2}$ 
gauge symmetry is to be accounted for, 
four more symmetry classes are permissible.

\medskip\noindent
\textbf{\textit{7.2.1.7\hskip 10 true pt Symmetry class D}}

\medskip\noindent
Impose a particle-hole symmetry through
\begin{subequations}
\label{eq: D 2x2}
\begin{equation}
\mathcal{H}=
-
\mathcal{H}^{*},
\label{eq: D 2x2 a}
\end{equation}
then
\begin{equation}
\mathcal{H}=
-
\tau^{\,}_{3}\,\mathrm{i}\partial^{\,}_{x}
+
m^{\,}_{2}(x)\,
\tau^{\,}_{2}
\label{eq: D 2x2 b}
\end{equation}
\end{subequations}
is said to belong to the symmetry class D.

\medskip\noindent
\textbf{\textit{7.2.1.8\hskip 10 true pt Symmetry class DIII}}

\medskip\noindent
Impose a particle-hole symmetry
and time-reversal symmetry through
\begin{subequations}
\label{eq: DIII 2x2}
\begin{equation}
\mathcal{H}=
-
\mathcal{H}^{*},
\qquad
\mathcal{H}=
+
\tau^{\,}_{2}\,
\mathcal{H}^{*}\,
\tau^{\,}_{2},
\label{eq: DIII 2x2 a}
\end{equation}
respectively, then
\begin{equation}
\mathcal{H}=
-
\tau^{\,}_{3}\,\mathrm{i}\partial^{\,}_{x}
\label{eq: DIII 2x2 b}
\end{equation}
\end{subequations}
is said to belong to the symmetry class DIII.

\medskip\noindent
\textbf{\textit{7.2.1.9\hskip 10 true pt Symmetry class C}}

\medskip\noindent
Impose a particle-hole symmetry through
\begin{subequations}
\label{eq: C 2x2}
\begin{equation}
\mathcal{H}=
-
\tau^{\,}_{2}\,
\mathcal{H}^{*}\,
\tau^{\,}_{2},
\label{eq: C 2x2 a}
\end{equation}
then
\begin{equation}
\mathcal{H}=
a^{\,}_{1}(x)\,
\tau^{\,}_{3}
+
m^{\,}_{2}(x)\,
\tau^{\,}_{2}
+
m^{\,}_{1}(x)\,
\tau^{\,}_{1}
\label{eq: C 2x2 b}
\end{equation}
\end{subequations}
is said to belong to the symmetry class C.
The Dirac kinetic energy is prohibited
for a $2\times2$ Dirac Hamiltonian
from the symmetry class C.

\medskip\noindent
\textbf{\textit{7.2.1.10\hskip 10 true pt Symmetry class CI}}

\medskip\noindent
Impose a particle-hole symmetry
and time-reversal symmetry through
\begin{subequations}
\label{eq: CI 2x2}
\begin{equation}
\mathcal{H}=
-
\tau^{\,}_{2}\,
\mathcal{H}^{*}\,
\tau^{\,}_{2},
\qquad
\mathcal{H}=
+
\tau^{\,}_{1}\,
\mathcal{H}^{*}\,
\tau^{\,}_{1},
\label{eq: CI 2x2 a}
\end{equation}
respectively,
then
\begin{equation}
\mathcal{H}=
m^{\,}_{2}(x)\,
\tau^{\,}_{2}
+
m^{\,}_{1}(x)\,
\tau^{\,}_{1}
\label{eq: CI 2x2 b}
\end{equation}
\end{subequations}
is said to belong to the symmetry class CI.
The Dirac kinetic energy is prohibited
for a $2\times2$ Dirac Hamiltonian
from the symmetry class CI.

\subsection{Symmetries for the case of two one-dimensional channels}
\label{subsec: Symmetries for the case of two one-dimensional channels}

Imagine two coupled linear chains along which non-interacting
spinless fermions are allowed to hop. If the two chains
are decoupled and the hopping is a uniform nearest-neighbor hopping
along any one of the two chains, 
then the low-energy and long-wave length effective single-particle
Hamiltonian in the vicinity of the chemical potential 
$\varepsilon^{\,}_{\mathrm{F}}=0$
is the tensor product of the massless Dirac
Hamiltonian (\ref{eq: def massless 2x2 Dirac Hamiltonian in 1d a})
with the $2\times2$ unit matrix $\sigma^{\,}_{0}$. 
Let the three Pauli matrices
$\bm{\sigma}$
act on the same vector space as $\sigma^{\,}_{0}$ does. 
For convenience, introduce the sixteen Hermitean 
$4\times4$ matrices
\begin{equation}
X^{\,}_{\mu\nu}\:=
\tau^{\,}_{\mu}
\otimes
\sigma^{\,}_{\nu},
\qquad
\mu,\nu=0,1,2,3.
\end{equation}

\medskip\noindent
\textbf{\textit{7.2.2.1\hskip 10 true pt Symmetry class A}} 

\medskip\noindent
The generic Dirac Hamiltonian of rank $r=4$ is
\begin{equation}
\mathcal{H}\:=
-
X^{\,}_{30}\,\mathrm{i}\partial^{\,}_{x}
+
a^{\,}_{1,\nu}(x)\,X^{\,}_{3\nu}
+
m^{\,}_{2,\nu}(x)\,X^{\,}_{2\nu}
+
m^{\,}_{1,\nu}(x)\,X^{\,}_{1\nu}
+
a^{\,}_{0,\nu}(x)\,X^{\,}_{0\nu}.
\label{eq: generic 4 by 4 Dirac Hamiltonian d=1}
\end{equation}
The summation convention over the repeated index $\nu=0,1,2,3$ is implied.
There are four real-valued parameters for the components
$a^{\,}_{1,\nu}$ with $\mu=0,1,2,3$ of an $U(2)$ vector potential,
eight for the components $m^{\,}_{1,\nu}$ and $m^{\,}_{2,\nu}$ with 
$\mu=0,1,2,3$ of two independent $U(2)$ masses,
and four for the components
$a^{\,}_{0,\nu}$ with $\mu=0,1,2,3$ of an $U(2)$ scalar potential.
As it should be there are 16 real-valued free parameters
(functions if one opts to break translation invariance).
If all components of the spinors solving the eigenvalue problem
\begin{equation}
\mathcal{H}\,\Psi(x)=
\varepsilon\,\Psi(x)
\end{equation}
are independent, the Dirac Hamiltonian
(\ref{eq: generic 4 by 4 Dirac Hamiltonian d=1})
belongs to the symmetry class A.

In addition to the conservation of the fermion number, one may
impose time-reversal symmetry on the Dirac Hamiltonian%
 (\ref{eq: generic 4 by 4 Dirac Hamiltonian d=1})
There are two possibilities to do so.

\medskip\noindent
\textbf{\textit{7.2.2.2\hskip 10 true pt Symmetry class AII}}

\medskip\noindent
If charge conservation holds together with time-reversal symmetry through
\begin{subequations}
\begin{equation}
\mathcal{H}=
+
X^{\,}_{12}\,
\mathcal{H}^{*}\,
\,X^{\,}_{12},
\end{equation}
then 
\begin{equation}
\mathcal{H}=
-
X^{\,}_{30}\,\mathrm{i}\partial^{\,}_{x}
+
\sum_{\nu=1,2,3}
a^{\,}_{1,\nu}(x)\,X^{\,}_{3\nu}
+
m^{\,}_{2,0}(x)\,X^{\,}_{20}
+
m^{\,}_{1,0}(x)\,X^{\,}_{10}
+
a^{\,}_{0,0}(x)\,X^{\,}_{00}
\end{equation}
\end{subequations}
is said to belong to the symmetry class AII.

\medskip\noindent
\textbf{\textit{7.2.2.3\hskip 10 true pt Symmetry class AI}}

\medskip\noindent
If charge conservation holds together with time-reversal symmetry through
\begin{subequations}
\begin{equation}
\mathcal{H}=
+
X^{\,}_{10}\,
\mathcal{H}^{*}\,
X^{\,}_{10},
\end{equation}
then
\begin{equation}
\mathcal{H}=
-
X^{\,}_{30}\,\mathrm{i}\partial^{\,}_{x}
+
a^{\,}_{1,2}(x)\,X^{\,}_{32}
+
\sum_{\nu=0,1,3}
\left[
m^{\,}_{2,\nu}(x)\,X^{\,}_{2\nu}
+
m^{\,}_{1,\nu}(x)\,X^{\,}_{1\nu}
+
a^{\,}_{0,\nu}(x)\,X^{\,}_{0\nu}
\right]
\end{equation}
\end{subequations}
is said to belong to the symmetry class AI.

The standard symmetry classes A, AII, and AI can be further constrained
by imposing the chiral symmetry. 
This gives the following three possibilities.

\medskip\noindent
\textbf{\textit{7.2.2.4\hskip 10 true pt Symmetry class AIII}}

\medskip\noindent
If charge conservation holds together with the chiral symmetry
\begin{subequations}
\begin{equation}
\mathcal{H}=
-
X^{\,}_{10}\,
\mathcal{H}\,
X^{\,}_{01},
\end{equation}
then
\begin{equation}
\mathcal{H}=
-
X^{\,}_{30}\,\mathrm{i}\partial^{\,}_{x}
+
a^{\,}_{1,\nu}(x)\,X^{\,}_{3\nu}
+
m^{\,}_{2,\nu}(x)\,X^{\,}_{2\nu}
\end{equation}
\end{subequations}
is said to belong to the symmetry class AIII.

\medskip\noindent
\textbf{\textit{7.2.2.5\hskip 10 true pt Symmetry class CII}}

\medskip\noindent
If charge conservation holds together with chiral symmetry and 
time-reversal symmetry
\begin{subequations}
\begin{equation}
\mathcal{H}=
-
X^{\,}_{10}\,
\mathcal{H}\,
X^{\,}_{10},
\qquad
\mathcal{H}=
+
X^{\,}_{12}\,
\mathcal{H}^{*}\,
X^{\,}_{12},
\end{equation}
respectively,
then
\begin{equation}
\mathcal{H}=
-
X^{\,}_{30}\,\mathrm{i}\partial^{\,}_{x}
+
\sum_{\nu=1,2,3}
a^{\,}_{1,\nu}(x)\,X^{\,}_{3\nu}
+
m^{\,}_{2,0}(x)\,X^{\,}_{20}
\end{equation}
\end{subequations}
is said to belong to the symmetry class CII.

\medskip\noindent
\textbf{\textit{7.2.2.6\hskip 10 true pt Symmetry class BDI}}

\medskip\noindent
If charge conservation holds together with chiral symmetry and 
time-reversal symmetry
\begin{subequations}
\begin{equation}
\mathcal{H}=
-
X^{\,}_{10}\,
\mathcal{H}\,
X^{\,}_{10},
\qquad
\mathcal{H}=
+
X^{\,}_{10}
\mathcal{H}^{*}\,
X^{\,}_{10},
\end{equation}
respectively, then
\begin{equation}
\mathcal{H}=
-
X^{\,}_{30}\,\mathrm{i}\partial^{\,}_{x}
+
a^{\,}_{1,2}(x)\,X^{\,}_{32}
+
\sum_{\nu=0,1,3}
m^{\,}_{2,\nu}(x)\,X^{\,}_{2\nu}
\end{equation}
\end{subequations}
is said to belong to the symmetry class BDI.

Now, we move to the four Bogoliubov-de-Gennes (BdG) symmetry classes
by relaxing the condition that all components of a spinor on which
the Hamiltonian acts be independent. This means that changing
each component of a spinor by a multiplicative global phase factor is not
legitimate anymore. However, changing each component of a spinor
by a global sign remains legitimate. The contraints among the components
of a spinor come about by imposing a particle-hole symmetry.

\medskip\noindent
\textbf{\textit{7.2.2.7\hskip 10 true pt Symmetry class D}}

\medskip\noindent
Impose particle-hole symmetry through
\begin{subequations}
\begin{equation}
\mathcal{H}=
-
\mathcal{H}^{*},
\end{equation}
then
\begin{equation}
\mathcal{H}=
-
X^{\,}_{30}\,\mathrm{i}\partial^{\,}_{x}
+
a^{\,}_{1,2}(x)\,X^{\,}_{32}
+
\sum_{\nu=0,1,3}
m^{\,}_{2,\nu}(x)\,X^{\,}_{2\nu}
+
m^{\,}_{1,2}(x)\,X^{\,}_{12}
+
a^{\,}_{0,2}(x)\,X^{\,}_{02}
\end{equation}
\end{subequations}
is said to belong to the symmetry class D.

\medskip\noindent
\textbf{\textit{7.2.2.8\hskip 10 true pt Symmetry class DIII}}

\medskip\noindent
Impose particle-hole symmetry and time-reversal symmetry through
\begin{subequations}
\begin{equation}
\mathcal{H}=
-
\mathcal{H}^{*},
\qquad
\mathcal{H}=
+
X^{\,}_{20}\,
\mathcal{H}^{*}\,
X^{\,}_{20},
\end{equation}
respectively, then
\begin{equation}
\mathcal{H}=
-
X^{\,}_{30}\,\mathrm{i}\partial^{\,}_{x}
+
a^{\,}_{1,2}(x)\,X^{\,}_{32}
+
m^{\,}_{1,2}(x)\,X^{\,}_{12}
\end{equation}
\end{subequations}
is said to belong to the symmetry class DIII.

\medskip\noindent
\textbf{\textit{7.2.2.9\hskip 10 true pt Symmetry class C}}

\medskip\noindent
Impose particle-hole symmetry through
\begin{subequations}
\begin{equation}
\mathcal{H}=
-
X^{\,}_{02}\,
\mathcal{H}^{*}\,
X^{\,}_{02},
\end{equation}
respectively, then 
\begin{equation}
\mathcal{H}=
-
X^{\,}_{30}\,\mathrm{i}\partial^{\,}_{x}
+
\sum_{\nu=1,2,3}
a^{\,}_{1,\nu}(x)\,X^{\,}_{3\nu}
+
m^{\,}_{2,0}(x)\,X^{\,}_{20}
+
\sum_{\nu=1,2,3}
\left[
m^{\,}_{1,\nu}(x)\,X^{\,}_{1\nu}
+
a^{\,}_{0,\nu}(x)\,X^{\,}_{0\nu}
\right]
\end{equation}
\end{subequations}
is said to belong to the symmetry class C.

\medskip\noindent
\textbf{\textit{7.2.2.10\hskip 10 true pt Symmetry class CI}}

\medskip\noindent
Impose particle-hole symmetry and time-reversal symmetry through
\begin{subequations}
\begin{equation}
\mathcal{H}=
-
X^{\,}_{02}\,
\mathcal{H}^{*}\,
X^{\,}_{02},
\qquad
\mathcal{H}=
+
X^{\,}_{10}\,
\mathcal{H}^{*}\,
X^{\,}_{10},
\end{equation}
respectively, then 
\begin{equation}
\mathcal{H}=
-
X^{\,}_{30}\,\mathrm{i}\partial^{\,}_{x}
+
a^{\,}_{1,2}(x)\,X^{\,}_{32}
+
m^{\,}_{2,0}(x)\,X^{\,}_{20}
+
m^{\,}_{1,1}(x)\,X^{\,}_{11}
+
m^{\,}_{1,3}(x)\,X^{\,}_{13}
+
a^{\,}_{0,3}(x)\,X^{\,}_{03}
\end{equation}
\end{subequations}
is said to belong to the symmetry class CI.

\subsection{Definition of the minimum rank}

In Sections
\ref{subsec: Symmetries for the case of one one-dimensional channel}
and
\ref{subsec: Symmetries for the case of two one-dimensional channels},
we have imposed ten symmetry restrictions corresponding to the
tenfold way introduced by Altland and Zirnbauer to Dirac Hamiltonians
with Dirac matrices of rank $r=2$ and $r=4$, respectively.
These Dirac Hamiltonians describe the propagation of single-particle states
in one-dimensional space. All ten symmetry classes shall be called the
Altland-Zirnbauer (AZ) symmetry classes.

Observe that some of the AZ symmetries can be very restrictive for
the Dirac Hamiltonians with Dirac matrices of small rank $r$.
For example, it is not possible to write down a Dirac Hamiltonian
of rank $r=2$ in the symmetry class CII, the symmetry classes C and CI
do not admit a Dirac kinetic energy of rank $r=2$, and the symmetry classes
AII and DIII do not admit Dirac masses in their Dirac Hamiltonians of rank $r=2$.

This observation suggests the definition of the minimum rank
$r^{\,}_{\mathrm{min}}$
for which the Dirac Hamiltonian describing the propagation in
$d$-dimensional space for a given AZ symmetry class admits a Dirac mass.
Hence, $r^{\,}_{\mathrm{min}}$ depends implicitly on the dimensionality of space 
and on the AZ symmetry class. 
In one-dimensional space,
we have found that
\begin{equation}
\begin{split}
&
r^{\mathrm{A}}_{\mathrm{min}}=2,
\qquad
r^{\mathrm{AII}}_{\mathrm{min}}=4,
\qquad
r^{\mathrm{AI}}_{\mathrm{min}}=2,
\\
&
r^{\mathrm{AIII}}_{\mathrm{min}}=2,
\qquad
r^{\mathrm{CII}}_{\mathrm{min}}=4,
\qquad
r^{\mathrm{BDI}}_{\mathrm{min}}=2,
\\
&
r^{\mathrm{D}}_{\mathrm{min}}=2,
\qquad
r^{\mathrm{DIII}}_{\mathrm{min}}=4,
\qquad
r^{\mathrm{C}}_{\mathrm{min}}=4,
\qquad
r^{\mathrm{CI}}_{\mathrm{min}}=4.
\end{split}
\end{equation}
This definition is useful for a number of reasons.

First, Anderson localization in a given AZ symmetry class 
is impossible for any random Dirac Hamiltonian 
with Dirac matrices of rank $r$ smaller than $r^{\,}_{\mathrm{min}}$. 
This is the case for symmetry classes AII and DIII for a Dirac Hamiltonian
of rank $r=2$ in one-dimensional space. The lattice realization of these 
Dirac Hamiltonians is along the boundary of a two-dimensional
insulator in the symmetry classes AII and DIII when the bulk
realizes a topologically non-trivial insulating phase owing to the fermion
doubling problem.
This is why an odd number of helical pairs of edge states in the symmetry 
class AII and an odd number of helical pairs of Majorana edge states in the
symmetry class DIII can evade Anderson localization. 
The limit $r=2$ for the Dirac Hamiltonians encoding 
one-dimensional propagation in the symmetry classes AII and DIII
are the signatures for the topologically non-trivial entries of the group 
$\mathbb{Z}^{\,}_{2}$ in column $d=2$ from Table \ref{periodic table}.
For the symmetry classes A and D, we can consider the limit $r=1$ 
as a special limit that shares with a Dirac Hamiltonian 
the property that it is a first-order differential operator in space, 
but, unlike a Dirac Hamiltonian, this limit does no treat right- 
and left-movers on equal footing (and thus breaks time-reversal symmetry
explicitly). Such a first-order differential operator encodes the
propagation of right movers on the inner boundary of a two-dimensional ring
(the Corbino geometry of Fig.\ \ref{Chiral edges states}) 
while its complex conjugate encodes the
propagation of left movers on the outer boundary of a two-dimensional ring
or vice versa. 
For the symmetry class C, one must consider two copies of opposite spins
of the $r=1$ limit of class D. The limit $r=1$
for Dirac Hamiltonians encoding one-dimensional propagation 
in the symmetry classes A, D, and C are realized on the boundary of 
two-dimensional insulating phases supporting the integer quantum Hall effect, 
the thermal integer quantum Hall effect,
and the spin-resolved thermal integer quantum Hall effect, respectively.
The limits $r=1$ for the Dirac Hamiltonians encoding 
one-dimensional propagation in the symmetry classes A, D, and C
are the signatures for the non-trivial 
entries $\pm1$ and $\pm2$ of the groups $\mathbb{Z}$ and $2\mathbb{Z}$, 
respectively, in column $d=2$ from Table \ref{periodic table}.

Second, one can always define the quasi-$d$-dimensional Dirac Hamiltonian
\begin{subequations}
\label{eq: def generic QdD Dirac}
\begin{equation}
\mathcal{H}(\bm{x})\:=
-\mathrm{i}
(\bm{\alpha}\otimes I)\cdot\frac{\partial}{\partial\bm{x}}
+
\mathcal{V}(\bm{x}),
\end{equation}
where $\bm{\alpha}$ and $\beta$ are a set of matrices
that anticommute pairwise and square to the unit 
$r^{\,}_{\mathrm{min}}\times r^{\,}_{\mathrm{min}}$
matrix, $I$ is a unit $N\times N$ matrix,
and
\begin{equation}
\mathcal{V}(\bm{x})=
m(\bm{x})\,\beta\otimes I
+
\cdots
\end{equation}
\end{subequations}
with `$\cdots$' representing all other masses, vector potentials, 
and scalar potentials allowed by the AZ symmetry class.
For one-dimensional space, the stationary eigenvalue problem
\begin{equation}
\mathcal{H}(x)\,\Psi(x;\varepsilon)=
\varepsilon\,
\Psi(x;\varepsilon)
\end{equation}
with the given `initial value' $\Psi(y;\varepsilon)$
is solved through the transfer matrix
\begin{subequations}
\label{eq: def Q1D transfer matrix}
\begin{equation}
\Psi(x;\varepsilon)=
\mathcal{M}(x|y;\varepsilon)\,
\Psi(y;\varepsilon)
\end{equation}
where
\begin{equation}
\mathcal{M}(x|y;\varepsilon)\:=
\mathcal{P}^{\,}_{x'}\,
\exp
\left(
\int\limits_{y}^{x}\mathrm{d}x'\,
\mathrm{i}(\alpha\otimes I)\,[\varepsilon-\mathcal{V}(x')]
\right).
\end{equation}
\end{subequations}
The symbol $\mathcal{P}^{\,}_{x'}$ represents path ordering.
The limit $N\to\infty$ with all entries of $\mathcal{V}$
independently and identically distributed (iid) up to
the AZ symmetry constraints,
(averaging over the disorder is denoted by an overline)
\begin{equation}
\overline{\mathcal{V}^{\,}_{ij}(x)}\propto \mathsf{v}^{\,}_{ij},
\quad
\overline{
\left[
\mathcal{V}^{\,}_{ij}(x)
-
\mathsf{v}^{\,}_{ij}
\right]
\left[
\mathcal{V}^{\,}_{kl}(y)
-
\mathsf{v}^{\,}_{kl}
\right]
}\propto 
\mathrm{g}^{2}\,
e^{-|x-y|/\xi^{\,}_{\mathrm{dis}}},
\end{equation}
for $i,j,k,l=1,\cdots,r^{\,}_{\mathrm{min}}\,N$
defines the thick quantum wire limit.

\begin{figure}
\begin{center}
\includegraphics[width=0.4\textwidth]{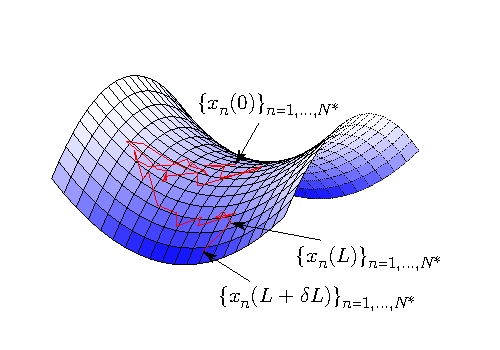}
\end{center}
\caption{
{[Colour online]}
The ``radial coordinate'' of the transfer matrix
$\mathcal{M}$ from Table \ref{Table: Butsuri}
is responsible for Brownian motion 
on an associated non-compact symmetric space.
(Taken from \protect\cite{Brouwer05}.)
\label{fig: Brown walk on symmetric spaces}
        }
\end{figure}

The consequences of Eq.\ (\ref{eq: def Q1D transfer matrix})
are the following. 
First, the local symmetries defining the symmetry classes A, AII, and AI
obeyed by  $\varepsilon-\mathcal{V}(x')$
carry through to the transfer matrix at any single-particle energy 
$\varepsilon$. The local unitary spectral symmetries defining the
symmetry classes AIII, CII, and BDI and the local anti-unitary spectral 
symmetries defining the symmetry classes D, DIII, C, and CI 
carry through to the transfer matrix at the single-particle energy 
$\varepsilon=0$. 
Second, the diagonal matrix entering the polar decomposition
of the transfer matrix at the band center $\varepsilon=0$
is related to the non-compact symmetric spaces from the
column $\mathcal{M}$ in Table \ref{Table: Butsuri}.
Third, the composition law obeyed by the transfer matrix
that encodes enlarging the length of a disordered wire coupled 
to perfect leads is matrix multiplication. It is then possible to
derive a Fokker-Planck equation for the joint probability obeyed 
by the radial coordinates on the  non-compact symmetric spaces from the
column $\mathcal{M}$ in Table \ref{Table: Butsuri}
as the length $L$ of of a disordered wire coupled 
to perfect leads is increased. In this way, the moments of
the dimensionless Landauer conductance $g$ in the columns
$\delta g$ and $-\overline{\ln g}$ can be computed
(see Table \ref{Table: Butsuri}).
An infinitesimal increase in the length of the disordered region 
for one of the ten symmetry classes induces an
infinitesimal Brownian motion 
(see Fig.~\ref{fig: Brown walk on symmetric spaces})
of the Lyapunov exponents 
that is solely controlled by the multiplicities of the
ordinary, long, and short roots of the corresponding
classical semi-simple Lie algebra under suitable assumptions
on the disorder (locality, weakness, and isotropy between all channels).
When the 
transfer matrix describes the stability of the metallic phase 
in the thick quantum wire limit of non-interacting fermions 
perturbed by static one-body random potentials with local correlations 
and of vanishing means in the bulk of a quasi-one-dimensional
lattice model, the multiplicities of the 
short root entering the Brownian motion of the Lyapunov exponents 
always vanish.
However, when the transfer matrix describes the quasi-one-dimensional
boundary of a two-dimensional topological band insulator moderately 
perturbed by static one-body random potentials with local correlations, 
the multiplicities of the short roots is nonvanishing in the Brownian motion 
of the Lyapunov exponents in the five AZ symmetry classes A, AII, 
D, DIII, and C.
Correspondingly, the conductance is of order one
along the infinitely long boundary, i.e., 
the insulating bulk supports extended edge states.
These extended edge states can be thought of as
realizing a quasi-one-dimensional
ballistic phase of quantum matter robust to disorder.

\subsection{Topological spaces for the normalized Dirac masses}

To study systematically the effects of static and local disorder on 
the insulating phases of quasi-$d$-dimensional phases, 
it is very useful to explore the topological properties of
the normalized Dirac masses entering a generic random Dirac Hamiltonian
of the form (\ref{eq: def generic QdD Dirac})
within any given AZ symmetry class.
This approach allows to construct the periodic Table \ref{periodic table}
\cite{Morimoto15}.
Deriving columns $d=1,\cdots,8$ from Table \ref{periodic table}
can be achieved by brute force if one constructs
the generic Dirac Hamiltonian with Dirac matrices of rank 
$r^{\,}_{\mathrm{min}}$ belonging to any one of the ten AZ symmetry classes
and repeat this construction with Dirac matrices or rank 
$2r^{\,}_{\mathrm{min}}$,
$3r^{\,}_{\mathrm{min}}$,
and so on. It then becomes apparent that for any dimension $d$:
\begin{enumerate}
\item
Five of the ten AZ symmetry classes accommodate one 
normalized Dirac matrix up to a sign when the Dirac matrices have the rank
$r=r^{\,}_{\mathrm{min}}$.
These are the symmetry classes that realize topologically distinct
localized phases of $d$-dimensional quantum matter.
\begin{enumerate}
\item
Three of these symmetry classes 
are characterized by having one normalized Dirac mass matrix that commutes
with all other Dirac matrices when $r=N\,r^{\,}_{\mathrm{min}}$ with
$N=2,3,\cdots$. These are the entries with the group
$\mathbb{Z}$ (or $2\mathbb{Z}$ when there is a degeneracy of 2)
in the periodic Table \ref{periodic table}. Mathematically,
the group $\mathbb{Z}$ is the zeroth homotopy group of the
normalized Dirac masses in the limit $N\to\infty$.
\item
Two of these symmetry classes 
are characterized by the fact that the sum of all mass terms
can be associated to a $2N\times 2N$ Hermitean and antisymmetric matrix
for any $r=r^{\,}_{\mathrm{min}}\,N$ 
with $N=1,2,\cdots$. The sign ambiguity of the Pfaffian of this matrix indexes
the two group elements in the entries with the group
$\mathbb{Z}^{\,}_{2}$ 
from the periodic Table \ref{periodic table}. Mathematically,
the group $\mathbb{Z}^{\,}_{2}$ is the zeroth homotopic group of the
normalized Dirac masses for $N$ sufficiently large.
\end{enumerate}

\item
The topological space of normalized Dirac masses is 
compact and path connected for the remaining five symmetry classes,
i.e., its zeroth homotopy group is the trivial one.
No Dirac mass is singled out. The localized phase of matter is topologically
trivial.

\end{enumerate}
The observed periodicity of two for the complex classes and of eight
for the real classes in Table \ref{periodic table}
follows from the Bott periodicity in K theory.

\section{Fractionalization from Abelian bosonization}
\label{sec: Fractionalization from Abelian bosonization}

\subsection{Introduction}

Abelian bosonization is attributed to Coleman \cite{Coleman75},
Mandelstam \cite{Mandelstam75},
and Luther and Peschel \cite{Luther75},
respectively. 
{However,
Abelian bosonization can be traced to the much older works of
Kronig \cite{Kronig35},
Tomonaga \cite{Tomonaga50},
and Luttinger \cite{Luttinger63},
in which it is observed that gapless fermions in one-dimensional space
are equivalent to gapless phonons.}
Here, we follow the more general formulation of
Abelian bosonization given by Haldane \cite{Haldane95},
as it lends itself to a description of 
one-dimensional quantum effective field theories
arising in the low-energy sector along the boundary in space of 
$(2+1)$-dimensional topological quantum field theories.

\subsection{Definition}

Let us define the quantum Hamiltonian (in units with the electric charge 
$e$, the characteristic speed, and $\hbar$ set to one)
\begin{subequations}
\label{eq: def chiral edge Hamiltonian}
\begin{equation}
\begin{split}
&
\widehat{H}\:=
\int\limits_{0}^{L}
\mathrm{d}x\,
\left[
\frac{1}{4\pi}\,
V^{\,}_{ij}\,
\left(D^{\,}_{x}\,\hat{u}^{\,}_{i}\right)
\left(D^{\,}_{x}\,\hat{u}^{\,}_{j}\right)
+
A^{\,}_{0}
\left(
\frac{q^{\,}_{i}}{2\pi}\,
K^{-1}_{ij}
\left(
D^{\,}_{x}\,\hat{u}^{\,}_{j}
\right)
\right)
\right](t,x),
\\
&
D^{\,}_{x}\,\hat{u}^{\,}_{i}(t,x)\:=
\left(
\partial^{\,}_{x}\,
\hat{u}^{\,}_{i}
+
q^{\,}_{i}\,
A^{\,}_{1}
\right)(t,x).
\end{split}
\label{eq: def chiral edge Hamiltonian a}
\end{equation}
The indices $i,j=1,\cdots,N$ label the bosonic modes.
Summation is implied for repeated indices.
The $N$ real-valued quantum fields 
$\hat{u}^{\,}_{i}(t,x)$ 
obey the equal-time commutation relations
\begin{equation}
\left[
\hat{u}^{\,}_{i}(t,x),
\hat{u}^{\,}_{j}(t,y)
\right]\:=
\mathrm{i}\pi
\left[
K^{\,}_{ij}\,
\mathrm{sgn}(x-y)
+
L^{\,}_{ij}
\right]
\label{eq: def chiral edge Hamiltonian b}
\end{equation}
for any pair $i,j=1,\cdots,N$.
The function $\mathrm{sgn}(x)=-\mathrm{sgn}(-x)$ 
gives the sign of the real variable $x$ and will be assumed
to be periodic with periodicity $L$.
The $N\times N$ matrix $K$ is symmetric, invertible, and
integer valued. Given the pair
$i,j=1,\cdots,N$,
any of its matrix elements thus obey 
\begin{equation}
K^{\,}_{ij}=
K^{\,}_{ji}
\in\mathbb{Z},
\qquad
K^{-1}_{ij}=
K^{-1}_{ji}
\in\mathbb{Q}.
\label{eq: def chiral edge Hamiltonian c}
\end{equation}
The $N\times N$ matrix $L$ is anti-symmetric 
\begin{equation}
L^{\,}_{ij}=
-L^{\,}_{ji}=
\left\{
\begin{array}{ll}
0,
&
\hbox{ if $i=j$},
\\&\\
\mathrm{sgn}(i-j)
\left(
K^{\,}_{ij}
+
q^{\,}_{i}\,
q^{\,}_{j}
\right),
&
\hbox{ otherwise,}
\end{array}
\right.
\label{eq: def chiral edge Hamiltonian d}
\end{equation}
for $i,j=1,\cdots,N$. The sign function $\mathrm{sgn}(i)$
of any integer $i$ is here not made periodic 
and taken to vanish at the origin of
$\mathbb{Z}$.
The external scalar gauge potential
$A^{\,}_{0}(t,x)$
and vector gauge potential
$A^{\,}_{1}(t,x)$
are real-valued functions of the time $t$ and space $x$ coordinates.
They are also chosen to be periodic under $x\mapsto x+L$.
The $N\times N$ matrix $V$ is symmetric and positive definite 
\begin{equation}
V^{\,}_{ij}=
V^{\,}_{ji}\in\mathbb{R},
\qquad
v^{\,}_{i}\,V^{\,}_{ij}\,v^{\,}_{j}>0,
\qquad
i,j=1,\cdots,N,
\label{eq: def chiral edge Hamiltonian e}
\end{equation}
for any nonvanishing vector $v=(v^{\,}_{i})\in\mathbb{R}^{N}$.
The charges $q^{\,}_{i}$ are integer valued and satisfy
\begin{equation}
(-1)^{K^{\,}_{ii}}=
(-1)^{q^{\,}_{i}},
\qquad
i=1,\cdots,N.
\label{eq: def chiral edge Hamiltonian f}
\end{equation}
Finally, we shall impose the boundary conditions
\begin{equation}
\hat{u}^{\,}_{i}(t,x+L)=
\hat{u}^{\,}_{i}(t,x)
+
2\pi n^{\,}_{i},
\qquad
n^{\,}_{i}\in\mathbb{Z},
\label{eq: def chiral edge Hamiltonian g}
\end{equation}
and
\begin{equation}
\left(
\partial^{\,}_{x}\,\hat{u}^{\,}_{i}
\right)(t,x+L)=
\left(
\partial^{\,}_{x}\,\hat{u}^{\,}_{i}
\right)(t,x),
\label{eq: def chiral edge Hamiltonian h}
\end{equation}
\end{subequations}
for any $i=1,\cdots,N$.

\subsection{Chiral equations of motion}

For any $i,j=1,\cdots,N$, 
one verifies with the help of
the equal-time commutation relations
\begin{equation}
\left[
\hat{u}^{\,}_{i}(t,x),
D^{\,}_{y}\,
\hat{u}^{\,}_{j}(t,y)
\right]=
-2\pi\mathrm{i}\,
K^{\,}_{ij}\,
\delta(x-y)
\label{eq: equal time commutation relation u partial u}
\end{equation}
that the equations of motions are 
\begin{align}
\mathrm{i}
\left(\partial^{\,}_{t}\,\hat{u}^{\,}_{i}\right)(t,x)\:=&\,
\left[
\hat{u}^{\,}_{i}(t,x),
\widehat{H}
\right]
\nonumber\\
=&\,
-
\mathrm{i}
K^{\,}_{ij}\,
V^{\,}_{jk}
\left(
\partial^{\,}_{x}\,\hat{u}^{\,}_{k}
+
q^{\,}_{k}\,
A^{\,}_{1}
\right)(t,x)
-
\mathrm{i}
q^{\,}_{i}\,
A^{\,}_{0}(t,x).
\label{eq: equations of motion bosonic chiral edge theory}
\end{align}
Introduce the covariant derivatives
\begin{equation}
D^{\,}_{\mu}\,\hat{u}^{\,}_{k}\:=
\left(
\partial^{\,}_{\mu}\,
\hat{u}^{\,}_{k}
+
q^{\,}_{k}\,A^{\,}_{\mu}
\right),
\qquad
\partial^{\,}_{0}\equiv
\partial^{\,}_{t},
\qquad
\partial^{\,}_{1}\equiv
\partial^{\,}_{x},
\end{equation}
for $\mu=0,1$ and $k=1,\cdots,N$.
The equations of motion 
\begin{equation}
0=
\delta^{\,}_{ik}\,
D^{\,}_{0}\,\hat{u}^{\,}_{k}
+
K^{\,}_{ij}\,
V^{\,}_{jk}\,
D^{\,}_{1}\,\hat{u}^{\,}_{k}
\label{eq: chiral equations of motion +}
\end{equation}
are chiral. Doing the substitutions 
$\hat{u}^{\,}_{i}\mapsto\hat{v}^{\,}_{i}$
and $K\mapsto-K$
everywhere in Eq.\ (\ref{eq: def chiral edge Hamiltonian}) 
delivers the chiral equations of motions
\begin{equation}
0=
\delta^{\,}_{ik}\,
D^{\,}_{0}\,\hat{v}^{\,}_{k}
-
K^{\,}_{ij}\,
V^{\,}_{jk}\,
D^{\,}_{1}\,\hat{v}^{\,}_{k},
\label{eq: chiral equations of motion -}
\end{equation}
with the opposite chirality.
Evidently, the chiral equations of motion 
(\ref{eq: chiral equations of motion +})
and
(\ref{eq: chiral equations of motion -})
are first-order differential equations,
as opposed to the Klein-Gordon equations of motion
obeyed by a relativistic quantum scalar field.

\subsection{Gauge invariance}

The chiral equations of motion 
(\ref{eq: chiral equations of motion +})
and
(\ref{eq: chiral equations of motion -})
are invariant under the local $U(1)$ gauge symmetry
\begin{subequations}
\label{eq: def gauge trsf for chiral edge}
\begin{equation}
\begin{split}
&
\hat{u}^{\,}_{i}(t,x)\ \=:\,
\hat{u}^{\prime}_{i}(t,x)
+
q^{\,}_{i}\,
\chi(t,x),
\\
&
A^{\,}_{0}(t,x)\=:\,
A^{\prime}_{0}(t,x)
-
\left(
\partial^{\,}_{t}\,\chi
\right)
(t,x),
\\
&
A^{\,}_{1}(t,x)\=:\,
A^{\prime}_{1}(t,x)
-
\left(
\partial^{\,}_{x}\,\chi
\right)
(t,x),
\end{split}
\label{eq: def gauge trsf for chiral edge a}
\end{equation}
for any real-valued function $\chi$
that satisfies the periodic boundary
conditions
\begin{equation}
\chi(t,x+L)=
\chi(t,x).
\end{equation}
\label{eq: def gauge trsf for chiral edge b}
\end{subequations}

Functional differentiation of 
Hamiltonian (\ref{eq: def chiral edge Hamiltonian a})
with respect to the gauge potentials allows to define
the two-current with the components
\begin{subequations}
\label{eq: def Jmu gauge invariant}
\begin{align}
\hat{J}^{0}(t,x)\:=&\,
\frac{\delta\widehat{H}}{\delta A^{\,}_{0}(t,x)}
=
\frac{1}{2\pi}\,
q^{\,}_{i}\,
K^{-1}_{ij}\,
\left(
D^{\,}_{1}\,\hat{u}^{\,}_{j}
\right)(t,x)
\label{eq: def Jmu gauge invariant a}
\end{align}
and
\begin{align}
\hat{J}^{1}(t,x)\:=&\,
\frac{\delta\widehat{H}}{\delta A^{\,}_{1}(t,x)}
=
\frac{1}{2\pi}\,
q^{\,}_{i}\,
V^{\,}_{ij}\,
\left(D^{\,}_{1}\,\hat{u}^{\,}_{j}\right)
(t,x)
+
\frac{1}{2\pi}\,
\left(
q^{\,}_{i}\,
K^{-1}_{ij}\,
q^{\,}_{j}
\right)
A^{\,}_{0}(t,x).
\label{eq: def Jmu gauge invariant b}
\end{align}
\end{subequations}
We introduce the short-hand notation
\begin{equation}
\sigma^{\,}_{\mathrm{H}}\:=
\frac{1}{2\pi}\,
\left(
q^{\,}_{i}\,
K^{-1}_{ij}\,
q^{\,}_{j}
\right)
\in
\frac{1}{2\pi}\,\mathbb{Q}
\label{eq: def sigma H in chairal bosonic theory}
\end{equation}
for the second term on the right-hand side of
Eq.\
(\ref{eq: def Jmu gauge invariant b}).
The subscript stands for Hall as we shall shortly interpret
$\sigma^{\,}_{\mathrm{H}}$ as a dimensionless
Hall conductance.

The transformation law of the two-current
(\ref{eq: def Jmu gauge invariant})
under the local gauge transformation
(\ref{eq: def gauge trsf for chiral edge}) 
is
\begin{subequations}
\label{eq: def Jmu gauge invariant bis}
\begin{align}
\hat{J}^{0}(t,x)=&\,
\hat{J}^{0\,\prime}(t,x)
\label{eq: def Jmu gauge invariant bis a}
\end{align}
and
\begin{align}
\hat{J}^{1}(t,x)=&\,
\hat{J}^{1\,\prime}(t,x)
-
\sigma^{\,}_{\mathrm{H}}\,
\left(
\partial^{\,}_{t}\chi
\right)(t,x).
\label{eq: def Jmu gauge invariant bis b}
\end{align}
\end{subequations}
The two-current
(\ref{eq: def Jmu gauge invariant})
is only invariant under gauge transformations
(\ref{eq: def gauge trsf for chiral edge})
that are static when $\sigma^{\,}_{\mathrm{H}}\neq0$. 

With the help of 
\begin{equation}
\left[
D^{\,}_{x}\,
\hat{u}^{\,}_{i}(t,x),
D^{\,}_{y}\,
\hat{u}^{\,}_{j}(t,y)
\right]=
-2\pi\mathrm{i}\,
K^{\,}_{ij}\,
\delta'(x-y)
\label{eq: equal time commutation relation partial u partial u}
\end{equation}
for $i,j=1,\cdots,N$,
one verifies 
that the {time} derivative of
$\hat{J}^{0}(t,x)$
is
\begin{align}
\frac{\partial\,\hat{J}^{0}}{\partial t}=&\,
-\mathrm{i}
\left[
\hat{J}^{0},
\widehat{H}
\right]
+
\sigma^{\,}_{H}\,
\frac{\partial\,A^{\,}_{1}}{\partial t}
\nonumber\\
=&\,
-
\frac{\partial\,\hat{J}^{1}}{\partial x}
+
\sigma^{\,}_{H}\,
\frac{\partial\,A^{\,}_{1}}{\partial t}.
\label{eq: total time derivative wideha J0}
\end{align}
There follows the continuity equation
\begin{equation}
\partial^{\,}_{\mu}\,\hat{J}^{\mu}=0
\label{eq: continuity equation when Amu static}
\end{equation}
provided $A^{\,}_{1}$ is time independent or $\sigma^{\,}_{H}=0$.
The continuity equation
(\ref{eq: continuity equation when Amu static})
delivers a conserved total charge if and only if
$A^{\,}_{0}$ and $A^{\,}_{1}$ are both static for arbitrary 
$\sigma^{\,}_{H}\neq0$. 

For any non-vanishing $\sigma^{\,}_{\mathrm{H}}$,
the continuity equation
\begin{equation}
\partial^{\,}_{\mu}\,\hat{J}^{\mu}=
\sigma^{\,}_{H}\,
\frac{\partial\,A^{\,}_{1}}{\partial t}
\label{eq: anomalous continuity equation when Amu time-dependent}
\end{equation}
is anomalous as soon as the vector gauge potential 
$A^{\,}_{1}$
is time dependent. The edge theory
(\ref{eq: def chiral edge Hamiltonian})
is said to be chiral when $\sigma^{\,}_{\mathrm{H}}\neq0$,
in which case the continuity equation
(\ref{eq: anomalous continuity equation when Amu time-dependent})
is anomalous. The anomalous continuity equation
(\ref{eq: anomalous continuity equation when Amu time-dependent})
is form covariant under any smooth gauge transformation
(\ref{eq: def gauge trsf for chiral edge}).
The choice of gauge may be fixed by the condition
\begin{subequations}
\label{eq: anomalous continuity equation when A0=0 gauge}
\begin{equation}
\frac{\partial\,A^{\,}_{0}}{\partial x}=0
\label{eq: anomalous continuity equation when A0=0 gauge a}
\end{equation}
for which the anomalous continuity equation
(\ref{eq: anomalous continuity equation when Amu time-dependent})
then becomes
\begin{equation}
\left(
\partial^{\,}_{\mu}\,\hat{J}^{\mu}
\right)(t,x)=
+
\sigma^{\,}_{H}\,E(t,x),
\label{eq: anomalous continuity equation when A0=0 gauge b}
\end{equation}
where 
\begin{equation}
E(t,x)\:=
+
\left(
\frac{\partial\,A^{\,}_{1}}{\partial t}
\right)(t,x)
\equiv
-
\left(
\frac{\partial\,A^{1}}{\partial t}
\right)(t,x)
\label{eq: anomalous continuity equation when A0=0 gauge c}
\end{equation}
\end{subequations}
represents the electric field in this gauge.

\begin{figure}[t]
\begin{center}
\includegraphics[angle=0,scale=0.35]{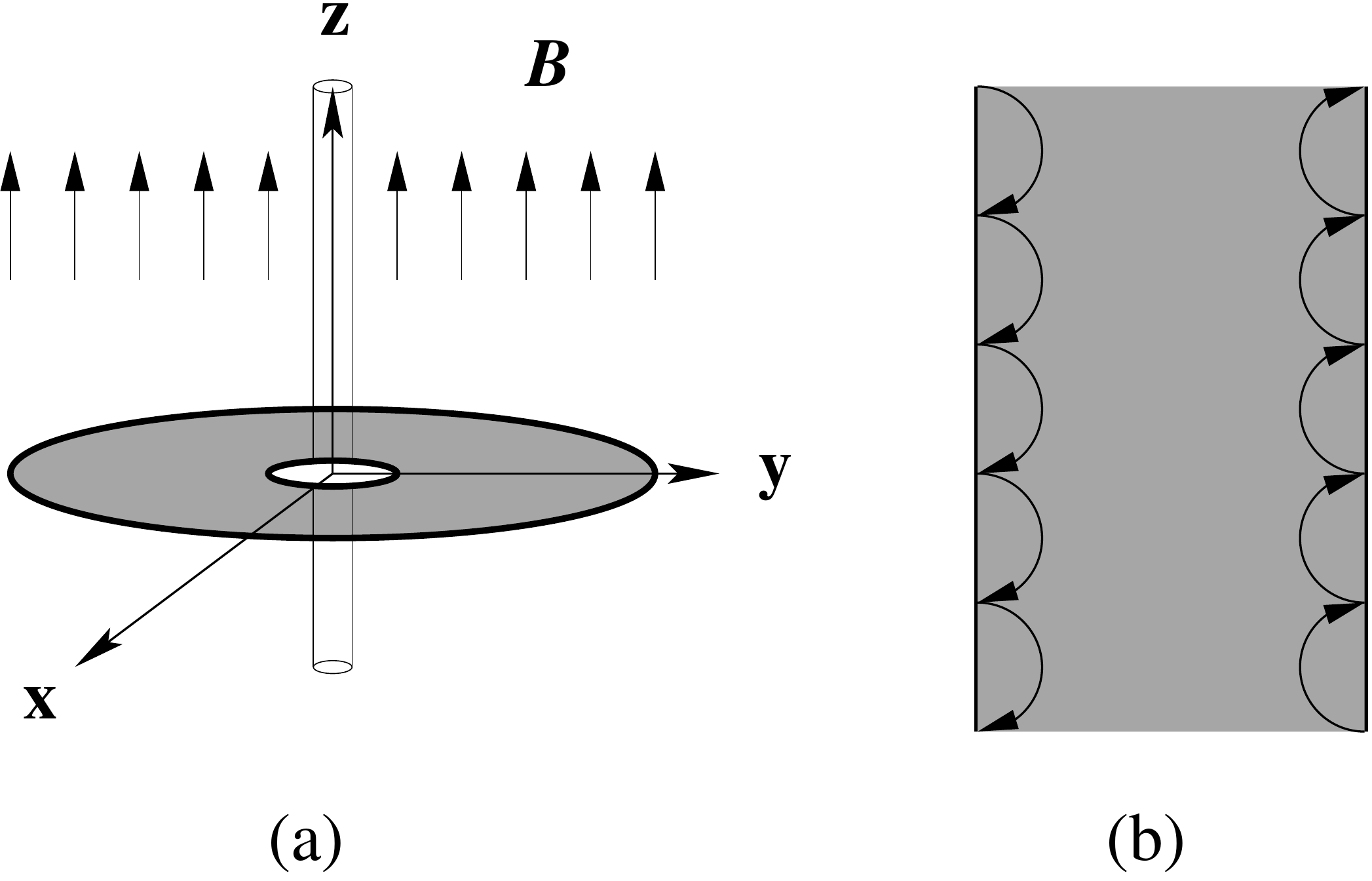}
\end{center}
\caption{
(a)
A ring of outer radius $R\equiv L/(2\pi)$ and inner radius $r$
in which electrons are confined. A uniform and static magnetic 
field $\bm{B}$ normal to the ring is present. The hierarchy
$\ell^{\,}_{B}\ll r\ll R$
of lengthscales is assumed,
where $\ell^{\,}_{B}\equiv \hbar\,c/|e\,B|$
is the magnetic length.
A time-dependent vector potential $\bm{A}(t,\bm{r})$
is induced by a time-dependent flux supported within
a solenoid of radius $r^{\,}_{\mathrm{sln}}\ll r$. 
This Corbino geometry has a cylindrical symmetry.
(b) The classical motion of electrons confined to a plane normal to
a uniform static magnetic field is circular. In the limit 
$R\to\infty$ holding $R/r$ fixed, the Corbino disk turns into a Hall bar.
An electron within a magnetic length of the boundary undergoes 
a classical skipping orbit. On quantization, 
a classical electron undergoing a skipping orbit 
turns into a chiral electron. On bosonization, a chiral electron
turns into a chiral boson.
        }
\label{fig: corbino-geometry}
\end{figure}

To interpret the anomalous continuity equation
(\ref{eq: anomalous continuity equation when A0=0 gauge})
of the bosonic chiral edge theory 
(\ref{eq: def chiral edge Hamiltonian}),
we recall that $x$ is a compact coordinate
because of the periodic boundary conditions
(\ref{eq: def chiral edge Hamiltonian g}),
(\ref{eq: def chiral edge Hamiltonian f}),
and
(\ref{eq: def gauge trsf for chiral edge b}). 
For simplicity, we assume
\begin{equation}
E(t,x)=E(t).
\end{equation}
The interval $0\leq x\leq L$ is thought of as a circle
of perimeter $L$ centered at the origin of the 
three-dimensional Euclidean space
as shown in Fig.\ \ref{fig: corbino-geometry}.
The vector potential $A^{1}(t)$ and the electric field
$E(t)=-\left(\frac{\partial\,A^{1}}{\partial t}\right)(t)$
along the circle of radius $R\equiv L/(2\pi)$
are then the polar components of a three-dimensional gauge
field $A^{\mu}(t,\bm{r})=(A^{0},\bm{A})(t,\bm{r})$
in a cylindrical geometry with the electro-magnetic fields 
\begin{equation}
\bm{E}(t,\bm{r})=
-(\bm{\nabla}\,A^{0})(t,\bm{r})
-(\partial^{\,}_{t}\,\bm{A})(t,\bm{r}),
\qquad
\bm{B}(t,\bm{r})=
(\bm{\nabla}\,\wedge\,\bm{A})(t,\bm{r}).
\end{equation}

The dimensionless Hall conductance 
$\sigma^{\,}_{\mathrm{H}}$
encodes the linear response of spin-polarized 
electrons confined to move along this circle in the presence of a
uniform and static magnetic field normal to the plane
that contains this circle. The time-dependent
anomalous term on the right-hand side of
the anomalous continuity equation
(\ref{eq: anomalous continuity equation when A0=0 gauge b})
is caused by a solenoid of radius $r^{\,}_{\mathrm{sln}}\ll r\ll R$ 
in a puncture of the plane that contains the circle of radius 
$r^{\,}_{\mathrm{sln}}$
supporting a time-dependent flux. The combination of this
time-dependent flux with the uniform static magnetic field
exerts a Lorentz force on spin-polarized electrons moving
along circles in the ring with the inner edge of radius $r$
and the outer edge of radius $R$. This Lorentz force causes
a net transfer of charge between the inner and outer edges
\begin{equation}
\frac{1}{L}\,\int\limits_{0}^{T}\mathrm{d}t\,Q(T)\:=
\int\limits_{0}^{T}\mathrm{d}t\,
\langle\hat{J}^{0}(t)\rangle=
\sigma^{\,}_{\mathrm{H}}\,
\int\limits_{0}^{T}\mathrm{d}t\,E(t)
\end{equation}
during the adiabatic evolution with period $T$ 
of the normalized many-body ground state of the outer edge,
provided we may identify the anomalous continuity equation
(\ref{eq: anomalous continuity equation when A0=0 gauge b})
with that of chiral spin-polarized electrons propagating 
along the outer edge in Fig.\
\ref{fig: corbino-geometry}. Hereto,
separating the many-body ground state
at the outer edge from all spin-polarized electrons 
supported between the inner and outer edge requires
the existence of an energy scale separating it from
many-body states in which these bulk 
spin-polarized electrons participate and by demanding
that the inverse of this energy scale, a length scale,
is much smaller than $R-r$. This energy scale
is brought about by the uniform and static magnetic field
$\bm{B}$ in Fig.\ \ref{fig: corbino-geometry}.
That none of this pumped charge is lost 
in the shaded region of the ring follows if it is assumed that
the spin-polarized electrons are unable to transport 
(dissipatively) a charge current 
across any circle of radius less than $R$ and greater than $r$.
The Hall conductance in the Corbino geometry of
Fig.\ \ref{fig: corbino-geometry}
is then a rank two anti-symmetric tensor proportional to
the rank two Levi-Civita anti-symmetric tensor
with $\sigma^{\,}_{\mathrm{H}}$ the proportionality constant
in units of $e^{2}/h$.
The charge density and current density for the ring obey
a continuity equation as full gauge invariance is restored
in the ring. 

The chiral bosonic theory 
(\ref{eq: def chiral edge Hamiltonian})
is nothing but a theory for chiral electrons at the outer edge
of the Corbino disk, as we still have to demonstrate.
Chiral fermions are a fraction of the original fermion
(a spin-polarized electron). More precisely, low-energy fermions
have been split into one half that propagate on the outer
edge and another half that propagate on the inner edge
of the Corbino disk. The price for this fractionalization 
is an apparent breakdown of gauge invariance and charge conservation,
when each chiral edge is treated independently from the other. 
Manifest charge conservation and gauge invariance are only restored 
if all low-energy degrees of freedom from the Corbino disk are treated 
on equal footing.

\subsection{Conserved topological charges}

Let us turn off the external gauge potentials
\begin{equation}
A^{\,}_{0}(t,x)=A^{\,}_{1}(t,x)=0.
\label{eq: no external scalar gauge potential}
\end{equation}
For any $i=1,\cdots,N$,
we define the operator
\begin{equation}
\begin{split}
\widehat{\mathcal{N}}^{\,}_{i}(t)\:=&\,
\frac{1}{2\pi}
\int\limits_{0}^{L}
\mathrm{d}x\,
\left(\partial^{\,}_{x}\hat{u}^{\,}_{i}\right)(t,x)
\\
=&\,
\frac{1}{2\pi}
\left[
\hat{u}^{\,}_{i}(t,L)
-
\hat{u}^{\,}_{i}(t,0)
\right].
\end{split}
\label{eq: def topo charge}
\end{equation}
This operator is conserved if and only if
\begin{equation}
\left(\partial^{\,}_{x}\hat{u}^{\,}_{i}\right)(t,x)=
\left(\partial^{\,}_{x}\hat{u}^{\,}_{i}\right)(t,x+L),
\qquad
0\leq x\leq L,
\end{equation}
for
\begin{equation} 
\begin{split}
\mathrm{i}
\left(
\partial^{\,}_{t}
\widehat{\mathcal{N}}^{\,}_{i}
\right)(t)
=&\,
-
\frac{\mathrm{i}}{2\pi}
K^{\,}_{ik}
V^{\,}_{kl}
\big[
\left(
\partial^{\,}_{x} \hat{u}^{\,}_{l}
\right)
(t,L)
-
\left(
\partial^{\,}_{x} \hat{u}^{\,}_{l}
\right)
(t,0)
\big].
\end{split}
\label{eq: commutator topo charge and chiral boson H}
\end{equation}
Furthermore, if we demand that there exists an 
$n^{\,}_{i}{\in\mathbb{Z}}$ such that
\begin{equation}
\hat{u}^{\,}_{i}(t,x+L)=
\hat{u}^{\,}_{i}(t,x)
-
2\pi n^{\,}_{i},
\end{equation}
it then follows that
\begin{equation}
\widehat{\mathcal{N}}^{\,}_{i}=
n^{\,}_{i}.
\label{eq: values taken by topological charges}
\end{equation}

{A corollary to Eq.\ 
(\ref{eq: values taken by topological charges})
is that the $N$ conserved topological charges
$\mathcal{N}^{\,}_{i}$ with $i=1,\cdots,N$
commute pairwise. The same conclusion follows from
the brute force manipulations}
\begin{equation}
\begin{split}
\left[
\widehat{\mathcal{N}}^{\,}_{i},
\widehat{\mathcal{N}}^{\,}_{j}
\right]=&\,
\frac{1}{2\pi}
\int\limits_{0}^{L}
\mathrm{d}y\,
\left[
\widehat{\mathcal{N}}^{\,}_{i},
\left(
\partial^{\,}_{y}\hat{u}^{\,}_{j}
\right)(y)
\right]
\\
=&\,
\frac{1}{2\pi}
\int\limits_{0}^{L}
\mathrm{d}y\,
\partial^{\,}_{y}
\left[
\widehat{\mathcal{N}}^{\,}_{i},
\hat{u}^{\,}_{j}(y)
\right],
\end{split}
\end{equation}
whereby $j=1,\cdots,N$ and 
\begin{equation} 
\begin{split}
\left[
\widehat{\mathcal{N}}^{\,}_{i},
\hat{u}^{\,}_{j}(y)
\right]
=&\,
\mathrm{i}
K^{\,}_{ij}
\end{split}
\label{eq: [ mathcal{N}, u]}
\end{equation}
is independent of $y$.

The local counterpart to the global conservation of
the topological charge is
\begin{subequations}
\label{eq: local tpo conservation laws}
\begin{equation}
\partial^{\,}_{t}\,
\hat{\rho}^{\text{top}}_{i}
+
\partial^{\,}_{x}\,
\hat{j}^{\text{top}}_{i}=0,
\label{eq: local tpo conservation laws a}
\end{equation}
where the local topological density operator is defined by
\begin{equation}
\hat{\rho}^{\text{top}}_{i}(t,x)\:=
\frac{1}{2\pi}\,
\left(\partial^{\,}_{x}\hat{u}^{\,}_{i}\right)(t,x)
\label{eq: local tpo conservation laws b}
\end{equation}
and the local topological current operator is defined by
\begin{equation}
\hat{j}^{\text{top}}_{i}(t,x)\:=
\frac{1}{2\pi}\,
K^{\,}_{ik}\,
V^{\,}_{kl}\,
\left(\partial^{\,}_{x}\hat{u}^{\,}_{l}\right)(t,x)
\label{eq: local tpo conservation laws c}
\end{equation}
\end{subequations}
for $i=1,\cdots,N$.
The local topological density operator obeys the
equal-time algebra
\begin{subequations}
\label{eq: topo current algebra}
\begin{equation}
\left[
\hat{\rho}^{\text{top}}_{i}(t,x),
\hat{\rho}^{\text{top}}_{j}(t,y)
\right]=
-
\frac{\mathrm{i}}{2\pi}\,
K^{\,}_{ij}\,
\partial^{\,}_{x}\delta(x-y)
\end{equation}
for any $i,j=1,\cdots,N$.
The local topological current operator obeys the
equal-time algebra
\begin{equation}
\left[
\hat{j}^{\text{top}}_{i}(t,x),
\hat{j}^{\text{top}}_{j}(t,y)
\right]=
-
\frac{\mathrm{i}}{2\pi}
K^{\,}_{ik }V^{\,}_{kl }\,
K^{\,}_{jk'}V^{\,}_{k'l'}\,
K^{\,}_{ll'}\,
\partial^{\,}_{x}\delta(x-y)
\end{equation}
for any $i,j=1,\cdots,N$.
Finally,
\begin{equation}
\left[
\hat{\rho}^{\text{top}}_{i}(t,x),
\hat{j}^{\text{top}}_{j}(t,y)
\right]=
-
\frac{\mathrm{i}}{2\pi}\,
K^{\,}_{jk}\,
V^{\,}_{kl}\,
K^{\,}_{il}\,
\partial^{\,}_{x}\delta(x-y)
\end{equation}
\end{subequations}
for any $i,j=1,\cdots,N$.

Let us introduce the local charges and currents
\begin{subequations}
\begin{equation}
\hat{\rho}^{\,}_{i}(t,x)\:=
K^{-1}_{ij}
\hat{\rho}^{\text{top}}_{j}(t,x)
\label{eq: local conservation laws a}
\end{equation}
and
\begin{equation}
\hat{j}^{\,}_{i}(t,x)\:=
K^{-1}_{ij}
\hat{j}^{\text{top}}_{j}(t,x),
\label{eq: local conservation laws b}
\end{equation}
respectively, for any $i=1,\cdots,N$.
The continuity equation (\ref{eq: local tpo conservation laws a})
is unchanged under this linear transformation,
\begin{equation}
\partial^{\,}_{t}\,
\hat{\rho}^{\,}_{i}
+
\partial^{\,}_{x}\,
\hat{j}^{\,}_{i}=0,
\label{eq: local conservation laws c}
\end{equation}
\end{subequations}
for any $i=1,\cdots,N$.
The topological current algebra (\ref{eq: topo current algebra})
transforms into
\begin{subequations}
\label{eq: current algebra}
\begin{eqnarray}
&&
\left[
\hat{\rho}^{\,}_{i}(t,x),
\hat{\rho}^{\,}_{j}(t,y)
\right]=
-
\frac{\mathrm{i}}{2\pi}\,
K^{-1}_{ij}\,
\partial^{\,}_{x}\delta(x-y),
\label{eq: current algebra a}\\
&&
\left[
\hat{j}^{\,}_{i}(t,x),
\hat{j}^{\,}_{j}(t,y)
\right]=
-
\frac{\mathrm{i}}{2\pi}\,
V^{\,}_{ik }\,
V^{\,}_{jl}\,
K^{\,}_{kl}\,
\partial^{\,}_{x}\delta(x-y),
\label{eq: current algebra b}\\
&&
\left[
\hat{\rho}^{\,}_{i}(t,x),
\hat{j}^{\,}_{j}(t,y)
\right]=
-
\frac{\mathrm{i}}{2\pi}\,
V^{\,}_{ij}\,
\partial^{\,}_{x}\delta(x-y),
\label{eq: current algebra c}
\end{eqnarray}
\end{subequations}
for any $i,j=1,\cdots,N$.

At last, if we contract the continuity equation
(\ref{eq: local conservation laws c})
with the integer-valued charge vector,
we obtain the flavor-global continuity equation
\begin{subequations}
\begin{equation}
\partial^{\,}_{t}\,
\hat{\rho}
+
\partial^{\,}_{x}\,
\hat{j}=0,
\label{eq: local global flavor conservation laws a}
\end{equation}
where the local flavor-global charge operator is
\begin{equation}
\hat{\rho}(t,x)\:=
q^{\,}_{i}\,
K^{-1}_{ij}\,
\hat{\rho}^{\text{top}}_{j}(t,x)
\label{eq: local global flavor conservation laws b}
\end{equation}
and the local flavor-global current operator is
\begin{equation}
\hat{j}(t,x)\:=
q^{\,}_{i}\,
K^{-1}_{ij}\,
\hat{j}^{\text{top}}_{j}(t,x).
\label{eq: local global flavor conservation laws c}
\end{equation}
\end{subequations}
The flavor-resolved current algebra~(\ref{eq: current algebra})
turns into the flavor-global  current algebra
\begin{subequations}
\label{eq: flavor-global current algebra}
\begin{align}
&
\left[
\hat{\rho}(t,x),
\hat{\rho}(t,y)
\right]=
-\frac{\mathrm{i}}{2\pi}
\left(
q^{\,}_{i}\,
K^{-1}_{ij}\,
q^{\,}_{j}
\right)\,
\partial^{\,}_{x}\delta(x-y),
\label{eq: flavor-global current algebra a}\\
&
\left[
\hat{j}(t,x),
\hat{j}(t,y)
\right]=
-\frac{\mathrm{i}}{2\pi}
\left(
q^{\,}_{i}\,
V^{\,}_{ik }\,
K^{\,}_{kl}\,
V^{\,}_{lj}\,
q^{\,}_{j}
\right)\,
\partial^{\,}_{x}\delta(x-y),
\label{eq: flavor-global current algebra b}\\
&
\left[
\hat{\rho}(t,x),
\hat{j}(t,y)
\right]=
-\frac{\mathrm{i}}{2\pi}
\left(
q^{\,}_{i}\,
V^{\,}_{ij}\,
q^{\,}_{j}
\right)\,
\partial^{\,}_{x}\delta(x-y).
\label{eq: flavor-global current algebra c}
\end{align}
\end{subequations}

\subsection{Quasi-particle and particle excitations}

When Eq.\ (\ref{eq: no external scalar gauge potential})
holds, there exist $N$ conserved global topological 
(i.e., integer valued) charges
$\widehat{\mathcal{N}}^{\,}_{i}$ with $i=1,\cdots,N$
defined in Eq.\ (\ref{eq: def topo charge})
that commute pairwise.
Let us define the $N$ global charges
\begin{equation}
\widehat{Q}^{\,}_{i}\:=
\int\limits_{0}^{L}
\mathrm{d}x\,
\hat{\rho}^{\,}_{i}(t,x)=
K^{-1}_{ij}\,
\widehat{\mathcal{N}}^{\,}_{j}, 
\qquad
i=1,\cdots,N.
\end{equation}
These charges shall shortly be interpreted as the elementary Fermi-Bose charges.

For any $i=1,\cdots,N$,
let us define the pair of vertex operators
\begin{subequations}
\label{eq: def vertex operators}
\begin{equation}
\widehat{\Psi}^{\dag}_{\text{q-p},i}(t,x)\:=
e^{-\mathrm{i} K^{-1}_{ij}\,\hat{u}^{\,}_{j}(t,x)}
\label{eq: def vertex operators a}
\end{equation}
and
\begin{equation}
\widehat{\Psi}^{\dag}_{\text{f-b},i}(t,x)\:=
e^{-\mathrm{i}\delta^{\,}_{ij}\,\hat{u}^{\,}_{j}(t,x)},
\label{eq: def vertex operators b}
\end{equation}
\end{subequations}
respectively.
The quasi-particle vertex operator 
$\widehat{\Psi}^{\dag}_{\text{q-p},i}(t,x)$
is multi-valued
under a shift by $2\pi$ of all $\hat{u}^{\,}_{j}(t,x)$
where $j=1,\cdots,N$.
The Fermi-Bose  vertex operator
$\widehat{\Psi}^{\dag}_{\text{f-b},i}(t,x)$
is single-valued
under a shift by $2\pi$ of all $\hat{u}^{\,}_{j}(t,x)$
where $j=1,\cdots,N$.

For any pair $i,j=1,\cdots,N$,
the commutator~(\ref{eq: [ mathcal{N}, u]})
delivers the identities
\begin{equation}
\left[
\widehat{\mathcal{N}}^{\,}_{i},
\widehat{\Psi}^{\dag}_{\text{q-p},j}(t,x)
\right]=
\delta^{\,}_{ij}\,
\widehat{\Psi}^{\dag}_{\text{q-p},j}(t,x),
\quad
\left[
\widehat{\mathcal{N}}^{\,}_{i},
\widehat{\Psi}^{\dag}_{\text{f-b},j}(t,x)
\right]=
K^{\,}_{ij}\,
\widehat{\Psi}^{\dag}_{\text{f-b},j}(t,x),
\end{equation}
and
\begin{equation}
\left[
\widehat{Q}^{\,}_{i},
\widehat{\Psi}^{\dag}_{\text{q-p},j}(t,x)
\right]=
K^{-1}_{ij}\,
\widehat{\Psi}^{\dag}_{\text{q-p},j}(t,x),
\quad\!\!
\left[
\widehat{Q}^{\,}_{i},
\widehat{\Psi}^{\dag}_{\text{f-b},j}(t,x)
\right]=
\delta^{\,}_{ij}\,
\widehat{\Psi}^{\dag}_{\text{f-b},j}(t,x),
\label{eq: flavor resolved Q trsf of Psi f-b}
\end{equation}
respectively.
The quasi-particle vertex operator 
$\widehat{\Psi}^{\dag}_{\text{q-p},i}(t,x)$
is an eigenstate of the topological number operator
$\widehat{\mathcal{N}}^{\,}_{i}$ with eigenvalue one.
The Fermi-Bose vertex operator
$\widehat{\Psi}^{\dag}_{\text{f-b},i}(t,x)$
is an eigenstate of the charge number operator
$\widehat{Q}^{\,}_{i}$ with eigenvalue one.

The Baker-Campbell-Hausdorff formula implies that
\begin{equation}
e^{\hat{A}}\,e^{\hat{B}}=
e^{\hat{A}+\hat{B}}\,e^{+(1/2)[\hat{A},\hat{B}]}=
e^{\hat{B}}\,e^{\hat{A}}\,e^{[\hat{A},\hat{B}]}
\label{eq: Baker-Campbell-Hausdorff}
\end{equation}
whenever two operators $\hat{A}$ and $\hat{B}$ have a $\mathbb{C}$-number
as their commutator. 

A first application of the
Baker-Campbell-Hausdorff formula
to any pair of quasi-particle vertex operator at equal time $t$ but
two distinct space coordinates $x\neq y$ gives
\begin{subequations}
\label{eq: flavor-resolved q-p statistics}
\begin{equation}
\begin{split}
\widehat{\Psi}^{\dag}_{\text{q-p},i}(t,x)\,
\widehat{\Psi}^{\dag}_{\text{q-p},j}(t,y)=&\,
e^{
-
\mathrm{i}\pi\, 
\Theta^{\text{q-p}}_{ij}
  }\,
\widehat{\Psi}^{\dag}_{\text{q-p},j}(t,y)\,
\widehat{\Psi}^{\dag}_{\text{q-p},i}(t,x),
\end{split}
\label{eq: flavor-resolved q-p statistics a}
\end{equation}
where
\begin{equation}
\Theta^{\text{q-p}}_{ij}\:=\,
K^{-1}_{ji}\,
\mathrm{sgn}(x-y)
+
\left(
K^{-1}_{ik}\,
K^{-1}_{jl}\,
K^{\,}_{kl}
+
q^{\,}_{k}\,
K^{-1}_{ik}\,
K^{-1}_{jl}\,
q^{\,}_{l}
\right)
\mathrm{sgn}(k-l).
\label{eq: flavor-resolved q-p statistics b}
\end{equation}
\end{subequations}
Here and below, it is understood that
\begin{equation}
\mathrm{sgn}(k-l)=0
\end{equation}
when $k=l=1,\cdots,N$.
Hence, the quasi-particle vertex operators obey neither bosonic nor
fermionic statistics since $K^{-1}_{ij}\in\mathbb{Q}$.

The same exercise applied to the Fermi-Bose
vertex operators yields
\begin{equation}
\begin{split}
\widehat{\Psi}^{\dag}_{\text{f-b},i}(t,x)\,
\widehat{\Psi}^{\dag}_{\text{f-b},j}(t,y)=&\,
\left\{
\begin{array}{ll}
(-1)^{K^{\,}_{ii}}\,
\widehat{\Psi}^{\dag}_{\text{f-b},i}(t,y)\,
\widehat{\Psi}^{\dag}_{\text{f-b},i}(t,x),
&
\hbox{ if $i=j$},
\\
&
\\
(-1)^{q^{\,}_{i}\,q^{\,}_{j}}\,
\widehat{\Psi}^{\dag}_{\text{f-b},j}(t,y)\,
\widehat{\Psi}^{\dag}_{\text{f-b},i}(t,x),
&
\hbox{ if $i\neq j$},
\end{array}
\right.
\end{split}
\label{eq: flavor-resolved f-b statistics}
\end{equation}
when $x\neq y$.
The self statistics of the Fermi-Bose vertex operators is
carried by the diagonal matrix elements $K^{\,}_{ii}\in\mathbb{Z}$.
The mutual statistics of any pair of
Fermi-Bose vertex operators labelled by $i\neq j$ is
carried by the product $q^{\  }_{i}\,q^{\  }_{j}\in\mathbb{Z}$ 
of the integer-valued charges
$q^{\  }_{i}$ and $q^{\  }_{j}$.
Had we not assumed that $K^{\,}_{ij}$ with $i\neq j$ are integers, 
the mutual statistics would not be Fermi-Bose because of the 
non-local term $K^{\,}_{ij}\mathrm{sgn}\,(x-y)$.

A third application of the Baker-Campbell-Hausdorff formula
allows to determine the boundary conditions
obeyed by the quasi-particle and Fermi-Bose vertex operators:
\begin{align}
&
\widehat{\Psi}^{\dag}_{\text{q-p},i}(t,x+L)=
\widehat{\Psi}^{\dag}_{\text{q-p},i}(t,x)\,
e^{-2\pi\mathrm{i}\,K^{-1}_{ij}\widehat{\mathcal{N}}^{\,}_{j}}\,
e^{-\pi\mathrm{i}\,K^{-1}_{ii}},
\label{eq: boundary conditions of flavor-resolved q-p}
\\
&
\widehat{\Psi}^{\dag}_{\text{f-b},i}(t,x+L)=
\widehat{\Psi}^{\dag}_{\text{f-b},i}(t,x)\,
e^{-2\pi\mathrm{i}\,\widehat{\mathcal{N}}^{\,}_{i}}\,
e^{-\pi\mathrm{i}\,K^{\,}_{ii}}.
\label{eq: boundary conditions of flavor-resolved f-b}
\end{align}
{
The quasi-particle vertex operators obey twisted boundary conditions.
The Fermi vertex operators obey anti-periodic boundary conditions.
The Bose vertex operators obey periodic boundary conditions.}

We close this discussion with the following definitions.
We introduce the operators
\begin{equation}
\widehat{Q}\:=
q^{\,}_{i}\,
\widehat{Q}^{\,}_{i},
\quad
\widehat{\Psi}^{\dag}_{\text{q-p},\bm{m}}\:=
e^{
-\mathrm{i} m^{\,}_{i}K^{-1}_{ij}\hat{u}^{\,}_{j}(t,x)
  },
\qquad
\widehat{\Psi}^{\dag}_{\text{f-b},\bm{m}}\:=
e^{
-\mathrm{i} m^{\,}_{i}\delta^{\,}_{ij}\hat{u}^{\,}_{j}(t,x)
  },
\end{equation}
where $\bm{m}\in\mathbb{Z}^{N}$ is the vector with the 
integer-valued components $m^{\,}_{i}$ for any $i=1,\cdots,N$.
The $N$ charges $q^{\,}_{i}$ with $i=1,\cdots,N$
that enter Hamiltonian~(\ref{eq: def chiral edge Hamiltonian a}) 
can also be viewed as the components of the vector $\bm{q}\in\mathbb{Z}^{N}$.
Let us define the functions
\begin{subequations}
\begin{equation}
\begin{split}
q:\mathbb{Z}^{N}&\rightarrow \mathbb{Z},
\\
\bm{m}&\mapsto
q(\bm{m})\:=
q^{\,}_{i}\,m^{\,}_{i}\equiv\bm{q}\cdot\bm{m},
\end{split}
\end{equation}
and
\begin{equation}
\begin{split}
K:\mathbb{Z}^{N}&\rightarrow \mathbb{Z},
\\
\bm{m}&\mapsto
K(\bm{m})\:=
m^{\,}_{i}\,
K^{\,}_{ij}\,
m^{\,}_{j}.
\end{split}
\end{equation}
\end{subequations}
On the one hand,
for any distinct pair of space coordinate
$x\neq y$, we deduce from 
Eqs.\ (\ref{eq: flavor resolved Q trsf of Psi f-b}),
(\ref{eq: flavor-resolved q-p statistics}),
and (\ref{eq: boundary conditions of flavor-resolved q-p})
that
\begin{subequations}
\begin{align}
&
\left[
\widehat{Q},
\widehat{\Psi}^{\dag}_{\text{q-p},\bm{m}}(t,x)
\right]=
\left(q^{\,}_{i}\,K^{-1}_{ij}\,m^{\,}_{j}\right)
\widehat{\Psi}^{\dag}_{\text{q-p},\bm{m}}(t,x),
\\
&
\widehat{\Psi}^{\dag}_{\text{q-p},\bm{m}}(t,x)\,
\widehat{\Psi}^{\dag}_{\text{q-p},\bm{n}}(t,y)=
e^{
-\mathrm{i}\pi\, m^{\,}_{i}\,\Theta^{\text{q-p}}_{ij}\,n^{\,}_{j} 
  }\,
\widehat{\Psi}^{\dag}_{\text{q-p},\bm{n}}(t,y)\,
\widehat{\Psi}^{\dag}_{\text{q-p},\bm{m}}(t,x),
\\
&
\widehat{\Psi}^{\dag}_{\text{q-p},\bm{m}}(t,x+L)=
\widehat{\Psi}^{\dag}_{\text{q-p},\bm{m}}(t,x)\,
e^{-2\pi\mathrm{i}\,m^{\,}_{i}\,K^{-1}_{ij}\,\widehat{\mathcal{N}}^{\,}_{j}}\,
e^{-\pi\mathrm{i}\,m^{\,}_{i}\,K^{-1}_{ij}\,m^{\,}_{j}},
\end{align}
\end{subequations}
respectively. 
On the other hand,
for any distinct pair of space coordinate
$x\neq y$, we deduce from Eqs.
(\ref{eq: flavor resolved Q trsf of Psi f-b}),
(\ref{eq: flavor-resolved f-b statistics}),
and (\ref{eq: boundary conditions of flavor-resolved f-b})
that
\begin{subequations}
\begin{align}
&
\left[
\widehat{Q},
\widehat{\Psi}^{\dag}_{\text{f-b},\bm{m}}(t,x)
\right]=
q(\bm{m})\,
\widehat{\Psi}^{\dag}_{\text{f-b},\bm{m}}(t,x),
\\
&
\widehat{\Psi}^{\dag}_{\text{f-b},\bm{m}}(t,x)\,
\widehat{\Psi}^{\dag}_{\text{f-b},\bm{n}}(t,y)=
e^{-\mathrm{i}\pi\,m^{\,}_{i}\,\Theta^{\text{f-b}}_{ij}\,n^{\,}_{j}}\,
\widehat{\Psi}^{\dag}_{\text{f-b},\bm{n}}(t,y)\,
\widehat{\Psi}^{\dag}_{\text{f-b},\bm{m}}(t,x),
\\
&
\widehat{\Psi}^{\dag}_{\text{f-b},\bm{m}}(t,x+L)=
\widehat{\Psi}^{\dag}_{\text{f-b},\bm{m}}(t,x)\,
e^{-2\pi\mathrm{i}\,m^{\,}_{i}\widehat{\mathcal{N}}^{\,}_{i}}\,
e^{-\pi\mathrm{i}\,m^{\,}_{i}K^{\,}_{ij}m^{\,}_{j}},
\end{align}
respectively, where
\begin{equation}
\Theta^{\text{f-b}}_{ij}\:=
K^{\,}_{ij}\,
\mathrm{sgn}(x-y)
+
\left(
K^{\,}_{ij}
+
q^{\,}_{i}\,
q^{\,}_{j}
\right)\,
\mathrm{sgn}(i-j).
\end{equation}
\end{subequations}
The integer quadratic form $K(\bm{m})$
is thus seen to dictate whether the vertex operator
$\widehat{\Psi}^{\dag}_{\text{f-b},\bm{m}}(t,x)$
realizes a fermion or a boson. 
The vertex operator
$\widehat{\Psi}^{\dag}_{\text{f-b},\bm{m}}(t,x)$
realizes a fermion if and only if
\begin{equation}
K(\bm{m}) \hbox{ is an odd integer}
\end{equation}
or a boson
if and only if
\begin{equation}
K(\bm{m}) \hbox{ is an even integer}.
\end{equation}
Because of assumption~(\ref{eq: def chiral edge Hamiltonian f}),
\begin{equation}
(-1)^{K(\bm{m})}=(-1)^{q(\bm{m})}.
\end{equation}
Hence, the vertex operator
$\widehat{\Psi}^{\dag}_{\text{f-b},\bm{m}}(t,x)$
realizes 
\begin{itemize}
\item
a fermion if and only if
$q(\bm{m})$ is an odd integer;
\item
a boson if and only if
$q(\bm{m})$ is an even integer.
\end{itemize}

\subsection{Bosonization rules}

We are going to relate the theory of chiral bosons
(\ref{eq: def chiral edge Hamiltonian})
without external gauge fields
to a massless Dirac Hamiltonian.
To this end, we proceed in three steps.

\textbf{Step 1.}
Make the following choices in Eq.\ (\ref{eq: def chiral edge Hamiltonian}): 
\begin{subequations}
\begin{equation}
N=2, 
\qquad
i,j=1,2\equiv-,+,
\end{equation}
 and
\begin{equation}
K\:=
\begin{pmatrix}
+1
&
0
\\
0
&
-1
\end{pmatrix},
\qquad
V\:=
\begin{pmatrix}
+1
&
0
\\
0
&
+1
\end{pmatrix},
\qquad
\bm{q}=
\begin{pmatrix}
1
\\
1
\end{pmatrix}.
\end{equation}
\end{subequations}
With these choices, 
the free bosonic Hamiltonian 
on the real line is
\begin{subequations}
\label{eq: def free bosonic chiral theory N=2}
\begin{equation}
\widehat{H}^{\,}_{\mathrm{B}}=
\int\limits_{\mathbb{R}}\mathrm{d}x\,
\frac{1}{4\pi}
\left[
(\partial^{\,}_{x}\,\hat{u}^{\,}_{-})^{2}
+
(\partial^{\,}_{x}\,\hat{u}^{\,}_{+})^{2}
\right],
\end{equation}
where
\begin{align}
&
[\hat{u}^{\,}_{-}(t,x),\hat{u}^{\,}_{-}(t,y)]=
+\mathrm{i}\pi\,\mathrm{sgn}(x-y),
\\
&
[\hat{u}^{\,}_{+}(t,x),\hat{u}^{\,}_{+}(t,y)]=
-\mathrm{i}\pi\,\mathrm{sgn}(x-y),
\\
&
[\hat{u}^{\,}_{-}(t,x),\hat{u}^{\,}_{+}(t,y)]=
+\mathrm{i}\pi.
\end{align}
\end{subequations}
From this, there follow the chiral equations of motion
[recall Eq.\ (\ref{eq: equations of motion bosonic chiral edge theory})]
\begin{equation}
\partial^{\,}_{t}\,\hat{u}^{\,}_{-}=
-
\partial^{\,}_{x}\,\hat{u}^{\,}_{-},
\qquad
\partial^{\,}_{t}\,\hat{u}^{\,}_{+}=
+
\partial^{\,}_{x}\,\hat{u}^{\,}_{+},
\label{eq: equations motion for u- and u+}
\end{equation}
obeyed by the right-mover $\hat{u}^{\,}_{-}$ and the left-mover
$\hat{u}^{\,}_{+}$
\begin{subequations}
\label{eq: flavor-global current algebra N=2}
\begin{align}
&
\left[\vphantom{\hat{j}}
\hat{\rho}(t,x),
\hat{\rho}(t,y)
\right]=
0,
\label{eq: flavor-global current algebra N=2 a}\\
&
\left[
\hat{j}(t,x),
\hat{j}(t,y)
\right]\,=
0,
\label{eq: flavor-global current algebra N=2 b}\\
&
\left[
\hat{\rho}(t,x),
\hat{j}(t,y)
\right]\,=
-\frac{\mathrm{i}}{\pi}
\partial^{\,}_{x}\delta(x-y),
\label{eq: flavor-global current algebra N=2 c}
\end{align}
obeyed by the density
\begin{equation}
\begin{split}
\hat{\rho}=&\,
+
\frac{1}{2\pi}
\left(
\partial^{\,}_{x}\,\hat{u}^{\,}_{-}
-
\partial^{\,}_{x}\,\hat{u}^{\,}_{+}
\right)
\equiv
\hat{j}^{\,}_{-}
+
\hat{j}^{\,}_{+}
\end{split}
\label{eq: flavor-global current algebra N=2 d}
\end{equation}
and the current density
\footnote{%
Notice that the chiral equations of motion implies that
$
\hat{j}=
-
\frac{1}{2\pi}
\left(
\partial^{\,}_{t}\,\hat{u}^{\,}_{-}
-
\partial^{\,}_{t}\,\hat{u}^{\,}_{+}
\right)
$.
          }
\begin{equation}
\begin{split}
\hat{j}=&\,
+
\frac{1}{2\pi}
\left(
\partial^{\,}_{x}\,\hat{u}^{\,}_{-}
+
\partial^{\,}_{x}\,\hat{u}^{\,}_{+}
\right)
\equiv
\hat{j}^{\,}_{-}
-
\hat{j}^{\,}_{+},
\end{split}
\label{eq: flavor-global current algebra N=2 e}
\end{equation}
\end{subequations}
and the identification of the pair of vertex operators 
[recall Eq.\ (\ref{eq: def vertex operators b})]
\begin{equation}
\hat{\psi}^{\dag}_{-}\:=
\sqrt{\frac{1}{4\pi\,\mathfrak{a}}}\,
e^{-\mathrm{i}\hat{u}^{\,}_{-}},
\qquad
\hat{\psi}^{\dag}_{+}\:=
\sqrt{\frac{1}{4\pi\,\mathfrak{a}}}\,
e^{+\mathrm{i}\hat{u}^{\,}_{+}},
\end{equation}
with a pair of creation operators for fermions.
The multiplicative prefactor $1/\sqrt{4\pi}$ is a matter of convention
and the constant $\mathfrak{a}$ carries the dimension of length, i.e.,
the fermion fields carries the dimension of $1/\sqrt{\hbox{length}}$.
By construction, the chiral currents
\begin{equation}
\hat{j}^{\,}_{-}\:=
+\frac{1}{2\pi}\,\partial^{\,}_{x}\,\hat{u}^{\,}_{-},
\qquad
\hat{j}^{\,}_{+}\:=
-\frac{1}{2\pi}\,\partial^{\,}_{x}\,\hat{u}^{\,}_{+},
\end{equation}
obey the chiral equations of motion
\begin{equation}
\partial^{\,}_{t}\,\hat{j}^{\,}_{-}=
-\partial^{\,}_{x}\,\hat{j}^{\,}_{-},
\qquad
\partial^{\,}_{t}\,\hat{j}^{\,}_{+}=
+\partial^{\,}_{x}\,\hat{j}^{\,}_{+},
\end{equation}
i.e., they depend solely on $(t-x)$ and $(t+x)$, respectively.
As with the chiral fields $\hat{u}^{\,}_{-}$ and $\hat{u}^{\,}_{+}$, the chiral
currents $\hat{j}^{\,}_{-}$ and $\hat{j}^{\,}_{+}$ are 
right-moving and left-moving solutions, respectively, 
of the Klein-Gordon equation
\begin{equation}
(\partial^{2}_{t}-\partial^{2}_{x})\,f(t,x)=
(\partial^{\,}_{t}-\partial^{\,}_{x})\,
(\partial^{\,}_{t}+\partial^{\,}_{x})\,f(t,x)=
0.
\end{equation}

\textbf{Step 2.}
Let us define the free Dirac Hamiltonian
\begin{subequations}
\label{eq: def free Dirac theory}
\begin{equation}
\widehat{H}^{\,}_{\mathrm{D}}\:=
{-}
\int\limits_{\mathbb{R}}\mathrm{d}x\,
\hat{\psi}^{\dag}\,
\gamma^{0}\,\gamma^{1}\,
\mathrm{i}\partial^{\,}_{x}\,
\hat{\psi}
\equiv
\int\limits_{\mathbb{R}}\mathrm{d}x\,
\hat{\bar{\psi}}\,
\gamma^{1}\,
\mathrm{i}\partial^{\,}_{x}\,
\hat{\psi},
\label{eq: def free Dirac theory a}
\end{equation}
where
\begin{equation}
\left\{
\hat{\psi}^{\,}_{\alpha}(t,x),
\hat{\psi}^{\dag}_{\beta}(t,y)
\right\}=
\delta^{\,}_{\alpha\beta}\,
\delta(x-y)
\label{eq: def free Dirac theory b}
\end{equation}
\end{subequations}
delivers the only non-vanishing equal-time anti-commutators.
If we define the chiral projections
($\gamma^{\,}_{5}\equiv-\gamma^{5}\equiv-\gamma^{0}\gamma^{1}$)
\begin{subequations}
\begin{equation}
\hat{\psi}^{\dag}_{\mp}\:=
\hat{\psi}^{\dag}\,
\frac{1}{2}(1\mp\gamma^{\,}_{5}),
\qquad
\hat{\psi}^{\,}_{\mp}\:=
\frac{1}{2}(1\mp\gamma^{\,}_{5})\,
\hat{\psi}^{\,},
\end{equation}
there follows the chiral equations of motion
\begin{equation}
\partial^{\,}_{t}\,\hat{\psi}^{\,}_{-}=
-
\partial^{\,}_{x}\,\hat{\psi}^{\,}_{-},
\qquad
\partial^{\,}_{t}\,\hat{\psi}^{\,}_{+}=
+
\partial^{\,}_{x}\,\hat{\psi}^{\,}_{+}.
\end{equation}
\end{subequations}
The annihilation operator
$\hat{\psi}^{\,}_{-}$
removes a right-moving fermion.
The annihilation operator
$\hat{\psi}^{\,}_{+}$
removes a left-moving fermion.
Moreover, the Lagrangian density
\begin{equation}
{\widehat{\mathcal{L}}}^{\,}_{\mathrm{D}}\:=
\hat{\psi}^{\dag}\,
\gamma^{0}\,
\mathrm{i}\gamma^{\mu}\,
\partial^{\,}_{\mu}\,
\hat{\psi}
\end{equation}
obeys the additive decomposition
\begin{equation}
{\widehat{\mathcal{L}}}^{\,}_{\mathrm{D}}=
\hat{\psi}^{\dag}_{-}\,
\mathrm{i}(\partial^{\,}_{0}+\partial^{\,}_{1})\,
\hat{\psi}^{\,}_{-}
+
\hat{\psi}^{\dag}_{+}\,
\mathrm{i}(\partial^{\,}_{0}-\partial^{\,}_{1})\,
\hat{\psi}^{\,}_{+}
\end{equation}
with the two independent chiral currents
\begin{subequations}
\begin{equation}
\hat{j}^{\,}_{\mathrm{D}-}\:=
2\,
\hat{\psi}^{\dag}_{-}\,
\hat{\psi}^{\,}_{-},
\qquad
\hat{j}^{\,}_{\mathrm{D}+}\:=
2\,
\hat{\psi}^{\dag}_{+}\,
\hat{\psi}^{\,}_{+},
\end{equation}
obeying the independent conservation laws
\begin{equation}
\partial^{\,}_{t}\,\hat{j}^{\,}_{\mathrm{D}-}=
-
\partial^{\,}_{x}\,\hat{j}^{\,}_{\mathrm{D}-},
\qquad
\partial^{\,}_{t}\,\hat{j}^{\,}_{\mathrm{D}+}=
+
\partial^{\,}_{x}\,\hat{j}^{\,}_{\mathrm{D}+}.
\end{equation}
\end{subequations}
Finally, it can be shown that if the chiral currents are normal ordered
with respect to the filled Fermi sea with a vanishing chemical potential,
then the only non-vanishing equal-time commutators are
\begin{subequations}
\label{eq: Dirac current algebra}
\begin{equation}
\left[\hat{j}^{\,}_{\mathrm{D}-}(t,x),\hat{j}^{\,}_{\mathrm{D}-}(t,y)\right]=
-\frac{\mathrm{i}}{2\pi}\,
\partial^{\,}_{x}\,\delta(x-y),
\end{equation}
and
\begin{equation}
\left[\hat{j}^{\,}_{\mathrm{D}+}(t,x),\hat{j}^{\,}_{\mathrm{D}+}(t,y)\right]=
+\frac{\mathrm{i}}{2\pi}\,
\partial^{\,}_{x}\,\delta(x-y).
\end{equation}
\end{subequations}

\textbf{Step 3.}
The Dirac chiral current algebra (\ref{eq: Dirac current algebra})
is equivalent to the bosonic chiral current algebra
(\ref{eq: flavor-global current algebra N=2}). 
This equivalence is interpreted as the fact that
(i) the bosonic theory 
(\ref{eq: def free bosonic chiral theory N=2})
is equivalent to the Dirac theory 
(\ref{eq: def free Dirac theory}),
and (ii) there is a one-to-one correspondence between the following
operators acting on their respective Fock spaces.
To establish this one-to-one correspondence,
we introduce the pair of bosonic fields
\begin{subequations}
\begin{align}
&
\hat{\phi}(x^{0},x^{1})\:=
\hat{u}^{\,}_{-}(x^{0}-x^{1})
+
\hat{u}^{\,}_{+}(x^{0}+x^{1}),
\\
&
\hat{\theta}(x^{0},x^{1})\:=
\hat{u}^{\,}_{-}(x^{0}-x^{1})
-
\hat{u}^{\,}_{+}(x^{0}+x^{1}).
\end{align}
\end{subequations}
Now, the relevant one-to-one correspondence between operators in
the Dirac theory for fermions and operators in
the chiral bosonic theory is given in Table 
\ref{table: Abelian bosonization in 1d}. 

\begin{table}[t]
\caption{\label{table: Abelian bosonization in 1d}
Abelian bosonization rules in two-dimensional Minkowski space.
The conventions with regard to
the scalar mass $\hat{\bar{\psi}}\,\hat{\psi}$ and the pseudoscalar
mass $\hat{\bar{\psi}}\,\gamma^{\,}_{5}\,\hat{\psi}$
are
$\hat{\bar{\psi}}=\hat{\psi}^{\dag}\,\gamma^{0}$
with $\hat{\psi}^{\dag}=(\hat{\psi}^{\dag}_{-},\hat{\psi}^{\dag}_{+})$,
whereby 
$\gamma^{0}=\tau^{\,}_{1}$ and
$\gamma^{1}=\mathrm{i}\tau^{\,}_{2}$
so that
$\gamma^{5}=-\gamma^{\,}_{5}=+\gamma^{0}\,\gamma^{1}=-\tau^{\,}_{3}$.
        }
\begin{center}
\begin{tabular}{l|cc}
&
Fermions
&
Bosons
\\
\hline
\\&&\\
\hbox{Kinetic energy}
&
$
\hat{\bar{\psi}}\,
\mathrm{i}\gamma^{\mu}\,\partial^{\,}_{\mu}\,
\hat{\psi}
$
&
$
\frac{1}{8\pi}\,
(\partial^{\mu}\,\hat{\phi})
(\partial^{\,}_{\mu}\,\hat{\phi})
$
\\&&\\
\hbox{Current}
&
$
\hat{\bar{\psi}}\,\gamma^{\mu}\,\hat{\psi}
$
&
$
\frac{1}{2\pi}\,
\epsilon^{\mu\nu}\,\partial^{\,}_{\nu}\,\hat{\phi}
$
\\&&\\
\hbox{Chiral currents}
&
$
2\,
\hat{\psi}^{\dag}_{\mp}\,
\hat{\psi}^{\,}_{\mp}
$
&
$
\pm
\frac{1}{2\pi}
\partial^{\,}_{x}\,\hat{u}^{\,}_{\mp}
$
\\&&\\
\hbox{Right and left movers}
&
$
\hat{\psi}^{\dag}_{\mp}
$
&
$
\sqrt{\frac{1}{4\pi\,\mathfrak{a}}}\,
e^{\mp\mathrm{i}\hat{u}^{\,}_{\mp}}
$
\\&&\\
\hbox{Backward scattering}
&
$
\hat{\psi}^{\dag}_{-}\,
\hat{\psi}^{\,}_{+}
$
&
$
\frac{1}{4\pi\,\mathfrak{a}}\,
e^{-\mathrm{i}\hat{\phi}}
$
\\&&\\
\hbox{Cooper pairing}
&
$
\hat{\psi}^{\dag}_{-}\,
\hat{\psi}^{\dag}_{+}
$
&
$
\frac{1}{4\pi\,\mathfrak{a}}\,
e^{-\mathrm{i}\hat{\theta}}
$
\\&&\\
\hbox{Scalar mass}
&
$
\hat{\psi}^{\dag}_{-}\,
\hat{\psi}^{\,}_{+}
+
\hat{\psi}^{\dag}_{+}\,
\hat{\psi}^{\,}_{-}
$
&
$
\frac{1}{2\pi\,\mathfrak{a}}\,\cos\hat{\phi}
$
\\&&\\
\hbox{Pseudo-scalar mass}
&
$
\hat{\psi}^{\dag}_{-}\,
\hat{\psi}^{\,}_{+}
-
\hat{\psi}^{\dag}_{+}\,
\hat{\psi}^{\,}_{-}
$
&
$
\frac{-\mathrm{i}}{2\pi\,\mathfrak{a}}\,\sin\hat{\phi}
$
\end{tabular}
\end{center}
\end{table}

\subsection{From the Hamiltonian to the Lagrangian formalism}

What is the Minkowski path integral
that is equivalent to the quantum theory defined by 
Eq.\ (\ref{eq: def chiral edge Hamiltonian})?
In other words, we seek the path integrals
\begin{subequations}
\begin{equation}
Z^{(\pm)}\:=
\int\mathcal{D}[u]\,
e^{\mathrm{i}S^{(\pm)}[u]}
\end{equation}
with the Minkowski action{s}
\begin{equation}
S^{(\pm)}[u]\:=
\int\limits_{-\infty}^{+\infty}\mathrm{d}t\,
L^{(\pm)}[u]\equiv
\int\limits_{-\infty}^{+\infty}\mathrm{d}t\,
\int\limits_{0}^{L}\mathrm{d}x\,
\mathcal{L}^{(\pm)}[u](t,x)
\end{equation}
\end{subequations}
such that one of the two Hamiltonians
\begin{equation}
H^{(\pm)}\:=
\int\limits_{0}^{L}\mathrm{d}x\,
\Big[
\Pi^{(\pm)}_{i}
\left(\partial^{\,}_{t}\,u^{\,}_{i}\right)
-
\mathcal{L}^{(\pm)}[u]
\Big]
\end{equation}
can be identified with $\widehat{H}$ in Eq.
(\ref{eq: def chiral edge Hamiltonian a})
after elevating the classical fields
\begin{subequations}
\begin{equation}
u^{\,}_{i}(t,x)
\end{equation}
and
\begin{equation}
\Pi^{(\pm)}_{i}(t,x)\:=
\frac{
\delta\,\mathcal{L}^{(\pm)}
     }
     {
\delta(\partial^{\,}_{t}\,u^{\,}_{i})(t,x)
     }
\end{equation}
entering $\mathcal{L}^{(\pm)}[u]$
to the status of quantum fields
$\hat{u}^{\,}_{i}(t,x)$
and
$\hat{\Pi }^{(\pm)}_{j}(t,y)$
upon imposing the equal-time commutation relations
\begin{equation}
\left[
\hat{u}^{\,}_{i}(t,x),
\hat{\Pi }^{(\pm)}_{j}(t,y)
\right]=
\pm
\frac{\mathrm{i}}{2}\,
\delta^{\,}_{ij}\,
\delta(x-y)
\label{eq: def chiral canonical quantization rule}
\end{equation}
\end{subequations}
for any $i,j=1,\cdots,N$. The unusual factor $\pm1/2$ 
(instead of $1$) on the right-hand side
of the commutator between pairs of canonically conjugate fields
arises because each scalar field $u^{\,}_{i}$ with $i=1,\cdots,N$ 
is chiral, i.e., it represents ``one-half'' of a canonical scalar field.

{Without loss of generality, we set $A^{\,}_{0}=A^{\,}_{1}=0$
in Eq.\ (\ref{eq: def chiral edge Hamiltonian a})}.
We try
\begin{subequations}
\label{eq: trial Haldane Lagrangian}
\begin{equation}
\mathcal{L}^{(\pm)}\:=
\frac{1}{4\pi}
\left[
\mp
\left(
\partial^{\,}_{x}\,u^{\,}_{i}
\right)
K^{-1}_{ij}
\left(
\partial^{\,}_{t}\,u^{\,}_{j}
\right)
-
\left(
\partial^{\,}_{x}\,u^{\,}_{i}
\right)
V^{\,}_{ij}
\left(
\partial^{\,}_{x}\,u^{\,}_{j}
\right)
\right]
\label{eq: trial Haldane Lagrangian a}
\end{equation}
with the chiral equations of motion
\begin{align}
0=&\,
\partial^{\,}_{\mu}\,
\frac{\delta\,\mathcal{L}^{(\pm)}}{\delta\,\partial^{\,}_{\mu}\,u^{\,}_{i}}
-
\frac{\delta\,\mathcal{L}^{(\pm)}}{\delta\,u^{\,}_{i}}
\nonumber\\
=&\,
\partial^{\,}_{t}\,
\frac{\delta\,\mathcal{L}^{(\pm)}}{\delta\,\partial^{\,}_{t}\,u^{\,}_{i}}
+
\partial^{\,}_{x}\,
\frac{\delta\,\mathcal{L}^{(\pm)}}{\delta\,\partial^{\,}_{x}\,u^{\,}_{i}}
-
\frac{\delta\,\mathcal{L}^{(\pm)}}{\delta\,u^{\,}_{i}}
\nonumber\\
=&\,
\frac{1}{4\pi}\,
\left(\vphantom{\Big(}
\mp
K^{-1}_{ji}\,
\partial^{\,}_{t}\,
\partial^{\,}_{x}
\mp
K^{-1}_{ij}\,
\partial^{\,}_{x}\,
\partial^{\,}_{t}
-
2V^{\,}_{ij}\,
\partial^{\,}_{x}\,
\partial^{\,}_{x}
\right)
u^{\,}_{j}
\nonumber\\
=&\,
\mp
\frac{K^{-1}_{ij}}{2\pi}\,
\partial^{\,}_{x}\,
\left(\vphantom{\Big(}
\delta^{\,}_{jl}\,
\partial^{\,}_{t}
\pm
K^{\,}_{jk}\,
V^{\,}_{kl}\,
\partial^{\,}_{x}
\right)
u^{\,}_{l}
\label{eq: trial Haldane Lagrangian b}
\end{align}
\end{subequations}
for any $i=1,\cdots,N$. Observe that the
term that mixes time $t$ and space $x$ derivatives only becomes imaginary
in Euclidean time $\tau=\mathrm{i}t$. 

\begin{proof}
The canonical momentum $\Pi^{(\pm)}_{i}$
to the field $u^{\,}_{i}$ is
\begin{equation}
\Pi^{(\pm)}_{i}(t,x)\:=
\frac{
\delta\,\mathcal{L}^{(\pm)}
     }
     {
\delta\,(\partial^{\,}_{t}\,u^{\,}_{i})(t,x)
     }
=
\mp
\frac{1}{4\pi}\,
K^{-1}_{ij}\,
\left(
\partial^{\,}_{x}\,u^{\,}_{j}
\right)(t,x)
\end{equation}
for any $i=1,\cdots,N$ owing to the symmetry of the matrix $K$.
Evidently, the Legendre transform
\begin{equation}
\mathcal{H}^{(\pm)}\:=
\Pi^{(\pm)}_{i}\,
\left(\partial^{\,}_{t}\,u^{\,}_{i}\right)
-
\mathcal{L}^{(\pm)}
\end{equation}
delivers
\begin{equation}
\mathcal{H}^{(\pm)}=
\frac{1}{4\pi}\,
\left(
\partial^{\,}_{x}\,u^{\,}_{i}
\right)\,
V^{\,}_{ij}\,
\left(
\partial^{\,}_{x}\,u^{\,}_{j}
\right).
\end{equation}
The right-hand side does not depend on the chiral index $\pm$.
We now quantize the theory by
elevating the classical fields $u^{\,}_{i}$
to the status of operators $\hat{u}^{\,}_{i}$
obeying the algebra
(\ref{eq: def chiral edge Hamiltonian b}).
This gives a quantum theory that meets all the demands
of the quantum chiral edge theory
(\ref{eq: def chiral edge Hamiltonian})
in all compatibility with the canonical quantization rules
(\ref{eq: def chiral canonical quantization rule}),
for
\begin{align}
\left[
\hat{u}^{\,}_{i}(t,x),
\hat{\Pi }^{(\pm)}_{j}(t,y)
\right]=&\,
\mp
\frac{1}{4\pi}\,
K^{-1}_{jk}\,
\partial^{\,}_{y}
\left[
\hat{u}^{\,}_{i}(t,x),
\hat{u}^{\,}_{k}(t,y)
\right]
\nonumber\\
\hbox{
\tiny
Eq.
(\ref{eq: def chiral edge Hamiltonian b})
     }
=&\,
\mp
\frac{1}{4\pi}\,
K^{-1}_{jk}\,
(\pi\mathrm{i})
K^{\,}_{ik}\,
(-2)\delta(x-y)
\nonumber\\
\hbox{\tiny
$K^{\,}_{ik}=K^{\,}_{ki}$
     }
=&\,
\pm
\frac{\mathrm{i}}{2}\,
K^{-1}_{jk}\,
K^{\,}_{ki}\,
\delta(x-y)
\nonumber\\
=&\,
\pm
\frac{\mathrm{i}}{2}\,
\delta^{\,}_{ij}\,
\delta(x-y),
\end{align}
where $i,j=1,\cdots,N$.
\end{proof}

Finally, analytical continuation to Euclidean time
\begin{subequations}
\begin{equation}
\tau=
\mathrm{i}t
\end{equation}
allows to define the finite-temperature quantum chiral theory through
the path integral
\begin{align}
&
Z^{(\pm)}_{\beta}\:=
\int\mathcal{D}[u]\,
\exp
\left(
-
\int\limits_{0}^{\beta}\mathrm{d}\tau\,
\int\limits_{0}^{L}\mathrm{d}x\,
\mathcal{L}^{(\pm)}
\right),
\\
&
\mathcal{L}^{(\pm)}\:=
\frac{1}{4\pi}\,
\left[
(\pm)
\mathrm{i}
\left(
\partial^{\,}_{x}\,u^{\,}_{i}
\right)\,
K^{-1}_{ij}\,
\left(
\partial^{\,}_{\tau}\,u^{\,}_{j}
\right)
+
\left(
\partial^{\,}_{x}\,u^{\,}_{i}
\right)\,
V^{\,}_{ij}\,
\left(
\partial^{\,}_{x}\,u^{\,}_{j}
\right)
\right]
\nonumber\\
&
\hphantom{\mathcal{L}^{(\pm)}\:=}
+
J
\left(
\frac{q^{\,}_{i}}{2\pi}\,
K^{-1}_{ij}\,(\partial^{\,}_{x}\,u^{\,}_{j})
\right),
\end{align}
\end{subequations}
in the presence of an external source field $J$ that couples
to the charges $q^{\,}_{i}$ like a scalar potential would do.

\subsection{Applications to polyacetylene}

Consider the Dirac Hamiltonian
\begin{subequations}
\label{eq: many-body 1D Dirac H0+H1}
\begin{equation}
\widehat{H}^{\,}_{\mathrm{D}}\:=
\widehat{H}^{\,}_{\mathrm{D}\,0}
+
\widehat{H}^{\,}_{\mathrm{D}\,1},
\label{eq: many-body 1D Dirac H0+H1 a}
\end{equation}
where the free-field and massless contribution is
\begin{equation}
\widehat{H}^{\,}_{\mathrm{D}\,0}\:=
\int\limits_{\mathbb{R}}\mathrm{d}x\,
\left(
\hat{\psi}^{\dag}_{+}\,
\mathrm{i}\partial^{\,}_{x}\,
\hat{\psi}^{\,}_{+}
-
\hat{\psi}^{\dag}_{-}\,
\mathrm{i}
\partial^{\,}_{x}\,
\hat{\psi}^{\,}_{-}
\right),
\label{eq: many-body 1D Dirac H0+H1 b} 
\end{equation}
while 
\begin{equation}
\widehat{H}^{\,}_{\mathrm{D}\,1}\:=
\int\limits_{\mathbb{R}}\mathrm{d}x\,
\left[
\phi^{\,}_{1}
\left(
\hat{\psi}^{\dag}_{-}\,
\hat{\psi}^{\,}_{+}
+
\hat{\psi}^{\dag}_{+}\,
\hat{\psi}^{\,}_{-}
\right)
+
\mathrm{i}
\phi^{\,}_{2}
\left(
\hat{\psi}^{\dag}_{-}\,
\hat{\psi}^{\,}_{+}
-
\hat{\psi}^{\dag}_{+}\,
\hat{\psi}^{\,}_{-}
\right)
\right]
\label{eq: many-body 1D Dirac H0+H1 c}
\end{equation}
\end{subequations}
couples the Dirac field to two real-valued and classical 
scalar fields $\phi^{\,}_{1}$ and $\phi^{\,}_{2}$. 
The only non-vanishing equal-time anti-commutators 
are given by Eq.\ (\ref{eq: def free Dirac theory b}).
{This Hamiltonian was considered by Goldstone and Wilczek
in their study \cite{Goldstone81}
of charge fractionalization for polyacetylene.}

According to the bosonization rules from Table
\ref{table: Abelian bosonization in 1d}
and with the help of the polar decomposition
\begin{equation}
\phi^{\,}_{1}(t,x)=|\bm{\phi}(t,x)|\,\cos\varphi(t,x),
\quad
\phi^{\,}_{2}(t,x)=|\bm{\phi}(t,x)|\,\sin\varphi(t,x),
\end{equation}
the many-body bosonic Hamiltonian that is equivalent 
to the Dirac Hamiltonian
(\ref{eq: many-body 1D Dirac H0+H1})
is
\begin{subequations}
\label{eq: many-body 1D boso H0+H1}
\begin{equation}
\widehat{H}^{\,}_{\mathrm{B}}\:=
\widehat{H}^{\,}_{\mathrm{B}\,0}
+
\widehat{H}^{\,}_{\mathrm{B}\,1},
\label{eq: many-body 1D boso H0+H1 a}
\end{equation}
where 
\begin{equation}
\widehat{H}^{\,}_{\mathrm{B}\,0}\:=
\int\limits_{\mathbb{R}}\mathrm{d}x\,
\frac{1}{8\pi}
\left[
\widehat{\Pi}^{2}
+
\left(\partial^{\,}_{x}\,\hat{\phi}\right)^{2}
\right],
\label{eq: many-body 1D boso H0+H1 b} 
\end{equation}
while 
\begin{equation}
\widehat{H}^{\,}_{\mathrm{B}\,1}\:=
\int\limits_{\mathbb{R}}\mathrm{d}x\,
\frac{1}{2\pi\,\mathfrak{a}}\,
|\bm{\phi}|\,
\cos
\left(
\hat{\phi}
-
\varphi
\right).
\label{eq: many-body 1D boso H0+H1 c}
\end{equation}
\end{subequations}
Here, the canonical momentum
\begin{subequations}
\label{eq: scalar field algebra}
\begin{equation}
\widehat{\Pi}(t,x)\:=
\left(\partial^{\,}_{t}\,\hat{\phi}\right)(t,x)
\end{equation} 
shares with $\hat{\phi}(t,x)$ the only non-vanishing
equal-time commutator
\begin{equation}
\left[\hat{\phi}(t,x),\widehat{\Pi}(t,y)\right]=
\mathrm{i}\delta(x-y).
\end{equation}
\end{subequations}

Hamiltonian (\ref{eq: many-body 1D boso H0+H1})
is interacting, and its interaction 
(\ref{eq: many-body 1D boso H0+H1 c})
can be traced to
the mass contributions in the non-interacting
Dirac Hamiltonian
(\ref{eq: many-body 1D Dirac H0+H1}).
The interaction
(\ref{eq: many-body 1D boso H0+H1 c})
is minimized when the operator identity
\begin{equation}
\hat{\phi}(t,x)=
\varphi(t,x)
+
\pi
\end{equation}
holds. This identity can only be met in the limit
\begin{equation}
|\bm{\phi}(t,x)|\to\infty
\label{eq: limits in which use bosonization is exact}
\end{equation}
for all time $t$ and position $x$ in view of the
algebra (\ref{eq: scalar field algebra})
and the competition between the contributions
(\ref{eq: many-body 1D boso H0+H1 b})
and
(\ref{eq: many-body 1D boso H0+H1 c}).

Close to the limit (\ref{eq: limits in which use bosonization is exact}),
the bosonization formula for the conserved current
\begin{equation}
\hat{\bar{\psi}}\,\gamma^{\mu}\,\hat{\psi}\to
\frac{1}{2\pi}\,
\epsilon^{\mu\nu}\,\partial^{\,}_{\nu}\hat{\phi}
\end{equation}
simplifies to
\begin{equation}
\frac{1}{2\pi}\,
\epsilon^{\mu\nu}\,\partial^{\,}_{\nu}\hat{\phi}\approx
\frac{1}{2\pi}\,
\epsilon^{\mu\nu}\,\partial^{\,}_{\nu}\varphi.
\end{equation}
On the one hand, the conserved charge 
\begin{equation}
\widehat{Q}\:=
\int\limits_{\mathbb{R}}\mathrm{d}x\,
\left(\hat{\bar{\psi}}\,\gamma^{0}\,\hat{\psi}\right)(t,x)
\to
\frac{\epsilon^{01}}{2\pi}\,
\left[
\hat{\phi}(t,x=+\infty)
-
\hat{\phi}(t,x=-\infty)
\right]
\end{equation}
for the static profile $\varphi(x)$
is approximately given by
\begin{equation}
\widehat{Q}\approx
\frac{\epsilon^{01}}{2\pi}\,
\left[
\varphi(x=+\infty)
-
\varphi(x=-\infty)
\right].
\label{eq: induced charge through gradient expansion bis}
\end{equation}
On the other hand,
the number of electrons per period $T=2\pi/\omega$
that flows across a point $x$
\begin{equation}
\hat{I}\:=
\int\limits_{0}^{T}\mathrm{d}t\,
\left(\hat{\bar{\psi}}\,\gamma^{1}\,\hat{\psi}\right)(t,x)\to
\frac{\epsilon^{10}}{2\pi}\,
\left[
\hat{\phi}(T,x)
-
\hat{\phi}(0,x)
\right]
\end{equation}
for the uniform profile $\varphi(t)=\omega\,t$
is approximately given
\begin{equation}
\hat{I}\approx
\frac{\epsilon^{10}}{2\pi}\,
\omega\,T=
\epsilon^{10}.
\label{eq: one electron pumped per cycle bis}
\end{equation}
Results 
(\ref{eq: induced charge through gradient expansion bis})
and
(\ref{eq: one electron pumped per cycle bis})
are sharp operator identities in the limit
(\ref{eq: limits in which use bosonization is exact}).
The small parameter in both expansions is $1/{m}$
where ${m}\:=\lim_{x\to\infty}|\bm{\phi}(t,x)|$.

\section{Stability analysis for the edge theory in the symmetry class AII}
\label{sec: Stability analysis for the edge theory in the symmetry class AII}

\begin{figure}[t]
\begin{center}
\includegraphics[width=0.4\textwidth]{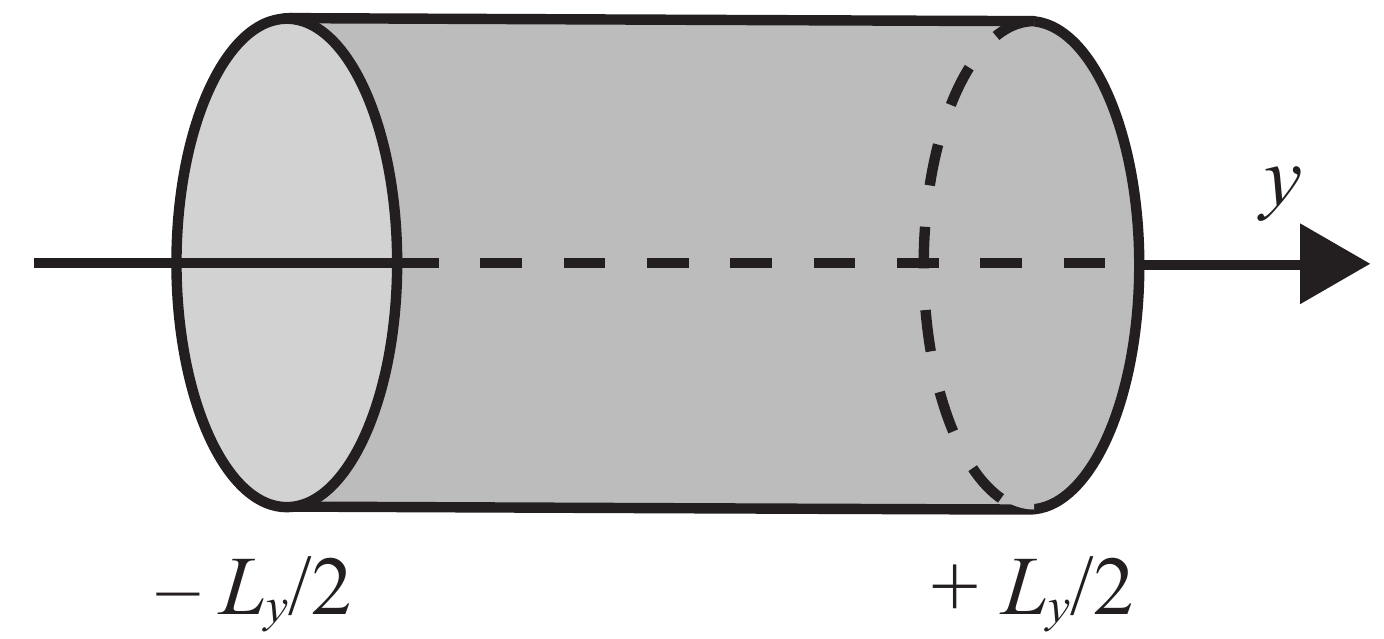}
\end{center}
\caption{
Cylindrical geometry for a two-dimensional band insulator.
The cylinder axis is labelled by the coordinate $y$. 
Periodic boundary conditions are
imposed in the transverse direction labelled by the coordinate $x$.
There is an edge at $y=-L^{\,}_{y}/2$ and another one at $y=+L^{\,}_{y}/2$.
Bulk states have support on the shaded surface of the cylinder.
Edge states are confined in the $y$ direction to the vicinity
of the edges $y=\pm L^{\,}_{y}/2$. 
Topological band insulators have the property
that there are edge states freely propagating in the $x$ direction
even in the presence of disorder with the mean free path  
$\ell$, provided the limit $\ell/L^{\,}_{y}\ll 1$ holds.
\label{fig: cylinder geometry} 
        }
\end{figure}

\subsection{Introduction}
\label{sec:intro}

The hallmark of the integer quantum effect (IQHE)
in an open geometry is the localized nature of 
all two-dimensional (bulk) states
while an integer number of chiral edge states freely propagate
along the one-dimensional boundaries \cite{Klitzing80,Laughlin81,Halperin82}.  
These chiral 
edge states are immune to the physics of Anderson localization as long as
backward scattering between edge states of opposite chiralities
is negligible \cite{Laughlin81,Halperin82}.

Many-body interactions among electrons can be treated
perturbatively in the IQHE provided the characteristic many-body energy
scale is less than the single-particle gap between Landau levels. 
This is not true anymore if the chemical potential lies within a Landau level
as the non-interacting many-body ground state 
is then macroscopically degenerate.
The lifting of this extensive degeneracy by the many-body interactions is
a non-perturbative effect.
At some ``magic'' filling fractions that deliver the fractional
quantum Hall effect (FQHE) \cite{Tsui82,Stormer83,Laughlin83,Haldane83c},
a screened Coulomb interaction
selects a finitely degenerate family of ground states, each of which
describes a featureless liquid separated from excitations by an energy gap
in a closed geometry. Such a ground state is called an incompressible 
fractional Hall liquid. The FQHE is an example of topological order
\cite{Wen90a,Wen90b,Wen90c}.
In an open geometry, there are branches of 
excitations that disperse across the spectral gap of the two-dimensional bulk,
but these excitations are localized along the direction normal to
the boundary while they  propagate freely along the boundary
\cite{Wen90c,Wen91a,Wen91b,Wen91e}.
Contrary to the IQHE, 
these excitations need not all share the same chirality. However,
they are nevertheless immune to the physics of Anderson localization 
provided scattering induced by the disorder between distinct edges
in an open geometry is negligible. 

The integer quantum Hall effect (IQHE) is the 
archetype of a two-dimensional topological band insulator.
The two-dimensional $\mathbb{Z}^{\,}_{2}$ 
topological band insulator is a close relative
of the IQHE that occurs in semi-conductors with sufficiently large
spin-orbit coupling but no breaking of time-reversal symmetry
\cite{Kane05a,Kane05b,Bernevig06a,Bernevig06b,Konig07}.
As with the IQHE, the smoking gun for the
$\mathbb{Z}^{\,}_{2}$ topological band insulator is the existence
of gapless Kramers degenerate pairs of edge states that are delocalized along
the boundaries of an open geometry as long as disorder-induced
scattering between distinct boundaries is negligible.
In contrast to the IQHE, it is the odd parity in the number of Kramers pairs
of edge states that is robust to the physics of Anderson localization.

A simple example of a two-dimensional $\mathbb{Z}^{\,}_{2}$
topological band insulator can be obtained by putting together two
copies of an IQHE system with opposite chiralities for up and down
spins. For instance, one could take two copies of Haldane's model
\cite{Haldane88},
each of which realizes an integer Hall effect on the honeycomb lattice,
but with Hall conductance differing by a sign.
In this case the spin current is conserved, a consequence of the
independent conservation of the up and down currents, and the spin
Hall conductance inherits its quantization from the IQHE of each spin
species. This example thus realizes an integer quantum spin Hall
effect (IQSHE). However, although simple, this example is not generic. The
$\mathbb{Z}^{\,}_{2}$ topological band insulator does not necessarily
have conserved spin currents, let alone quantized responses.

Along the same line of reasoning, two copies of a FQHE system put
together, again with opposite chiralities for up and down particles,
would realize a fractional quantum spin Hall effect
(FQSHE), as proposed by Bernevig and Zhang \cite{Bernevig06a}
(see also \cite{Freedman04} and \cite{Hansson04}).
Levin and Stern \cite{Levin09} proposed 
to characterize two-dimensional fractional topological 
liquids supporting the FQSHE by the criterion that
their edge states are stable against disorder 
provided that they do not break time-reversal symmetry spontaneously.

In the discussion here, the condition that
projection about some quantization axis 
of the electron spin from the underlying microscopic model
is a good quantum number is \textit{not} imposed.
Only time-reversal symmetry is assumed to hold.
The generic cases of fractional topological liquids 
with time-reversal symmetry from the special
cases of fractional topological liquids with time-reversal symmetry
\textit{and} with residual spin-1/2 $U(1)$ rotation symmetry
will thus be distinguished.
In the former cases, the electronic spin is not a good quantum number. 
In the latter cases, conservation of spin allows the FQSHE.

The subclass of incompressible time-reversal-symmetric liquids 
that we construct here is closely related to
Abelian Chern-Simons theories. Other possibilities
that are not discussed, may include non-Abelian
Chern Simons theories \cite{Frohlich90,Moore91},
or theories that include, additionally, 
conventional local order parameters (Higgs fields) \cite{Ryu09}.

The relevant effective action for the
Abelian Chern-Simons theory is of the form
\cite{Wen90b,Wen90c,Wen91a,Wen91e}
\begin{subequations}
\label{eq: def generic 2N Abelian CS}
\begin{equation}
S\:=
S^{\,}_{0}
+
S^{\,}_{e}
+
S^{\,}_{s},
\label{eq: def generic 2N Abelian CS a}
\end{equation}
where
\begin{equation}
S^{\,}_{0}\:=
-
\int\,\mathrm{d}t\, \mathrm{d}^2\bs{x}\; 
\epsilon^{\mu\nu\rho}\,
\frac{1}{4\pi}
K^{\,}_{ij}\;a^{i}_{\mu}\;\partial^{\,}_{\nu}\,a^{j}_{\rho},
\label{eq: def generic 2N Abelian CS b}
\end{equation}
\begin{equation}
S^{\,}_{e}\:=
\int\,\mathrm{d}t\, \mathrm{d}^2\bs{x}\; 
\epsilon^{\mu\nu\rho}\,
\frac{e}{2\pi}\;
Q^{\,}_{i}\,A^{\,}_{\mu}\partial^{\,}_{\nu} \,a^{i}_{\rho},
\label{eq: def generic 2N Abelian CS c}
\end{equation}
and
\begin{equation}
S^{\,}_{s}\:=
\int\,\mathrm{d}t\, \mathrm{d}^2\bs{x}\; 
\epsilon^{\mu\nu\rho}\,
\frac{s}{2\pi}\;
S^{\,}_{i}\,B^{\,}_{\mu}\partial^{\,}_{\nu} \,a^{i}_{\rho}.
\label{eq: def generic 2N Abelian CS d}
\end{equation}
The indices $i$ and $j$ run from 1 to $2N$
and any pair thereof labels an integer-valued matrix element
$K^{\,}_{ij}$ of the symmetric and invertible $2N\times2N$ matrix $K$.
The indices $\mu,\nu$, and $\rho$ run from 0 to 2.
They either label the component $x^{\,}_{\mu}$
of the coordinates $(t,\bm{x})$
in (2+1)-dimensional space and time
or the component
$A^{\,}_{\mu}(t,\bs{x})$
of an external electromagnetic gauge potential,
or the component $B^{\,}_{\mu}(t,\bs{x})$
of an external gauge potential that couples to the spin-1/2 degrees of freedom
along some quantization axis,
or the components of $2N$ flavors of dynamical Chern-Simons
fields $a^{i}_{\mu}(t,\bs{x})$.
The integer-valued component $Q^{\,}_{i}$ 
of the $2N$-dimensional vector $Q$ represents the $i$-th electric charge 
in units of the electronic charge $e$
and obeys the compatibility condition
\begin{equation}
(-1)^{Q^{\,}_{i}}=
(-1)^{K^{\,}_{ii}}
\end{equation}
\end{subequations}
for any $i=1,\cdots,2N$ in order for bulk quasiparticles
or, in an open geometry, quasiparticles
on edges to obey a consistent statistics.
The integer-valued component $S^{\,}_{i}$ 
of the $2N$-dimensional vector $S$ represents the $i$-th spin charge 
in units of the spin charge $s$ along some conserved quantization axis.
The operation of time reversal is the map
\begin{subequations}
\begin{align}
&
A^{\,}_{\mu}(t,\bs{x})\mapsto 
+g^{\mu\nu}\,A^{\,}_{\nu}(-t,\bs{x}),
\\
&
B^{\,}_{\mu}(t,\bs{x})\mapsto
-g^{\mu\nu}\,b^{\,}_{\nu}(-t,\bs{x}),
\\
&
a^{i}_{\mu}(t,\bs{x})\mapsto
-g^{\mu\nu}a^{i+N}_{\nu}(-t,\bs{x}),
\end{align}
\end{subequations}
for $i=1,\cdots,N$.
Here, $g^{\,}_{\mu\nu}=\mathrm{diag}(+1,-1,-1)$ is the Lorentz metric
in $(2+1)$-dimensional space and time.
It will be shown that time-reversal symmetry imposes that the matrix
$K$ is of the block form
\begin{subequations}
\label{decomposition K}
\begin{eqnarray}
&&
K=\left(
\begin{matrix}
    \kappa&\;\Delta\\
    \Delta^{\!\mathsf{T}}&-\kappa
\end{matrix}
\right),
\label{eq:intro-K-matrix-a}
\\
\nonumber\\
&&
\kappa^{\!\mathsf{T}}=\kappa,\quad \Delta^{\!\mathsf{T}}=-\Delta,
\label{eq:intro-K-matrix-b}
\end{eqnarray}
where $\kappa$ and $\Delta$ are $N\times N$ matrices, 
while the integer-charge vectors $Q$ and $S$ are of the block forms
\begin{eqnarray}
&&
Q=
\left(
\begin{matrix}
\varrho\\
\varrho
\end{matrix}
\right),
\qquad
S=
\left(
\begin{matrix}
\varrho\\
-\varrho
\end{matrix}
\right).
\label{eq:intro-K-matrix-c}
\end{eqnarray}
\end{subequations}
 
The $K$ matrix together with the charge vector $Q$ 
and spin vector $S$ 
that characterize the topological field theory with the action
(\ref{eq: def generic 2N Abelian CS a})
define the charge filling fraction, a rational number,
\begin{subequations}
\label{eq: def nu}
\begin{equation}
\nu^{\,}_{e}\:=
Q^{\mathsf{T}}\;K^{-1}\;Q
\label{eq: def nu a}
\end{equation}
and the spin filling fraction, another rational number,
\begin{equation}
\nu^{\,}_{s}\:=
\frac{1}{2}\,
Q^{\mathsf{T}}\;K^{-1}\;S,
\label{eq: def nuspin a}
\end{equation}
respectively.
The block forms
of $K$ and $Q$ 
in Eq.\ (\ref{decomposition K})
imply that
\begin{equation}
\nu^{\,}_{e}=0.
\label{eq: def nu b}
\end{equation}
The  ``zero charge filling fraction''~(\ref{eq: def nu b}) 
states nothing but the fact that there is no charge Hall conductance when
time-reversal symmetry holds. On the other hand, time-reversal symmetry
of the action
(\ref{eq: def generic 2N Abelian CS a})
is compatible with a non-vanishing FQSHE as measured by the non-vanishing
quantized spin-Hall conductance
\begin{equation}
\sigma^{\,}_{\mathrm{sH}}\:= 
\frac{e}{2\pi}\,\times\,
\nu^{\,}_{s}.
\label{eq: def nu spin b}
\end{equation}
\end{subequations}
The origin of the FQSHE in the action
(\ref{eq: def generic 2N Abelian CS a})
is the $U(1)\times U(1)$ gauge symmetry when 
$(2+1)$-dimensional space and time has the same topology as
a manifold without boundary. It is always assumed that
the $U(1)$ symmetry associated with charge conservation holds
in this lecture. However, we shall not do the same with the $U(1)$
symmetry responsible for the conservation of the ``spin''
quantum number.
 
The special cases of the FQSHE treated in
\cite{Bernevig06a} and {}\cite{Levin09} 
correspond to imposing the condition
\begin{equation}
\Delta=0
\end{equation}
on the $K$ matrix in Eq.\ (\ref{eq:intro-K-matrix-a}).
This restriction is, however, not necessary 
to treat either the FQSHE or the generic case when there is
no residual spin-1/2 $U(1)$ symmetry in the underlying microscopic
model.

The effective topological field theory
(\ref{eq: def generic 2N Abelian CS}) 
with the condition for time-reversal symmetry
(\ref{decomposition K})
is made of $2N$ Abelian Chern-Simons fields.
As is the case with the FQHE, when two-dimensional space is a
manifold without boundary of genus one, i.e., when two-dimensional space
is topologically equivalent to a torus,
it is characterized by distinct topological sectors.
\cite{Wen90a,Wen90b,Wen90c}
All topological sectors are in one-to-one correspondence 
with a finite number 
$\mathcal{N}^{\,}_{\mathrm{GS}}$
of topologically degenerate ground states
of the underlying microscopic theory.
\cite{Wen90a,Wen90b,Wen90c}
This degeneracy is nothing but the magnitude of the determinant
$K$ in Eq.\ (\ref{eq: def generic 2N Abelian CS a}), 
which is, because of the block structure
(\ref{eq:intro-K-matrix-a}),
in turn given by 
\begin{eqnarray}
\label{eq:intro-degeneracy}
\mathcal{N}^{\,}_{\mathrm{GS}}&=&
\left|
\mathrm{det}
\begin{pmatrix}
\kappa
&
\Delta
\\
\Delta^{\!\mathsf{T}}
&
-\kappa
\end{pmatrix}
\right|
\nonumber\\
&=&
\left|
\det
\left(
\begin{matrix}
\Delta^{\!\mathsf{T}}
&
-\kappa
\\
\kappa
&
\;\Delta
\end{matrix}
\right)
\right|
\nonumber\\
&=&
\left|
{\rm Pf}
\left(
\begin{matrix}
\Delta^{\!\mathsf{T}}
&
-\kappa\\
\kappa
&
\;\Delta
\end{matrix}
\right)
\right|^2
\nonumber\\
&=&
\left(
\mathrm{integer}
\right)^2.
\end{eqnarray}
To reach the last line, the fact that the $K$ matrix is integer valued
was used. It is thus predicted that the class of two-dimensional 
time-reversal-symmetric fractional topological liquids, 
whose universal properties are captured by 
Eqs.\ (\ref{eq: def generic 2N Abelian CS}) 
and~(\ref{decomposition K}),
are characterized by a topological ground state degeneracy that
is always the square of an integer, even if $\Delta\ne 0$,
when space is topologically equivalent to a torus. (Notice
that the condition that $\Delta$ is anti-symmetric implies that
non-vanishing $\Delta$ can only occur for $N>1$.)

The stability of the edge states associated with
the bulk Chern-Simons action
~(\ref{eq: def generic 2N Abelian CS}) 
obeying the condition for the time-reversal symmetry
(\ref{decomposition K}) is discussed in detail.
A single one-dimensional edge is considered
and an interacting quantum field theory for 
$1\leq N^{\,}_{\mathrm{K}}\leq N$
pairs of Kramers degenerate electrons subject to
strong disorder that preserves time-reversal symmetry
is constructed.
(The integer $2N^{\,}_{\mathrm{K}}$ is the number of odd charges
entering the charge vector $Q$.
\footnote{%
More precisely,
to guarantee that there are $N^{\,}_{\mathrm{K}}$ 
Kramers degenerate pairs of
electrons in the theory, demand that there exists a 
space and time independent transformation
$
O=
+\Sigma^{\,}_{1}\,O\,\Sigma^{\,}_{1}
\in\mathrm{SL}(2N,\mathbb{Z})
$
such that
$
K\mapsto O^{\mathsf{T}}\,K\,O,
$
$
V\mapsto O^{\mathsf{T}}\,V\,O,
$
and
$
Q\mapsto O^{\mathsf{T}}\,Q
$
with the transformed charge vector 
containing $2N^{\,}_{\mathrm{K}}$ odd integers.
\label{footnote-on-Kramers-pairs}
          }
The conditions under which at least one Kramers
degenerate pair of electrons remains gapless in spite of the
interactions and disorder are identified.   
This approach is here inspired by the
stability analysis of the edge states performed for the single-layer
FQHE by Haldane in \cite{Haldane95} (see also
\cite{Kane94} and \cite{Moore98}), by Naud et al.\ in
\cite{Naud00} and \cite{Naud01} for the bilayer
FQHE, and specially that by Levin and Stern in
\cite{Levin09} for the FQSHE and that in \cite{Neupert11b}. 
As for the FQSHE, our analysis departs from the analysis of
Haldane in that we impose time-reversal symmetry. 
In this lecture, we also depart from \cite{Levin09}
by considering explicitly the effects of the  off-diagonal elements 
$\Delta$ in the $K$-matrix. Such terms 
are generically present for any realistic
underlying microscopic model independently
of whether this underlying microscopic model supports or not the FQSHE. 
When considering the stability of the edge theory, 
we allow the residual spin-1/2 $U(1)$ symmetry 
responsible for the FQSHE to be broken
by interactions among the edge modes 
or by a disorder potential. Hence, we seek a criterion for
the stability of the edge theory that does not rely on
the existence of a quantized spin Hall conductance in the bulk
as was done in \cite{Levin09}.

The stability of the edge states against disorder hinges on whether
the integer
\begin{equation}
R\:=
r\,
\varrho^{\mathsf{T}}\,
(\kappa-\Delta)^{-1}\,
\varrho
\label{eq: def R}
\end{equation}
is odd (stable) or even (unstable). 
The vector $\varrho$ together with the matrices
$\kappa$ and $\Delta$ were defined in Eq.
(\ref{decomposition K}).
The integer $r$ is the smallest integer
such that all the $N$ components of the vector 
$r\,(\kappa-\Delta)^{-1}\, \varrho$ are integers. 
One can quickly check a few simple examples. 
First, observe that, in the limit $\Delta=0$, 
we recover the criterion derived 
in \cite{Levin09}. 
Second, when we impose a residual spin-1/2 $U(1)$ symmetry
by appropriately restricting the interactions between
edge channels,
$\nu^{\,}_{\uparrow}=
-\nu^{\,}_{\downarrow}=
\varrho^{\mathsf{T}}(\kappa-\Delta)^{-1}\,\varrho$ 
can be interpreted as the Hall conductivity 
$\sigma^{\,}_{\mathrm{xy}}$ in units of $e^{2}/h$
for each of the separately conserved spin components
along the spin quantization axis. 
The integer $r$ has the interpretation of the number of fluxes 
needed to pump a unit of charge, or the inverse of the ``minimum charge'' 
of \cite{Levin09}. Further restricting to the case when
$\kappa=\myopenone^{\,}_{N}$ gives $R=N$, i.e., we
have recovered the same criterion as for the two-dimensional
non-interacting $\mathbb{Z}^{\,}_{2}$ topological band insulator.

When there is no residual spin-$1/2$ $U(1)$ symmetry, 
one can no longer relate the index
$R$ to a physical spin Hall conductance. 
Nevertheless, the index $R$ defined in
Eq.\ (\ref{eq: def R}) discriminates in all cases whether there is or
not a remaining branch of gapless modes dispersing along the edge.

\subsection{Definitions}
\label{sec: Quantum chiral edge theory with time-reversal symmetry}

Consider an interacting model for electrons
in a two-dimensional cylindrical geometry
as is depicted in Fig.~\ref{fig: cylinder geometry}.
Demand that (i) charge conservation and time-reversal symmetry 
are the only intrinsic symmetries of the microscopic quantum Hamiltonian,
(ii) neither are broken spontaneously by the many-body ground state,
and (iii), if periodic boundary conditions are assumed
along the $y$ coordinate in Fig.~\ref{fig: cylinder geometry},
then there is at most a finite number of
degenerate many-body ground states and each 
many-body ground state is separated from 
its tower of many-body excited states by an energy gap.
Had the condition that time-reversal symmetry holds been relaxed,
the remaining assumptions would be realized for the FQHE.

In the open geometry of Fig.~\ref{fig: cylinder geometry},
the only possible excitations with an energy smaller than the
bulk gap in the closed geometry of a torus
must be localized along the $y$ coordinate in the vicinities
of the edges at $\pm L^{\,}_{y}/2$. If $L^{\,}_{y}$ 
is much larger than the characteristic linear extension into the bulk
of edge states,
the two edges decouple from each other.
It is then meaningful to define a low-energy and long-wavelength
quantum field theory for the edge states propagating along 
any one of the two boundaries
in Fig.~\ref{fig: cylinder geometry}, which we take to be of
length $L$ each. 

The low-energy and long-wavelength effective quantum field 
theory for the edge
that we are going to construct is inspired by the construction by
Wen of the chiral Luttinger edge theory for the FQHE.%
\cite{Wen91a,Wen90b,Wen91e}
As for the FQHE, this time-reversal symmetric 
boundary quantum field theory has a correspondence to the effective 
time-reversal symmetric bulk topological quantum-field theory 
built out of $2N$ Abelian Chern-Simon fields.

The simplest class of quantum Hamiltonians that fulfills requirements
(i)--(iii) can be represented in terms of $2N$ real-valued chiral
scalar quantum fields $\widehat{\Phi}^{\,}_{i}(t,x)$ with $i=1,\ldots,2N$
that form the components of the quantum vector field
$\widehat{\Phi}(t,x)$.
After setting the electric charge $e$, the characteristic speed, 
and $\hbar$ to unity, the Hamiltonian for the system is given by
\footnote{%
Do the linear transformation
$\hat{u}^{\,}_{i}\equiv K^{\,}_{ij}\,\widehat{\Phi}^{\,}_{j}$
in Eq.\ (\ref{eq: def chiral edge Hamiltonian}),
where we recall that the matrix $K$ and its inverse $K^{-1}$ 
are symmetric so that we may write
$\Theta^{\,}_{ij}\equiv K^{-1}_{ii'}\,L^{\,}_{i'j'}\,K^{-1}_{j'j}$.
          }
\begin{subequations}
\label{eq:Mdef quantum edge theory}
\begin{eqnarray}
\widehat{H}\:=
\widehat{H}^{\,}_{0}
+
\widehat{H}^{\,}_{\mathrm{int}},
\label{eq:Mdef quantum edge theory a}
\end{eqnarray}
where 
\begin{equation}
\widehat{H}^{\,}_{0}\:=
\int\limits_{0}^{L}
\mathrm{d}x\,
\frac{1}{4\pi}
\left(\partial^{\,}_{x}\widehat{\Phi}^{\mathsf{T}}\right)(t,x)\,
V(x)\,
\left(\partial^{\,}_{x}\widehat{\Phi}\right)(t,x),
\label{eq:Mdef quantum edge theory b}
\end{equation}
with $V(x)$ a $2N\times2N$ symmetric and positive definite matrix that
accounts, in this bosonic representation, for the screened
density-density interactions between electrons. 
The theory is quantized according to the equal-time commutators
\begin{equation}
\begin{split}
&
\left[
\widehat{\Phi}^{\,}_{i}(t,x ),
\widehat{\Phi}^{\,}_{j}(t,x')
\right]=
-
\mathrm{i}\pi
\Big[
K^{-1}_{ij}
\;\mathrm{sgn}(x-x')
+
\Theta_{ij}
\Big],
\end{split}
\label{eq: def quantum edge theory e}
\end{equation}
where $K$ is a $2N\times2N$ symmetric and invertible matrix with
integer-valued matrix elements, and the $\Theta$ matrix accounts for
Klein factors that ensure that charged excitations in the theory
(vertex operators) satisfy the proper commutation relations. 
Fermionic or bosonic charged excitations are
represented by the normal ordered vertex operators
\begin{equation}
\widehat{\Psi}^{\dag}_{T}(t,x)
\:=\ 
:e^{-\mathrm{i}\,T_{i}\,K^{\,}_{ij}\,\widehat{\Phi}^{\,}_{j}(t,x)}:,
\label{eq:vertex-def}
\end{equation}
where the integer-valued $2N$-dimensional vector $T$ determines 
the charge (and statistics) of the operator. 
The operator that measures the total charge density is
\begin{equation}
\hat{\rho}(t,x)=
\frac{1}{2\pi}\,
Q^{\,}_{i}\,
\left(\partial^{\,}_{x}\widehat{\Phi}^{\,}_{i}\right)(t,x),
\label{eq:charge-density-def}
\end{equation}
where the integer-valued $2N$-dimensional charge vector $Q$, 
together with the
$K$-matrix, specify the universal properties of the edge theory. 
The charge $q^{\,}_{T}$ of the vertex operator in
Eq.\ (\ref{eq:vertex-def}) follows from its commutation with the charge
density operator in Eq.\ (\ref{eq:charge-density-def}), yielding
$q^{\,}_{T}=T^{\mathsf{T}}\,Q$.

Tunneling of electronic charge among the different edge branches is
accounted for by
\begin{equation}
\widehat{H}^{\,}_{\mathrm{int}}\:=
-
\int\limits_{0}^{L}
\mathrm{d}x
\sum_{T\in\mathbb{L}}
h^{\,}_{T}(x)
:
\cos
\Big(
T^{\mathsf{T}} K\,\widehat{\Phi}(t,x)
+
\alpha^{\,}_{T}(x)
\Big)
:.
\label{eq:Mdef quantum edge theory c}
\end{equation}
The real functions $h^{\,}_{T}(x)\geq0$ and
$0\leq\alpha^{\,}_{T}(x)\leq2\pi$ encode information about 
the disorder along the edge when position dependent. 
The set 
\begin{equation}
\mathbb{L}\:=
\left\{
T\in\mathbb{Z}^{2N}\left|T^{\mathsf{T}}\,Q=0\right.
\right\},
\label{eq: def mathbb L}
\end{equation} 
\end{subequations}
encodes all the possible charge neutral
tunneling processes (i.e., those that just rearrange charge among the
branches). This charge neutrality condition implies that the operator 
$\hat{\Psi}^{\dag}_{T}(t,x)$ 
that makes up Eq.\ (\ref{eq:Mdef quantum edge theory c})
is bosonic, for it has even charge.
Observe that set $\mathbb{L}$ forms a lattice. 
Consequently, if $T$ belongs to $\mathbb{L}$ so does
$-T$. In turn, relabeling $T$ to $-T$ in 
$\widehat{H}^{\,}_{\mathrm{int}}$
implies that
$h^{\,}_{T}(x)=+h^{\,}_{-T}(x)$
whereas
$\alpha^{\,}_{T}(x)=-\alpha^{\,}_{-T}(x)$.

The theory~\eqref{eq:Mdef quantum edge theory}
is inherently encoding interactions.
The terms $\widehat{H}^{\,}_0$
and $\widehat{H}^{\,}_{\mathrm{int}}$
encode single-particle 
\textit{as well as} many-body interactions
with matrix elements that preserve and break translation symmetry, 
respectively.
Recovering the single-particle kinetic energy
of $N$ Kramers degenerate pairs of electrons
from Eq.\ (\ref{eq:Mdef quantum edge theory b})
corresponds to choosing the matrix $V$
to be 
proportional to the unit $2N\times2N$ matrix
with the proportionality constant fixed by the condition
that the scaling dimension of each electron is $1/2$
at the bosonic free-field fixed point defined by
Hamiltonian $\widehat{H}^{\,}_{0}$. Of course, to implement
the fermionic statistics for all $2N$ fermions,
one must also demand that all diagonal entries of $K$ are
odd integers in some basis.
\footnote{%
See footnote \ref{footnote-on-Kramers-pairs}.
          }

\subsection{Time-reversal symmetry of the edge theory}

The operation of time-reversal on the $\widehat{\Phi}$ fields is defined by
\begin{subequations}
\label{eq:Mdef time reversal trsf}
\begin{equation}
\mathcal{T}\,\widehat{\Phi}(t,x)\,\mathcal{T}^{-1}\:=
\Sigma^{\,}_{1}\,
\widehat{\Phi}(-t,x)
+
\pi
K^{-1}\,
\Sigma^{\,}_{\downarrow}\,
Q,
\label{eq:Mdef time reversal trsf b}
\end{equation}
where
\begin{equation}
\Sigma^{\,}_{1}=
\left(  
\begin{matrix}
0
&
\myopenone
\\
\myopenone
&
0
\end{matrix}
\right)
\qquad
\Sigma^{\,}_{\downarrow}=
\left(  
\begin{matrix}
0
&
0
\\
0
&
\myopenone
\end{matrix}
\right).
\label{eq:sigma-def}
\end{equation}
\end{subequations}
This definition ensures that the fermionic and bosonic vertex operators 
defined in Eq.\ (\ref{eq:vertex-def}) 
are properly transformed under reversal of time.
More precisely, one can then construct a pair of fermionic operators
$\hat{\Psi}^{\dag}_{1}$ and $\hat{\Psi}^{\dag}_{2}$ 
of the form~(\ref{eq:vertex-def}) by suitably choosing a pair 
of vectors $T^{\,}_{1}$ and $T^{\,}_{2}$, respectively, in such a way
that the operation of time-reversal maps
$\hat{\Psi}^{\dag}_{1}$ into $+\hat{\Psi}^{\dag}_{2}$
whereas it maps
$\hat{\Psi}^{\dag}_{2}$ into $-\hat{\Psi}^{\dag}_{1}$.
Thus, it is meaningful to interpret
the block structure displayed
in Eq.\ (\ref{eq:sigma-def}) 
as arising from the upper or lower projection along some 
spin-1/2 quantization axis.

Time-reversal symmetry on the chiral edge theory
(\ref{eq:Mdef quantum edge theory})
demands that
\begin{subequations}
\label{eq:Mconditions for TRS on parameters}
\begin{equation}
V=
+\Sigma^{\,}_{1}\,V\,\Sigma^{\,}_{1},
\label{eq:Mconditions for TRS on parameters a}
\end{equation}
\vfill
\begin{equation}
K=
-\Sigma^{\,}_{1}\,K\,\Sigma^{\,}_{1},
\label{eq:Mconditions for TRS on parameters b}
\end{equation}
\vfill
\begin{equation}
Q=
\Sigma^{\,}_{1}\,Q,
\label{eq:Mconditions for TRS on parameters c}
\end{equation}
\vfill
\begin{equation}
h^{\,}_{T}(x)
=h^{\,}_{\Sigma^{\,}_{1}T}(x),
\label{eq:Mconditions for TRS on parameters d}
\end{equation}
\vfill
\begin{equation}
\alpha^{\,}_{T}(x)=
\left(
-\alpha^{\,}_{\Sigma^{\,}_{1}\,T}(x)
+
\pi T^{\mathsf{T}}\,\Sigma^{\,}_{\downarrow}\,Q
\right)
\!\!\!\!
\mod 2\pi.
\label{eq:Mconditions for TRS on parameters e}
\end{equation}
\end{subequations}

\begin{proof}
The first two conditions,
Eqs.\ (\ref{eq:Mconditions for TRS on parameters a})
and (\ref{eq:Mconditions for TRS on parameters b}), 
follow from the requirement that $\widehat{H}^{\,}_{0}$ 
be time-reversal invariant. 
In particular, the decomposition
\begin{equation}
K=\left(
\begin{matrix}
\kappa&
\Delta
\\
\Delta^{\!\mathsf{T}}&
-\kappa
\end{matrix}
\right)
\qquad
\kappa^{\!\mathsf{T}}=
\kappa,\quad \Delta^{\!\mathsf{T}}=-\Delta,
\label{eq: intro-K-matrix-b}
\end{equation}
where $\kappa$ and $\Delta$ are $N\times N$ matrices, 
follows from Eq.\ (\ref{eq:Mconditions for TRS on parameters b}) and
$K=K^{\mathsf{T}}$.

The third condition,
Eq.\ (\ref{eq:Mconditions for TRS on parameters c}),
states that the charge density is invariant under time reversal. 
In particular, the decomposition 
\begin{equation}
Q=
\left(
\begin{matrix}
\varrho\\
\varrho
\end{matrix}
\right)
\label{eq: intro-K-matrix-c}
\end{equation}
follows.

Finally, 
$\mathcal{T} \widehat{H}^{\,}_{\mathrm{int}}
\mathcal{T}^{-1}=\widehat{H}^{\,}_{\mathrm{int}}$ 
requires that
\begin{eqnarray}
&&
\sum_{T\in\mathbb{L}}\;
h^{\,}_{T}(x)\,
\cos
\left(
T^{\mathsf{T}}\,K\,\widehat{\Phi}(t,x)\vphantom{\Big[}
+
\alpha^{\,}_{T}(x)
\right)=
\nonumber\\
&&
\sum_{T\in\mathbb{L}}\;
\mathcal{T}\, 
\left[
h^{\,}_{T}(x)\,
\cos
\left(
T^{\mathsf{T}}\,K\,\widehat{\Phi}(t,x)\vphantom{\Big[}
+
\alpha^{\,}_{T}(x)
\right)
\right]
\mathcal{T}^{-1}=
\nonumber\\
&&
\sum_{T\in\mathbb{L}}\;
h^{\,}_{T}(x)\,
\cos
\left(
-
\left(\Sigma^{\,}_{1}\,T\right)^{\mathsf{T}}
K\,
\widehat{\Phi}(-t,x)
+
\alpha^{\,}_{T}(x)
-\pi
T^{\mathsf{T}}\,
\Sigma^{\,}_{\downarrow}\,
Q
\right)=
\nonumber\\
&&
\sum_{T\in\mathbb{L}}\;
h^{\,}_{\Sigma^{\,}_{1}\,T}(x)\,
\cos
\left(
-
T^{\mathsf{T}}
K\,
\widehat{\Phi}(-t,x)
+
\alpha^{\,}_{\Sigma^{\,}_{1}\,T}(x)
-\pi
(\Sigma^{\,}_{1}\, T)^{\mathsf{T}}\,
\Sigma^{\,}_{\downarrow}\,
Q
\right)=
\nonumber\\
&&
\sum_{T\in\mathbb{L}}\;
h^{\,}_{\Sigma^{\,}_{1}T}(x)\,
\cos
\left(
T^{\mathsf{T}}
K\,
\widehat{\Phi}(-t,x)
-
\alpha^{\,}_{\Sigma^{\,}_{1}\,T}(x)
+\pi
(\Sigma^{\,}_{1} T)^{\mathsf{T}}\,
\Sigma^{\,}_{\downarrow}\,
Q
\right),
\end{eqnarray}
as the conditions needed to match the two trigonometric expansions.
This leads to the last two relations,
Eqs.\ (\ref{eq:Mconditions for TRS on parameters d})
and (\ref{eq:Mconditions for TRS on parameters e}).
\end{proof}

Disorder parametrized by
$h^{\,}_{T}(x)=+h^{\,}_{-T}(x)$ 
and 
$\alpha^{\,}_{T}(x)=-\alpha^{\,}_{-T}(x)$
and for which the matrix $T$ obeys
\begin{subequations}
\label{eq: T in mathbb{T}--}
\begin{equation}
\Sigma^{\,}_{1}\,T=-T,
\end{equation}
and
\begin{equation}
T^{\mathsf{T}}\,\Sigma^{\,}_{\downarrow}\,Q
\hbox{ is an odd integer},
\end{equation}
\end{subequations}
cannot satisfy the condition
(\ref{eq:Mconditions for TRS on parameters e})
for time-reversal symmetry. Such disorder is thus prohibited
to enter $\widehat{H}^{\,}_{\mathrm{int}}$ in Eq.
(\ref{eq:Mdef quantum edge theory c}), 
for it would break
explicitly time-reversal symmetry otherwise. Moreover, 
we also prohibit any ground state that provides
$\exp\Big(\mathrm{i}T^{\mathsf{T}}\,K\,\widehat{\Phi}(t,x)\Big)$
with an expectation value when 
$T$ satisfies Eq.\ (\ref{eq: T in mathbb{T}--}), 
for it would break spontaneously 
time-reversal symmetry otherwise.

\medskip
\subsection{Pinning the edge fields with disorder potentials: the Haldane criterion}
\label{subsec:pinning}

Solving the interacting theory
(\ref{eq:Mdef quantum edge theory})
is beyond the scope of this lecture.
What can be done, however, is to identify those fixed
points of the interacting theory
(\ref{eq:Mdef quantum edge theory})
that are pertinent to the question of whether or not some edge modes
remain extended along the edge in the limit of strong disorder
$h^{\,}_{T}(x)\to\infty$ for all tunneling matrices $T\in\mathbb{L}$
entering the interaction
(\ref{eq:Mdef quantum edge theory c}).

This question is related to the one posed and answered by
Haldane in \cite{Haldane95} for Abelian FQH states and which,
in the context of this lecture, would be as follows.
Given an interaction potential caused by 
\textit{weak} disorder on the edges
as defined by Hamiltonian~\eqref{eq:Mdef quantum edge theory c},
what are the tunneling vectors $T\in\mathbb{L}$ that 
can, in principle, describe relevant perturbations that will cause
the system to flow to a strong coupling fixed point characterized by
$h^{\,}_{T} \rightarrow \infty$
away from the fixed point $\widehat{H}^{\,}_{0}$?
(See \cite{Xu06}
for an answer to this weak-coupling question
in the context of the IQSHE and
$\mathbb{Z}^{\,}_{2}$ topological band insulators.)
By focusing on the strong coupling limit from the outset, we
avoid the issue of following the renormalization group flow
from weak to strong coupling. Evidently, this point of view
presumes that the strong coupling fixed point is stable
and that no intermediary fixed point prevents it from being reached.

To identify the fixed points of the interacting theory
(\ref{eq:Mdef quantum edge theory})
in the strong coupling limit (strong disorder limit)
$h^{\,}_{T}\to\infty$,
we ignore the contribution 
$\widehat{H}^{\,}_{0}$ 
and restrict the sum over the tunneling matrices in
$\widehat{H}^{\,}_{\mathrm{int}}$ 
to a subset 
$\mathbb{H}$ of $\mathbb{L}$
($\mathbb{H} \subset \mathbb{L}$) 
with a precise definition of $\mathbb{H}$ that will
follow in Eq.~\eqref{eq:def-Hset}.
For any choice of $\mathbb{H}$,
there follows the strong-coupling fixed point Hamiltonian
\begin{equation}
\widehat{H}^{\,}_{\mathbb{H}}\:=
-
\int\limits_{0}^{L}
\mathrm{d}x
\sum_{T\in\mathbb{H}}
h^{\,}_{T}(x)
:
\cos
\Big(
T^{\mathsf{T}} K\,\widehat{\Phi}(x)
+
\alpha^{\,}_{T}(x)
\Big)
:.
\label{eq: def strong coupling fixed point}
\end{equation}
Assume that a fixed point Hamiltonian
(\ref{eq: def strong coupling fixed point})
is stable if and only if
the set $\mathbb{H}$ is ``maximal''. 
The study of the renormalization group flows
relating the weak, moderate (if any), and the strong
fixed points in the infinite-dimensional parameter space
spanned by the non-universal data $V$, $h^{\,}_{T}(x)$, and
$\alpha^{\,}_{T}(x)$ is again beyond the scope of this lecture.   

One might wonder why we cannot simply choose
$\mathbb{H}=\mathbb{L}$. This is a consequence of the 
chiral equal-time commutation relations
(\ref{eq: def quantum edge theory e}),
as emphasized by Haldane in \cite{Haldane95},
that prevent the simultaneous locking of the phases of all the cosines through
the condition
\begin{equation}
\partial_{x}\left(T^{\mathsf{T}}\,K\,\widehat{\Phi}(t,x)+\alpha^{\,}_{T}(x)\right)
=
C^{\,}_{T}(x)
\label{eq: locking condition}
\end{equation}
for some time-independent and real-valued function $C^{\,}_{T}(x)$
on the canonical momentum 
\begin{equation}
(4\pi)^{-1}\,K\,
(\partial^{\,}_{x}\widehat{\Phi})(t,x)
\end{equation}
that is conjugate to $\widehat{\Phi}(t,x)$,
when applied to the ground state. 
The locking condition~(\ref{eq: locking condition})
removes a pair of chiral bosonic modes with opposite chiralities
from the gapless degrees of freedom of the theory. However,
even in the strong-coupling limit, there are quantum fluctuations
as a consequence of the  chiral equal-time commutation relations
(\ref{eq: def quantum edge theory e})
that prevent minimizing the interaction $\widehat{H}^{\,}_{\mathrm{int}}$
by minimizing separately each contribution to the trigonometric
expansion
(\ref{eq:Mdef quantum edge theory c}).
Finding the ground state in the strong coupling limit 
is a strongly frustrated problem of optimization.

To construct a maximal set $\mathbb{H}$,
demand that any $T\in\mathbb{H}$ must satisfy
the locking condition
(\ref{eq: locking condition}).
Furthermore, require that the phases of the cosines 
entering the fixed point Hamiltonian
(\ref{eq: def strong coupling fixed point})   
be constants of motion 
\begin{equation}
\label{eq:constant of motion}
\left[
\partial_{x}\left(T^{\mathsf{T}}\,K\,\widehat{\Phi}(t,x)\right), \widehat{H}_{\mathbb{H}}
\right] = 0\;.	
\end{equation}
To find the tunneling vectors $T\in\mathbb{H}$, one thus needs to
consider the following commutator
\begin{equation}
\begin{split}
&
\int\limits_{0}^{L}\mathrm{d}x'\,
\left[
\partial^{\,}_{x}\left(T^{\mathsf{T}}\,K\,\widehat{\Phi}(t,x)\right),\
h^{\,}_{T'}(x')\,
\cos\left(T'^{\,\mathsf{T}}\,K\,\widehat{\Phi}(t,x')+\alpha^{\,}_{T'}(x')\right)
\right]=
\\
&\qquad\qquad\qquad\qquad
-\mathrm{i}\,2\pi\;
T^{\mathsf{T}}\,K\,T^{\prime}\;\,h^{\,}_{T'}(x)\;
\sin\left(T'^{\,\mathsf{T}}\,K\,\widehat{\Phi}(t,x)+\alpha^{\,}_{T'}(x)\right),
\end{split}
\label{eq: quantum fluctuations between T and T'-eom}
\end{equation}
and demand that it vanishes. This is achieved if
$T^{\mathsf{T}}\,K\,T^{\prime}=0$.
Equation~(\ref{eq: quantum fluctuations between T and T'-eom}) implies
that any set $\mathbb{H}$ is composed of the charge neutral
vectors satisfying 
\begin{equation}
T^{\mathsf{T}}\,K\,T'= 0.
\end{equation}
It is by choosing a set $\mathbb{H}$ to be ``maximal'' that
we shall obtain the desired Haldane criterion for stability.

\medskip
\subsection{Stability criterion for edge modes}
\label{sec: Stability criterion for edge modes}

Section~\ref{sec:intro} presented and briefly discussed
the criteria for at least one branch of edge
excitations to remain delocalized even in the presence of
strong disorder. Here these criteria are proved. 

The idea is to count the maximum possible number
of edge modes that can be pinned (localized) along the edge
by tunneling processes. 
The set of pinning processes must satisfy
\begin{equation}
T^{\mathsf{T}}\,Q=0
\qquad
T^{\mathsf{T}}\, K\, T'=0,
\label{eq:def-Hset}
\end{equation}
which defines a set $\mathbb{H}$ introduced in Section~\ref{subsec:pinning}.
(Note, however, that $\mathbb H$ is not uniquely determined from 
this condition.)
Let us define the real extension $\mathbb V$ of a set $\mathbb H$, by
allowing the tunneling vectors $T$ that satisfy Eq.\ (\ref{eq:def-Hset}) 
to take real values instead of integer values. Notice that $\mathbb V$ 
is a vector space over the real numbers.
Demand that $\mathbb{H}$ forms a lattice that is as dense
as the lattice $\mathbb{L}$ by imposing
\begin{equation}
\mathbb{V}\cap\mathbb{L}=\mathbb{H}.
\end{equation}

For any vector $T\in\mathbb V$, consider the vector $K\,T$. It follows
from Eq.\ (\ref{eq:def-Hset}) that $K\,T\perp T', \forall T'\in\mathbb{V}$. 
So $K$ maps the space $\mathbb{V}$ into an orthogonal space
$\mathbb{V}^{\perp}$. Since $K$ is invertible, we have 
$\mathbb{V}^{\perp}=K\,\mathbb{V}$ as well as 
$\mathbb{V}=K^{-1}\,\mathbb{V}^{\perp}$, and
thus $\dim\,\mathbb{V}=\dim\,\mathbb{V}^{\perp}$. Since 
$\dim\mathbb{V}+\dim\mathbb{V}^{\perp}\le 2N$, it follows that 
$\dim\mathbb{V}\le N$. 
Therefore (as could be anticipated physically) the maximum number
of Kramers pairs of edge modes that can be pinned is $N$. If that
happens, the edge has no gapless delocalized mode.

Next, we look at the conditions for which the maximum dimension
$N$ is achieved in order to establish a contradiction.

Assume that $\dim\mathbb V=\dim\mathbb V^{\perp}=N$. It
follows that $ \mathbb{V}\oplus\mathbb{V}^{\perp}=\mathbb R^{2N}$,
exhausting the space of available vectors. In this case the
charge vector $Q\in\mathbb{V}^{\perp}$ because of
Eq.\ (\ref{eq:def-Hset}). Consequently, $K^{-1}\,Q\in\mathbb{V}$. 
We can then construct an integer vector 
$\bar{T}\parallel K^{-1}\,Q$ by scaling
$K^{-1}\,Q$ with the minimum integer $r$ that accomplishes this.
(This is always possible because $K^{-1}$ is a matrix with rational 
entries and $Q$ is a vector of integers.) 
Because the inverse of $K$ is not known,
it seems hopeless to write $K^{-1}\,Q$ in closed form. However,
$K^{-1}$ must anticommutes with $\Sigma^{\,}_{1}$ given that
$K$ anticommutes with $\Sigma^{\,}_{1}$, while $\Sigma^{\,}_{1}$ 
squares to the unit matrix. Hence, $K^{-1}\,Q$ is an eigenstate of
$\Sigma^{\,}_{1}$ with eigenvalue $-1$. Now,
\begin{equation}
\bar{T}\:=
r
\begin{pmatrix}
+(\kappa-\Delta)^{-1}\,
\varrho\\
-(\kappa-\Delta)^{-1}\,
\varrho
\end{pmatrix}
\label{eq:Adef T0}
\end{equation}
is also an eigenstate of $\Sigma^{\,}_{1}$ with eigenvalue $-1$.
This suggests that we may use $\bar{T}$ instead of $K^{-1}\,Q$. Indeed,
the existence of $(\kappa-\Delta)^{-1}$ follows from $\mathrm{det}\,K\neq0$
and 
\begin{equation}
\mathrm{det}\, K
=\,
(-1)^{N}
\left[\mathrm{det}(\kappa-\Delta)\right]^{2}.
\end{equation}
Moreover, one verifies that $\bar{T}$ is orthogonal to the charge vector $Q$
and that $K\,\bar{T}$ is  orthogonal to  $\bar{T}$.

Equipped with $\bar{T}$, we construct the integer
\begin{equation}
R\:=
-\bar T^{\,\mathsf{T}}\,\Sigma^{\,}_{\downarrow}\,Q.
\label{eq:Adef R}
\end{equation}
It is the parity of this integer number 
that will allow us to establish a contradiction, i.e.,
it is the parity of $R$ that determines if it is possible or not to 
localize all the modes with the $N$ tunneling operators. 
To establish the contradiction, we employ
Eq.\ (\ref{eq:Mconditions for TRS on parameters e})
together with the fact that $\Sigma^{\,}_{1}\,\bar{T}=-\bar{T}$. 
In other words,
\begin{equation}
\begin{split}
\pi R=&
-\pi \bar T^{\mathsf{T}}\,\Sigma^{\,}_{\downarrow}\,Q\\
&=
\Big(
-\alpha^{\,}_{\bar T}(x)
-\alpha^{\,}_{\Sigma^{\,}_{1}\,\bar T}(x)
\Big)
\!\!\!
\mod 2\pi
\\
&=
\Big(
-\alpha^{\,}_{\bar T}(x)
-\alpha^{\,}_{-\bar T}(x)
\Big)
\!\!\!\mod 2\pi
\\
&
=0
\!\!\!\mod 2\pi,
\label{eq:barT-cond}
\end{split}
\end{equation}
where in the last line
$\alpha^{\,}_{T}(x)=-\alpha^{\,}_{-T}(x)$ for
all $T\in\mathbb{L}$ was used.
If $\bar{T}$ satisfies Eq.~\eqref{eq:barT-cond},
then $R$ must be an even integer.
If Eq.~\eqref{eq:barT-cond} is violated
(i.e., $R$ is an odd integer)
then $\bar{T}$ is not allowed to enter
$\widehat{H}^{\,}_{\mathrm{int}}$ for it would otherwise
break time-reversal symmetry 
[thus $h^{\,}_{\bar{T}}(x)=0$ must always hold in this case 
to prevent $\bar{T}$ from entering $\widehat{H}^{\,}_{\mathrm{int}}$].
One therefore arrives at the condition that 

\vspace{0.3cm}
$\bullet$ If the maximum number of edge modes are localized or
gaped, then $R$ must be even.

\vspace{0.3cm}
A corollary is that

\vspace{0.3cm}
$\bullet$ If $R$ is odd, at least one edge branch 
is gapless and delocalized.

\vspace{0.3cm} It remains to prove that if $R$ is even, then
one can indeed reach the maximum dimension $N$ for the space of
pinning vectors. This is done by construction. Take all eigenvectors
of $\Sigma^{\,}_{1}$ with $+1$ eigenvalue. Choose $(N-1)$ of such vectors,
all those orthogonal to $Q$. For the last one, choose $\bar T$. One
can check that these $N$ vectors satisfy Eq.\ (\ref{eq:def-Hset}) with
the help of $\Sigma^{\,}_{1}\,K\,\Sigma^{\,}_{1}=-K$ 
[listed in
Eq.\ (\ref{eq:Mconditions for TRS on parameters b})] and of 
$\bar T \parallel K^{-1}Q$. Now, the $(N-1)$ vectors 
$\Sigma^{\,}_{1}T=+T$ are of the
form $T^\mathsf{T}=(t^\mathsf{T},t^\mathsf{T})$, where we need to satisfy
$T^{\mathsf{T}}Q=2t^\mathsf{T}\varrho=0$. This leads to 
$T^{\mathsf{T}}\,\Sigma^{\,}_{\downarrow}\,Q$ even, and then
Eq.\ (\ref{eq:Mconditions for TRS on parameters e})
brings no further conditions whatsoever. So we can take all these $(N-1)$ 
tunneling vectors. Finally, we take $\bar T$ as constructed above, which is
a legitimate choice since $R$ is assumed even and thus consistent with 
Eq.\ (\ref{eq:barT-cond}). Hence, we have constructed the $N$ tunneling vectors
that gap or localize all edge modes, and can state that

\vspace{0.3cm}
$\bullet$ If $R$ is even, then the maximum number of edge modes
are localized or gaped.

As a by-product, we see that it is always possible to
localize along the boundary at least all but one Kramers degenerate pair 
of edge states via the $(N-1)$ tunneling vectors that 
satisfy $\Sigma^{\,}_{1}T=+T$. Thus, either one or no Kramers degenerate
pair of edge state remains delocalized along the boundary 
when translation invariance is strongly broken along
the boundary.

{
We close Section\ \ref{sec: Stability criterion for edge modes}
by demonstrating
\footnote{%
{Private communication from Jyong-Hao Chen.}
          }
that the spin-Hall conductance
(\ref{eq: def nu spin b}) 
is related to the pair of integers $r$ and $R$ defined by Eqs.\
(\ref{eq:Adef T0}) and (\ref{eq:Adef R}),
respectively, through
\begin{equation}
\sigma^{\,}_{\mathrm{sH}}\:= 
\frac{e}{2\pi}\,\times\,
\frac{R}{r}
\Longleftrightarrow
\nu^{\,}_{\mathrm{s}}=
\frac{R}{r}.
\label{eq: Levin-Stern relation for AII}
\end{equation}
If we interpret 
\begin{equation}
e^{\star}\:=\frac{e}{r}
\end{equation}
as the minimal quasi-particle charge, we can state the
stability criterion for the helical edge states 
in the symmetry class AII to interactions or 
static disorder as the condition that the spin-Hall conductance divided by
$e^{\star}/(2\pi)$ must be an odd integer. This criterion
was first established by Levin and Stern in \cite{Levin09}
for the FQSHE when the spin-1/2 rotation-symmetry is broken up to
its $U(1)$ subgroup. We have shown that the same criterion holds 
even in the absence of any $U(1)$ residual symmetry of 
the spin-$SU(2)$ symmetry group.
\begin{proof}
By the definitions of the charge vector $Q$, the spin vector $S$,
and the matrix $\bar{T}$ 
[recall Eqs.\ 
(\ref{eq:intro-K-matrix-c})
and
(\ref{eq:Adef T0})],
\begin{equation}
\bar{T}^{\mathsf{T}}\,S=
r\, Q^{\mathsf{T}}\,K^{-1}\,S.
\end{equation}
The dimensionless spin-Hall conductance defined by
Eq.\ (\ref{eq: def nuspin a}) 
is thus nothing but
\begin{align}
\nu^{\,}_{\mathrm{s}}=&\,
\frac{1}{2\,r}\, S^{\mathsf{T}}\,\bar{T}
\nonumber\\
=&\,
\rho^{\mathsf{T}}\,\left(\kappa-\Delta\right)^{-1}\,\rho.
\end{align}
From the definition of $R$ in Eq.\ (\ref{eq:Adef R}),
\begin{align}
R=&\,
-\begin{pmatrix}0\\ \rho\end{pmatrix}^{\mathsf{T}}\,\bar{T}
\nonumber\\
=&\,
+r\,
\rho^{\mathsf{T}}\,
\left(\kappa-\Delta\right)^{-1}\,\rho
\nonumber\\
=&\,
+r\,\nu^{\,}_{\mathrm{s}}.
\end{align}
\end{proof}
    }

\subsection{The stability criterion for edge modes in the FQSHE}
\label{subsec: Stability criterion for the FQSHE}

What is the fate of the stability criterion
when we impose the residual spin-$1/2$ $U(1)$ symmetry
in the model so as to describe an underlying
microscopic model that supports
the FQSHE?
The residual spin-$1/2$ $U(1)$ symmetry is imposed
on the interacting theory
(\ref{eq:Mdef quantum edge theory})
by positing the existence of a spin vector 
$S=-\Sigma^{\,}_{1}\,S\in\mathbb{Z}^{2N}$ 
associated to a conserved $U(1)$ spin current. 
This spin vector is the counterpart to the charge vector  
$Q=+\Sigma^{\,}_{1}\,Q\in\mathbb{Z}^{2N}$.
The condition
\begin{subequations}
\label{eq: bosonic version $U(1)$ spin 1/2 symmetry}
\begin{equation}
S=-\Sigma^{\,}_{1}\, S
\label{eq: compatibility condition on S}
\end{equation}
is required for compatibility with time-reversal symmetry
and is the counterpart to Eq.
(\ref{eq:Mconditions for TRS on parameters c}).
Compatibility  with time-reversal symmetry
of $Q$ and $S$ thus imply that they are orthogonal,
$Q^{\mathsf{T}}\,S=0$.
If we restrict the interaction
(\ref{eq:Mdef quantum edge theory c})
by demanding that the tunneling matrices obey
\begin{equation}
T^{\mathsf{T}}\,S=0,
\label{eq: T S = 0}
\end{equation}
\end{subequations}
we probe the stability of the FQSHE described by
$\hat{H}^{\,}_{0}$
when perturbed by
$\hat{H}^{\,}_{\mathrm{int}}$.
\footnote{%
It is important to observe that the quadratic Hamiltonian
(\ref{eq:Mdef quantum edge theory b})
has a much larger symmetry group than the
interacting Hamiltonian
(\ref{eq:Mdef quantum edge theory c}).
For example, $\widehat{H}^{\,}_{0}$
commutes with the transformation
$
\widehat{\Phi}(t,x)
\rightarrow
\widehat{\Phi}(t,x)
+
\pi
K^{-1}\,
\Sigma^{\,}_{\downarrow}\,
S.
$
One verifies that the transformation law
of a Kramers doublet of fermions
under this transformation is the one expected from
a rotation about the quantization axis of
the residual spin-1/2 U(1) symmetry
provided the parities of the components of $S$ are the same
as those of $Q$. Hence, $\widehat{H}^{\,}_{0}$ has,
by construction, the residual spin-1/2 U(1) symmetry 
even though a generic
microscopic model with time-reversal symmetry 
does not. This residual spin-1/2 U(1) symmetry of
$\widehat{H}^{\,}_{0}$ is broken by
$\widehat{H}^{\,}_{\mathrm{int}}$, unless
one imposes the additional constraint
(\ref{eq: T S = 0})
on the tunneling matrices $T\in\mathbb{L}$ 
allowed to enter the interacting theory defined in Eq.
(\ref{eq:Mdef quantum edge theory}).
}

To answer this question we supplement the condition 
$T^{\mathsf{T}}Q=0$ on
tunneling vectors that belong to $\mathbb{L}$ and $\mathbb{H}$,
by $T^{\mathsf{T}}S=0$.
By construction, $S$
is orthogonal to $Q$.
Hence, it remains true that 
$\mathbb{H}$
is made of at most $N$ linearly
independent tunneling vectors.

The strategy for establishing the condition for the strong coupling limit of
$\hat{H}^{\,}_{\mathrm{int}}$
to open a mobility gap for all the extended modes of
$\hat{H}^{\,}_{0}$
thus remains to construct the largest
set $\mathbb{H}$ out of as few 
tunneling vectors with $T=-\Sigma^{\,}_{1}\,T$
as possible, 
since these tunneling vectors might spontaneously 
break time-reversal symmetry.

As before, there are $(N-1)$ linearly independent tunneling vectors
with $T=+\Sigma^{\,}_{1}\,T$,
while the tunneling matrix
$\bar{T}$ from Eq.\ (\ref{eq:Adef T0})
must belong to any $\mathbb{H}$
with $N$ linearly independent tunneling vectors.

At this stage, we need to distinguish the case
\begin{subequations}
\label{eq:Atwo option for T(0) if FQSHE}
\begin{equation}
\bar{T}^{\mathsf{T}}\,S=0
\label{eq:Atwo option for T(0) if FQSHE a}
\end{equation}
from the case
\begin{equation}
\bar{T}^{\mathsf{T}}\,S\neq 0.
\label{eq:Atwo option for T(0) if FQSHE b}
\end{equation}
\end{subequations}
In the former case, 
the spin neutrality condition
(\ref{eq: T S = 0})
holds for $\bar{T}$ and thus
the stability criterion
is unchanged for the FQSHE.
In the latter case,
the spin neutrality condition
(\ref{eq: T S = 0}) 
is violated 
so that $\hat{H}^{\,}_{\mathrm{int}}$ is independent
of any tunneling matrix proportional to
$\bar{T}$. Thus, when Eq.
(\ref{eq:Atwo option for T(0) if FQSHE b})
holds, as could be the case when
$\kappa\propto\myopenone^{\,}_{N}$ and $\Delta=0$ say, 
the FQSHE carried by at least one Kramers pair of
edge states of $\hat{H}^{\,}_{0}$
is robust to the strong coupling limit of
the  time-reversal symmetric and residual spin-$1/2$ 
U(1) symmetric
perturbation $\hat{H}^{\,}_{\mathrm{int}}$.

\section{Construction of two-dimensional topological phases from coupled wires}
\label{sec: Construction of two-dimensional topological phases from coupled wires}

\subsection{Introduction}

One accomplishment in the study of topological phases of matter 
has been the theoretical prediction and experimental discovery of two-dimensional
topological insulators \cite{Kane05a,Kane05b,Bernevig06a,Bernevig06b,Konig07}. 
The integer quantum Hall effect (IQHE) is an early example of how states 
can be classified into distinct topological classes using an integer, 
the Chern number, to express the quantized Hall conductivity
\cite{Klitzing80,Laughlin81,Thouless82}.
In the IQHE, the number of delocalized
edge channels is proportional to the quantized Hall conductivity
through the Chern number. More recently, it has been
found that the symmetry under reversal of time
protects the parity in the number of edge modes in (bulk)
insulators with strong spin-orbit interactions in two and
three dimensions \cite{Kane05a,Fu07a}.
Correspondingly, these systems are
characterized by a $\mathbb{Z}^{\,}_{2}$ topological invariant.

The discovery of $\mathbb{Z}^{\,}_{2}$ topological insulators initiated
a search for a classification of phases of 
fermionic matter that are distinct by some topological attribute. 
For non-interacting electrons, a complete classification, the tenfold way, 
has been accomplished in arbitrary dimensions
\cite{Schnyder08,Schnyder09,Ryu10,Kitaev09} .
In this scheme, three discrete symmetries that act locally in position space
-- time-reversal symmetry (TRS), 
particle-hole symmetry (PHS), 
and chiral or sublattice symmetry (SLS) -- 
play a central role when defining the quantum numbers that
identify the topological insulating fermionic phases of matter
within one of the ten symmetry classes 
(see the first three columns in Table \ref{table: main table}).

The tenfold way is believed to be robust to a perturbative treatment
of short-ranged electron-electron interactions for the following
reasons.  First, the unperturbed ground state in the clean limit and
in a closed geometry is non-degenerate. It is given by the filled bands 
of a band insulator. The band gap provides a small expansion parameter, namely
the ratio of the characteristic interacting energy scale to the band
gap. Second, the quantized topological invariant that characterizes
the filled bands, provided its definition and topological character
survives the presence of electron-electron interactions 
as is the case for the symmetry class A in two spatial dimensions,
cannot change in a perturbative treatment of 
short-range electron-electron interactions \cite{Gurarie11}.

On the other hand, the fate of the tenfold way when electron-electron
interactions are strong is rather subtle
\cite{Fidkowski10,Gurarie11,Manmana12,Wang12}. 
For example, short-range interactions can drive the system through a 
topological phase transition at which the energy gap closes
\cite{Raghu08,Budich12}.
They may also break spontaneously a defining symmetry of the topological
phase. Even when short-range interactions neither spontaneously break
the symmetries nor close the gap, it may be that two phases from the
non-interacting tenfold way cease to be distinguishable in the
presence of interactions. Indeed, it was shown for the symmetry class
BDI in one dimension by Fidkowski and Kitaev that the non-interacting
$\mathbb{Z}$ classification is too fine in that it must be replaced
by a $\mathbb{Z}^{\,}_{8}$ classification when generic short-range
interactions are allowed. How to construct a counterpart
to the tenfold way for interacting fermion (and boson) systems has
thus attracted a lot of interest
~\cite{Gu09,Pollmann10,Cirac11,Chen11a,Chen11b,Fidkowski11,Turner11,Gu12,Lu12,Chen13,Lu13}.

The fractional quantum Hall effect (FQHE) is the paradigm for a
situation by which interactions select topologically ordered ground states
of a very different kind than the non-degenerate ground states from
the tenfold way. On a closed two-dimensional manifold of genus $g$, 
interactions can stabilize incompressible many-body ground states 
with a $g$-dependent degeneracy. 
Excited states in the bulk must then carry fractional
quantum numbers (see \cite{Wen91b} and references therein).
Such phases of matter, that follow the FQHE paradigm,
appear in the literature under different names: fractional topological
insulators, long-range entangled phases, topologically ordered phases,
or symmetry enriched topological phases. In this section, 
the terminology long-range entangled (LRE) phase is used for all phases 
with nontrivial $g$-dependent ground state degeneracy.  
All other phases, 
i.e., those that follow the IQHE paradigm, are
called short-range entangled (SRE) phases.
(In doing so, the terminology of \cite{Lu12} is borrowed. 
It differs slightly from the one used in \cite{Chen11a}. 
The latter counts all chiral phases irrespective of their ground state 
degeneracy as LRE.)

While there are nontrivial SRE and LRE phases in the absence of any symmetry 
constraint, many SRE and LRE phases are defined by some symmetry they obey. 
If this symmetry is broken, the topological attribute of the phase 
is not well defined any more.
However, there is a sense in which LRE phases are more robust
than SRE phases against a weak breaking of the defining symmetry.
The topological attributes of LRE phases are not confined to the
boundary in space between two distinct topological realizations of
these phases, as they are for SRE phases. They also characterize 
intrinsic bulk properties such as the existence of gapped deconfined 
fractionalized excitations. 
Hence, whereas gapless edge states are gapped by any breaking of the
defining symmetry, topological bulk properties are robust to a weak breaking
of the defining symmetry as long as the characteristic energy scale
for this symmetry breaking is small compared to the bulk gap
in the LRE phase, for a small breaking of the protecting
symmetry does not wipe out the gapped deconfined fractionalized 
bulk excitations.

The purpose of this section is to implement a classification scheme for
interacting electronic systems in two spatial dimensions that treats 
SRE and LRE phases on equal footing. To this end, a coupled
wire construction for each of the symmetry classes from the tenfold
way is used. This approach has been pioneered in 
\cite{Yakovenko91} and \cite{Lee94}
for the IQHE and in 
\cite{Kane02} 
and 
\cite{Teo14}
for the FQHE
(see also the related works  
\cite{Sondhi01,gangof11,Klinovaja13a,Klinovaja14,Seroussi14,Vaezi14}).

The main idea is here the following.
To begin with, non-chiral Luttinger liquids are placed in a periodic
array of coupled wires. In doing so, forward-scattering two-body interactions 
are naturally accounted for within each wire. 
Back-scatterings (i.e., tunneling) within a given wire or
between neighboring wires are assumed to be the dominant energy scales.
Imposing symmetries constrains these allowed tunnelings.  
Whether a given arrangement of
tunnelings truly gaps out all bulk modes, except for some non-gapped edge
states on the first and last wire, is verified with the help of a
condition that applies to the limit of strong tunneling.
This condition is nothing but the Haldane criterion
of Section \ref{subsec:pinning} \cite{Haldane95}. 
It will be shown that, for a proper choice of the tunnelings, 
all bulk modes are gapped. Moreover, 
in five out of the ten symmetry classes of the tenfold way, there remain
gapless edge states in agreement with the tenfold way.  
It is the character of the tunnelings that
determines if this wire construction selects a SRE or a LRE phase.
Hence, this construction, predicated as it is on the strong tunneling
limit, generalizes the tenfold way for SRE phases to LRE phases.  
Evidently, this edge-centered classification
scheme does not distinguish between LRE phases of matter that do not
carry protected gapless edge modes at their interfaces. For example,
some fractional, time-reversal-symmetric, incompressible and topological phases
of matter can have fractionalized excitations in the bulk, 
while not supporting protected gapless modes at their boundaries
\cite{Wen91a,Sachdev91,Mudry94a}.

\begin{figure}[t]
\begin{center}
(a)
\includegraphics[scale=0.3,page=1]{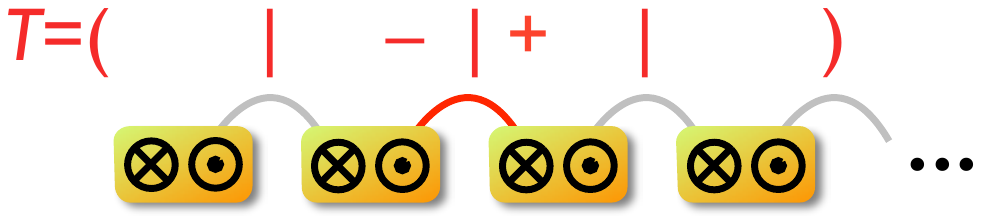}
(b)
\includegraphics[scale=0.3,page=2]{fig20.pdf}
\\ \vskip 20 true pt
(c)
\includegraphics[scale=0.3,page=3]{fig20.pdf}
(d)
\includegraphics[scale=0.3,page=4]{fig20.pdf}
\\ \vskip 20 true pt
(e)
\includegraphics[scale=0.3,page=7]{fig20.pdf}
(f)
\includegraphics[scale=0.3,page=8]{fig20.pdf}
\\ \vskip 20 true pt
(g)
\includegraphics[scale=0.3,page=5]{fig20.pdf}
(h)
\includegraphics[scale=0.3,page=6]{fig20.pdf}
\\ \vskip 20 true pt
(i)
\includegraphics[scale=0.3,page=9]{fig20.pdf}
(j)
\includegraphics[scale=0.3,page=10]{fig20.pdf}
\end{center}
\caption{
{[Colour online]} 
Figures for Table \ref{table: main table}.
(Reprinted with permission from [86]. Copyright 2014
by the American Physical Society.)
        }
\label{Figures for Table wire construction}
\end{figure}

The discussion in this section is inspired by \cite{Neupert14}.
It is organized as follows. The array of Luttinger
liquids is defined in Section \ref{sec: Definitions}.
The Haldane criterion, which plays an essential role for 
the stability analysis of the edge theory,
is reviewed in Section 7.5.3.3.
The five SRE entries and five LRE entries in 
\ref{table: main table}/Fig.\ \ref{Figures for Table wire construction} 
are derived in
Sections
\ref{sec: Reproducing the tenfold way} 
and 
\ref{sec: Fractionalized phases}, respectively.

The main result of this section are presented in Table \ref{table: main table}
and Fig.\ \ref{Figures for Table wire construction}.
For each of the symmetry classes A, AII, D, DIII, and C shown there,
the ground state supports propagating gapless edge modes localized
on the first and last wire that are immune to local and
symmetry-preserving perturbations. The first column labels the
symmetry classes according to the Cartan classification of symmetric
spaces. The second column dictates if the operations for reversal of
time 
($\widehat{\Theta}$ with the single-particle representation 
$\Theta$), 
exchange of particles and holes
($\widehat{\Pi}$ with the single-particle representation 
$\Pi$), 
and reversal of chirality 
($\widehat{C}$ with the single-particle representation 
$C$)
are the generators of symmetries 
with their single-particle representations
squaring to $+1$,
$-1$, or are not present in which case the entry $0$ is used.
(A chiral symmetry is present if
there exists a chiral operator $\widehat{C}$ that is
antiunitary and commutes with the Hamiltonian. The
single-particle representation $C$ of $\widehat{C}$
is a unitary operator that anticommutes with the
single-particle Hamiltonian.  In a basis in which
$C$ is strictly block off diagonal, $C$ reverses the
chirality.  This chirality is unrelated to the
direction of propagation of left and right movers
which is also called chirality in these lectures.)
The third column is the set to which the topological index from
the tenfold way, defined as it is in the non-interacting limit, belongs to.
The fourth column is a pictorial representation of the interactions 
(a set of tunnelings vectors $T$)
for the two-dimensional array of quantum wires that delivers short-range
entangled (SRE) gapless edge states. 
In these pictorial representations, 
shown in Fig.\ \ref{Figures for Table wire construction}(a, c, e, g, i),
a wire is represented by a shaded [coloured online]
box with the minimum number of channels compatible with the
symmetry class. Each channel in a wire is either a right mover
($\otimes$) or a left mover ($\odot$) that may or may not carry a
spin quantum number ($\uparrow,\downarrow$) or a particle (yellow
color) or hole (black color) attribute. The lines describe
tunneling processes within a wire or between consecutive wires 
in the array that are of one-body type 
when they do not carry an arrow
or of strictly many-body type when they carry an arrow.
Arrows point toward the sites on which creation operators act 
and away from the sites on which annihilation operators act.
For example in the symmetry class A, the single line connecting two
consecutive wires in the SRE column represents a one-body backward scattering
by which left and right movers belonging to consecutive wires are
coupled. The lines have been omitted for the fifth (LRE) column, 
only the tunneling vectors are specified.

\begin{table*}[t]
\caption{
Realization of a two-dimensional array of
quantum wires in each symmetry class of the tenfold way.
        }
\begin{center}
\begin{tabular}{| l | c c c |  c | l | l |}
\hline
$\vphantom{\Bigg[}$
&$\Theta^{2}$ &$\Pi^{2}$&$C^{2}$& &
SRE topological phase
&
LRE topological phase
\\
\hline
\hline
$\vphantom{\Bigg[}$
A
&$0$ &$0$&$0$&$\mathbb{Z}^{\,}$&
\begin{tabular}{c}
Fig.~\ref{Figures for Table wire construction}(a)
\end{tabular}
&
\begin{tabular}{c}
Fig.~\ref{Figures for Table wire construction}(b)
\end{tabular}
\\
\hline
$\vphantom{\Bigg[}$
AIII
&$0$ &$0$&$+$& & NONE &\\
\hline
\hline
$\vphantom{\Bigg[}$
AII
&$-$ &$0$&$0$&$\mathbb{Z}^{\,}_{2}$&
\begin{tabular}{c}
Fig.~\ref{Figures for Table wire construction}(c)
\end{tabular}
&
\begin{tabular}{c}
Fig.~\ref{Figures for Table wire construction}(d)
\end{tabular}
\\
\hline
$\vphantom{\Bigg[}$
DIII
&$-$ &$+$&$-$&$\mathbb{Z}^{\,}_{2}$&
\begin{tabular}{c}
Fig.~\ref{Figures for Table wire construction}(e)
\end{tabular}
&
\begin{tabular}{c}
Fig.~\ref{Figures for Table wire construction}(f)
\end{tabular}
\\
$\vphantom{\Bigg[}$
D
&$0$ &$+$&$0$&$\mathbb{Z}^{\,}$&
\begin{tabular}{c}
Fig.~\ref{Figures for Table wire construction}(g)
\end{tabular}
&
\begin{tabular}{c}
Fig.~\ref{Figures for Table wire construction}(h)
\end{tabular}
\\
$\vphantom{\Bigg[}$
BDI
&$+$ &$+$&$+$& & NONE &\\
\hline
$\vphantom{\Bigg[}$
AI
&$+$ &$0$&$0$& & NONE &\\
\hline
$\vphantom{\Bigg[}$
CI
&$+$ &$-$&$-$& & NONE &\\
$\vphantom{\Bigg[}$
C
&$0$ &$-$&$0$&$\mathbb{Z}^{\,}$&
\begin{tabular}{c}
Fig.~\ref{Figures for Table wire construction}(i)
\end{tabular}
&
\begin{tabular}{c}
Fig.~\ref{Figures for Table wire construction}(j)
\end{tabular}
\\
CII
&$-$ &$-$&$+$& & NONE &\\
\hline
\end{tabular}
\end{center}
\label{table: main table}
\end{table*}%

\begin{figure}[t]
\begin{center}
\includegraphics[width=84mm, page=1]{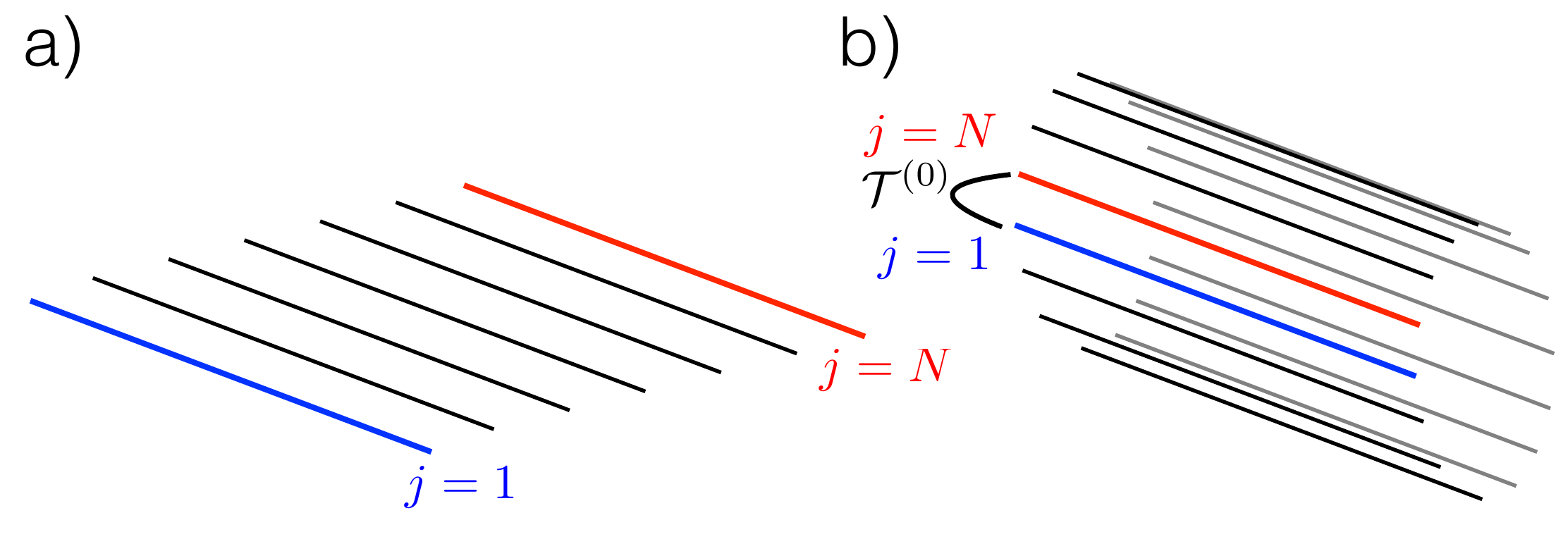}
\end{center}
\caption{
{[Colour online]}
The boundary conditions determine whether a topological phase 
has protected gapless modes or not.
(a)
With open boundary conditions, gapless modes exist near the wires $j=1$ 
and $j=N$, 
the scattering between them is forbidden by imposing locality in the limit 
$N\to\infty$.
(b)
Periodic boundary conditions allow the scattering vector $\mathcal{T}^{(0)}$ 
that gaps modes which were protected by locality before.
(Reprinted with permission from \protect\cite{Neupert14}. 
Copyright 2014 by the American Physical Society.)
        }
\label{fig: BoundaryConditions}
\end{figure}

\subsection{Definitions}
\label{sec: Definitions}

Consider an array of $N$ parallel wires that stretch along the $x$
direction of the two-dimensional embedding Euclidean space
(see Fig.~\ref{fig: BoundaryConditions}).
Label a wire by the Latin letter $i=1,\cdots,N$. Each wire supports fermions
that carry an even integer number $M$ of internal degrees of freedom
that discriminate between left- and right-movers, the projection along
the spin-$1/2$ quantization axis, and particle-hole quantum numbers,
among others (e.g., flavors). Label these internal degrees of
freedom by the Greek letter $\gamma=1,\cdots,M$. Combine those two
indices in a collective index $\mathsf{a}\equiv
(i,\gamma)$. Correspondingly, introduce the $M\times N$ pairs of
creation $\hat{\psi}^{\dag}_{\mathsf{a}}(x)$ and annihilation
$\hat{\psi}^{\,}_{\mathsf{a}}(x)$ field operators obeying the
fermionic equal-time algebra
\begin{subequations}
\label{eq: def hat psi's}
\begin{equation}
\left\{
\hat{\psi}^{\,}_{\mathsf{a}}(x),
\hat{\psi}^{\dag}_{\mathsf{a}'}(x')
\right\}=
\delta^{\,}_{\mathsf{a},\mathsf{a}'}\,
\delta(x-x')
\label{eq: def hat psi's a}
\end{equation}
with all other anticommutators vanishing and
the collective labels $\mathsf{a},\mathsf{a}'=1,\cdots,M\times N$.
The notation
\begin{equation}
\widehat{\Psi}^{\dag}(x)\equiv
\begin{pmatrix}
\hat{\psi}^{\dag}_{1}(x)&\cdots&\hat{\psi}^{\dag}_{MN}(x)
\end{pmatrix},
\quad
\widehat{\Psi}(x)\equiv
\begin{pmatrix}
\hat{\psi}^{\,}_{1}(x)\\ \vdots\\ \hat{\psi}^{\,}_{MN}(x)
\end{pmatrix},
\label{eq: def hat psi's b}
\end{equation}
\end{subequations}
is used for the operator-valued row ($\widehat{\Psi}^{\dag}$) 
and column ($\widehat{\Psi}^{\,}$) vector fields.
Assume that the many-body quantum dynamics of the fermions supported by 
this array of wires is governed by the Hamiltonian
$\hat{H}$, whereby interactions within each wire
are dominant over interactions between wires so that $\hat{H}$ 
may be represented as $N$ coupled Luttinger liquids,
each one of which is composed of $M$ interacting fermionic channels.

By assumption, the $M\times N$ 
fermionic channels making up the array may thus be bosonized
as was explained in chapters
\ref{sec: Fractionalization from Abelian bosonization}
and
\ref{sec: Stability analysis for the edge theory in the symmetry class AII}.
Within Abelian bosonization
\cite{Neupert11},
this is done by postulating first 
the $MN\times MN$ matrix 
\begin{subequations}
\label{eq: def mathcal K}
\begin{equation}
\mathcal{K}\equiv
\left(\mathcal{K}^{\,}_{\mathsf{a}\mathsf{a}'}\right)
\label{eq: def mathcal K a}
\end{equation}
to be symmetric with integer-valued entries.
Because an array of identical wires --
each of which having its quantum dynamics governed by
that of a Luttinger liquid -- is assumed, it is
natural to choose $\mathcal{K}$ to be reducible,
\begin{equation}
\mathcal{K}^{\,}_{\mathsf{a}\mathsf{a}'}=
\delta^{\,}_{ii'}\,
K^{\,}_{\gamma\gamma'},
\quad
\gamma,\gamma'=1,\cdots,M,
\quad
i,i'=1,\cdots,N.
\label{eq: def mathcal K c}
\end{equation}
\end{subequations}
A second $MN\times MN$ matrix is then defined by
\begin{subequations}
\label{eq: def mathcal L}
\begin{equation}
\mathcal{L}\equiv
\left(\mathcal{L}^{\,}_{\mathsf{a}\mathsf{a}'}\right)
\label{eq: def mathcal L a}
\end{equation}
where
\begin{equation}
\mathcal{L}^{\,}_{\mathsf{a}\mathsf{a}'}\:=
\mathrm{sgn}(\mathsf{a}-\mathsf{a}')
\left(
\mathcal{K}^{\,}_{\mathsf{a}\mathsf{a}'}
+
\mathcal{Q}^{\,}_{\mathsf{a}}\,
\mathcal{Q}^{\,}_{\mathsf{a}'}
\right),
\qquad
\mathsf{a},\mathsf{a}'=1,\cdots,MN,
\label{eq: def mathcal L bb}
\end{equation}
depends on the integer-valued
charge vector $\mathcal{Q}\equiv(\mathcal{Q}^{\,}_{\mathsf{a}})$ 
in addition to the matrix 
$\mathcal{K}\equiv \left(\mathcal{L}^{\,}_{\mathsf{a}\mathsf{a}'}\right)$.
The $MN$ compatibility conditions
\begin{equation}
(-1)^{\mathcal{K}^{\,}_{\mathsf{a}\mathsf{a}}}=
(-1)^{\mathcal{Q}^{\,}_{\mathsf{a}}},
\qquad
\mathsf{a}=1,\cdots,MN,
\label{eq: def mathcal L bbb}
\end{equation}
must hold. As we are after the effects of interactions between electrons,
we choose $\mathcal{Q}^{\,}_{\mathsf{a}}=1$ so that
\begin{equation}
\mathcal{L}^{\,}_{\mathsf{a}\mathsf{a}'}\:=
\mathrm{sgn}(\mathsf{a}-\mathsf{a}')
\left(
\mathcal{K}^{\,}_{\mathsf{a}\mathsf{a}'}
+
1
\right),
\qquad
\mathsf{a},\mathsf{a}'=1,\cdots,MN.
\label{eq: def mathcal L b}
\end{equation}
\end{subequations}
Third, one verifies that,
for any pair $\mathsf{a},\mathsf{a}'=1,\cdots,MN$,
the Hermitean fields
$\hat{\phi}^{\,}_{\mathsf{a}}$ 
and
$\hat{\phi}^{\,}_{\mathsf{a}'}$, 
defined by the Mandelstam formula
\begin{subequations}
\label{eq: def hat phi's}
\begin{equation}
\hat{\psi}^{\,}_{\mathsf{a}}(x)\equiv\;\;
:
\exp
\left(
+\mathrm{i}
\mathcal{K}^{\,}_{\mathsf{a}\mathsf{a}'}\,
\hat{\phi}^{\,}_{\mathsf{a}'}(x)
\right)
:
\label{eq: def hat phi's a}
\end{equation}
as they are,
obey the bosonic equal-time algebra
\begin{equation}
\left[
\hat{\phi}^{\,}_{\mathsf{a} }(x ),
\hat{\phi}^{\,}_{\mathsf{a}'}(x')
\right]=
-\mathrm{i}\pi
\left[
\mathcal{K}^{-1}_{\mathsf{a}\mathsf{a}'}\,
\mathrm{sgn}(x-x')
+
\mathcal{K}^{-1}_{\mathsf{a}\mathsf{b}}\,
\mathcal{L}^{\,}_{\mathsf{b}\mathsf{c}}\,
\mathcal{K}^{-1}_{\mathsf{c}\mathsf{a}'}
\right].
\label{eq: def hat phi's b}
\end{equation}
Here, the notation $:(\cdots):$ stands for normal ordering of the
argument $(\cdots)$ and the summation convention over repeated indices
is implied. In line with Eq.\ (\ref{eq: def hat psi's b}), 
the notation
\begin{equation}
\widehat{\Phi}^{\mathsf{T}}(x)\equiv
\begin{pmatrix}
\hat{\phi}^{\,}_{1}(x)&\cdots&\hat{\phi}^{\,}_{MN}(x)
\end{pmatrix},
\quad
\widehat{\Phi}(x)\equiv
\begin{pmatrix}
\hat{\phi}^{\,}_{1}(x)\\ \vdots\\ \hat{\phi}^{\,}_{MN}(x)
\end{pmatrix},
\label{eq: def hat phi's c}
\end{equation}
for the operator-valued row 
(i.e., $\widehat{\Phi}^{\mathsf{T}}$)
and column 
(i.e., $\widehat{\Phi}$)
vector fields is used.
Periodic boundary conditions along the $x$ direction
parallel to the wires are imposed by demanding that
\begin{equation}
\mathcal{K}\,\widehat{\Phi}(x+L)=
\mathcal{K}\,\widehat{\Phi}(x)
+
2\pi\,\mathcal{N},
\qquad
\mathcal{N}\in\mathbb{Z}^{MN}.
\end{equation}
\end{subequations}

Equipped with 
Eqs.\ (\ref{eq: def mathcal K})--(\ref{eq: def hat phi's}),
the many-body Hamiltonian $\widehat{H}$  
for the $MN$ interacting fermions all carrying the 
same electric charge $e$ and
propagating on the array of wires 
is decomposed additively into 
\begin{subequations}
\label{eq: hat H bosonized}
\begin{equation}
\widehat{H}=
\widehat{H}^{\,}_{\mathcal{V}}
+
\widehat{H}^{\,}_{\{\mathcal{T}\}}
+
\widehat{H}^{\,}_{\{\mathcal{Q}\}}.
\label{eq: hat H bosonized a}
\end{equation}
Hamiltonian
\begin{equation}
\widehat{H}^{\,}_{\mathcal{V}}\:=
\int\mathrm{d}x\,
\left(\partial^{\,}_{x}\widehat{\Phi}^{\mathsf{T}}\right)(t,x)\;
\mathcal{V}(x)\;
\left(\partial^{\,}_{x}\widehat{\Phi}\right)(t,x),
\label{eq: hat H bosonized b}
\end{equation}
even though quadratic in the bosonic field,
encodes both local one-body terms as well as
contact many-body interactions between
the $M$ fermionic channels in any given wire from the array 
through the block-diagonal, real-valued, and  symmetric $MN\times MN$ matrix
\begin{equation}
\label{eq: def mathcal V}
\mathcal{V}(x)\:=
\Big(
\mathcal{V}^{\,}_{\mathsf{a}\mathsf{a}'}(x)
\Big)\equiv
\left(
\mathcal{V}^{\,}_{(i,\gamma)(i',\gamma')}(x)
\right)=
\myopenone^{\,}_{N}\otimes
\Big(
V^{\,}_{\gamma\gamma'}(x)
\Big).
\end{equation}
Hamiltonian
\begin{align}
\widehat{H}^{\,}_{\{\mathcal{T}\}}\:=&\,
\int\mathrm{d}x\,
\sum_{\mathcal{T}}
\frac{h^{\,}_{\mathcal{T}}(x)}{2}\,
\left(
e^{+\mathrm{i}\alpha^{\,}_{\mathcal{T}}(x)}
\prod_{\mathsf{a}=1}^{MN}
\hat{\psi}^{\mathcal{T}^{\,}_{\mathsf{a}}}_{\mathsf{a}}(t,x)
+
\mathrm{H.c.}
\right)
\nonumber\\
=&\,
\int\mathrm{d}x\,
\sum_{\mathcal{T}}
h^{\,}_{\mathcal{T}}(x)\,
:
\cos
\left(
\mathcal{T}^{\mathsf{T}}\,
\mathcal{K}\,
\widehat{\Phi}(t,x)
+
\alpha^{\,}_{\mathcal{T}}(x)
\right):
\label{eq: hat H bosonized c}
\end{align}
is not quadratic in the bosonic fields. With the understanding that
the operator-multiplication of identical fermion fields at the same
point $x$ along the wire requires point splitting, and with the
short-hand notation
$\hat{\psi}^{-1}_{\mathsf{a}}(x)\equiv\hat{\psi}^{\dag}_{\mathsf{a}}(x)$,
$\widehat{H}^{\,}_{\{\mathcal{T}\}}$ is interpreted as 
a sum of all (possibly many-body) tunnelings between the fermionic channels. 
The set $\{\mathcal{T}\}$ comprises here  of \textit{all} integer-valued
tunneling vectors
\begin{equation}
\mathcal{T}\equiv
\left(\mathcal{T}^{\,}_{\mathsf{a}}\right)
\label{eq: def mathcal T a}
\end{equation}
obeying the condition
\begin{equation}
\sum_{\mathsf{a}=1}^{MN}
\mathcal{T}^{\,}_{\mathsf{a}}=
\begin{cases}
0\hbox{ mod }2,
&
\hbox{ for D, DIII, C, and CI,}
\\
&
\\
0,
&
\hbox{ otherwise.}
\end{cases} 
\label{eq: def mathcal T b}
\end{equation}
Moreover,
each $\mathcal{T}$ from the set $\{\mathcal{T}\}$
is assigned the real-valued functions
\begin{equation}
h^{\,}_{\mathcal{T}}(x)=
h^{* }_{\mathcal{T}}(x)\geq0
\label{eq: def mathcal T c}
\end{equation}
and
\begin{equation}
\alpha^{\,}_{\mathcal{T}}(x)=
\alpha^{* }_{\mathcal{T}}(x).
\label{eq: def mathcal T d}
\end{equation}
The condition~(\ref{eq: def mathcal T b}) ensures that
these tunneling events preserve the parity of the total fermion number
for the superconducting symmetry classes
(symmetry classes D, DIII, C, and CI in Table \ref{table: main table}), 
while they preserve the total fermion number for the non-superconducting 
symmetry classes (symmetry classes A, AIII, AI, AII, BDI, and CII in Table 
\ref{table: main table}).
The integer
\begin{equation}
q\:= 
\sum_{\mathsf{a}=1}^{MN}
\frac{|\mathcal{T}^{\,}_{\mathsf{a}}|}{2}
\label{eq:q-def}
\end{equation}
dictates that $\mathcal{T}$ encodes a $q$-body interaction
in the fermion representation. Hamiltonian
\begin{equation}
\widehat{H}^{\,}_{\{\mathcal{Q}\}}\:=
\int\mathrm{d}x\,
\frac{1}{2\pi}\,
A^{\,}_{0}(x)\,
\mathcal{Q}^{\mathsf{T}}\,
\left(
\partial^{\,}_{x}
\widehat{\Phi}
\right)(t,x)
\end{equation}
encodes the response to a static scalar potential 
$A^{\,}_{0}$
through the charge vector $\mathcal{Q}$ chosen to be
\begin{equation}
\mathcal{Q}=
\begin{pmatrix}1&\cdots&1\end{pmatrix}^{\mathsf{T}}
\end{equation}
\end{subequations}
in units of the electron charge $e$.

Hamiltonian (\ref{eq: hat H bosonized}) 
and the commutators (\ref{eq: def hat phi's b})
are form invariant under the transformation
\begin{subequations}
\label{eq: general linear trsf}
\begin{align}
&
\widehat{\Phi}(t,x)\=:
\mathcal{W}\,
\widetilde{\Phi}(t,x),
\label{eq: general linear trsfa}
\\
&
\widetilde{\mathcal{V}}(x)\:=
\mathcal{W}^{\mathsf{T}}\,
\mathcal{V}(x)\,
\mathcal{W},
\label{eq: general linear trsfb}
\\
&
\widetilde{\mathcal{K}}\:=
\mathcal{W}^{\mathsf{T}}\,
\mathcal{K}\,
\mathcal{W},
\label{eq: general linear trsfc}
\\
&
\widetilde{\mathcal{T}}\:=
\mathcal{W}^{-1}\,
\mathcal{T},
\label{eq: general linear trsfd}
\\
&
\widetilde{\mathcal{Q}}\:=
\mathcal{W}^{\mathsf{T}}\,
\mathcal{Q},
\label{eq: general linear trsfe}
\end{align}
\end{subequations}
where the $MN\times MN$ 
integer-valued matrix $\mathcal{W}$ is assumed to be invertible,
but not necessarily orthogonal!
Observe that the tunneling and charge vectors transform differently
whenever $\mathcal{W}^{-1}\neq\mathcal{W}^{\mathsf{T}}$
as they enter the Hamiltonian (\ref{eq: hat H bosonized})
with and without the matrix $\mathcal{K}$, respectively.
\footnote{%
Alternatively, 
$\mathcal{K}$ and 
$\mathcal{Q}\,\mathcal{Q}^{\mathsf{T}}$
must transform in the same way because of the Klein factors
(\ref{eq: def mathcal L bb}). 
          }

Even if the deviation of the matrix $\mathcal{W}$ from the 
$MN\times MN$ unit matrix is small, 
the relationship between the vertex operators
\begin{equation}
\begin{split}
\tilde{\psi}^{\,}_{\tilde{\mathsf{a}}}(t,x)\equiv&\,\;
:
\exp
\left(
+\mathrm{i}
\left(
\widetilde{\mathcal{K}}\,
\widetilde{\Phi}
\right)^{\,}_{\tilde{\mathsf{a}}}(t,x)
\right)
:\;
\\
=&\,\;
:
\exp
\left(
+\mathrm{i}
\left(
\mathcal{W}^{\mathsf{T}}\,
\mathcal{K}\,
\widehat{\Phi}
\right)^{\,}_{\tilde{\mathsf{a}}}(t,x)
\right)
:,
\quad
\tilde{\mathsf{a}}=1,\cdots,MN,
\end{split}
\label{eq: def tilde phi's}
\end{equation}
and the vertex operators (\ref{eq: def hat phi's a})
is non-perturbative. Performing a transformation of the form
(\ref{eq: general linear trsf}) to interpret a specific choice of
interactions encoded by the tunneling matrices $\{\mathcal{T}\}$
will play an essential role below. 

Because of the transformation laws
(\ref{eq: general linear trsfc}) and (\ref{eq: general linear trsfe}),
the dimensionless Hall conductivity
is invariant under the (not necessarily orthogonal)
transformation (\ref{eq: general linear trsf}). Indeed,
\begin{subequations}
\begin{equation}
\sigma^{\,}_{\mathrm{H}}\:=
\frac{1}{2\pi}\,
\left(
\mathcal{Q}^{\mathsf{T}}\,
\mathcal{K}^{-1}\,
\mathcal{Q}
\right)=
\frac{1}{2\pi}\,
\left(
\widetilde{\mathcal{Q}}^{\mathsf{T}}\,
\left(\mathcal{W}^{-1}\,\mathcal{W}\right)\,
\widetilde{\mathcal{K}}^{-1}\,
\left(\mathcal{W}^{-1}\,\mathcal{W}\right)^{\mathsf{T}}\,
\widetilde{\mathcal{Q}}
\right)
\end{equation}
equals
\begin{equation}
\widetilde{\sigma}^{\,}_{\mathrm{H}}\:=
\frac{1}{2\pi}\,
\left(
\widetilde{\mathcal{Q}}^{\mathsf{T}}\,
\widetilde{\mathcal{K}}^{-1}\,
\widetilde{\mathcal{Q}}
\right).
\end{equation}
\end{subequations}

In the sequel, we shall choose a non-orthogonal integer-valued $\mathcal{W}$ 
with $|\mathrm{det}\,\mathcal{W}|=1$ when
studying SRE phases of matter outside of the tenfold way, 
while we shall choose a non-orthogonal integer-valued 
$\mathcal{W}$ with $|\mathrm{det}\,\mathcal{W}|\neq1$
in order to construct LRE phases of matter.

\subsection{Strategy for constructing topological phases}
\label{sec: big strategy section}

The many-body Hamiltonian 
$\widehat{H}^{\,}_{\mathcal{V}}+\widehat{H}^{\,}_{\{\mathcal{T}\}}$
defined in Eq.\ (\ref{eq: hat H bosonized}) is to be chosen
so that
(i) it belongs to any one of the ten symmetry classes from 
the tenfold way (with the action of symmetries defined in Section
\ref{sec: representation of symmetries})
and (ii) all excitations in the bulk are gapped by a \textit{specific}
choice of the tunneling vectors $\{\mathcal{T}\}$ entering 
$\widehat{H}^{\,}_{\{\mathcal{T}\}}$
(with the condition for a spectral gap given in 
Section~\ref{sec: Conditions for a spectral gap}).
The energy scales in
$\widehat{H}^{\,}_{\{\mathcal{T}\}}$ 
are assumed sufficiently large compared to those in $\widehat{H}^{\,}_{\mathcal{V}}$ 
so that it is
$\widehat{H}^{\,}_{\mathcal{V}}$
that may be thought of as a perturbation of 
$\widehat{H}^{\,}_{\{\mathcal{T}\}}$ and not the converse. 

It will be shown that, for five of the ten symmetry classes, there can be 
protected gapless edge states because of locality and symmetry.
Step (ii) for each of the five symmetry classes 
supporting gapless edge states is represented pictorially
as is shown in the fourth and fifth columns 
of Table~\ref{table: main table}.
In each symmetry class, topologically trivial states that do not
support protected gapless edge states
in the tenfold classification can be constructed by gapping all 
states in each individual wire from the array.

\medskip\noindent
\textbf{\textit{7.5.3.1\hskip 10 true pt Representation of symmetries}}
\label{sec: representation of symmetries}

\medskip\noindent
The classification is based on the 
presence or the absence of the TRS 
and the PHS that 
are represented by the antiunitary many-body operator $\widehat{\Theta}$ 
and the unitary many-body operator $\widehat{\Pi}$, respectively.
Each of $\widehat{\Theta}$ and $\widehat{\Pi}$
can exist in two varieties such that their single-particle
representations $\Theta$ and $\Pi$
square to the identity operator
up to the multiplicative factor $\pm 1$,
\begin{equation}
\Theta^{2}=\pm 1,
\qquad
\Pi^{2}=\pm 1,
\label{eq: widehat Theta and widehat Phi square to pm1}
\end{equation} 
respectively. By assumption,
the set of all degrees of freedom in each given wire 
is invariant under the actions of $\widehat{\Theta}$ and $\widehat{\Pi}$.
If so, the actions of
$\widehat{\Theta}$ and $\widehat{\Pi}$
on the fermionic fields can be represented in two steps.
First, two $M\times M$-dimensional
matrix representations $P^{\,}_{\Theta}$ and $P^{\,}_{\Pi}$ 
of the permutation group of $M$ elements,
which are combined into the block-diagonal $MN\times MN$ 
real-valued and orthogonal matrices
\begin{subequations}
\label{eq: def reps widehat Theta and widehat Pi}
\begin{equation}
\mathcal{P}^{\,}_{\Theta}\:=\myopenone^{\,}_{N}\otimes P^{\,}_{\Theta},
\qquad
\mathcal{P}^{\,}_{\Pi}\:=\myopenone^{\,}_{N}\otimes P^{\,}_{\Pi},
\label{eq: def reps widehat Theta and widehat Pi a}
\end{equation}
where $\myopenone^{\,}_{N}$ is the $N\times N$ unit matrix
and 
$P^{\,}_{\Theta}$ and $P^{\,}_{\Pi}$
represent products of transpositions so that
\begin{equation}
P^{\,}_{\Theta}= 
P^{-1}_{\Theta}=
P^{\mathsf{T}}_{\Theta},
\qquad
P^{\,}_{\Pi}=
P^{-1}_{\Pi}=
P^{\mathsf{T}}_{\Pi},
\label{eq: def reps widehat Theta and widehat Pi a bis}
\end{equation}
are introduced.
Second, two column vectors 
$I^{\,}_{\Theta}\in\mathbb{Z}^{M}$ 
and 
$I^{\,}_{\Pi}\in\mathbb{Z}^{M}$,
which are combined into the two column vectors 
\begin{equation}
\mathcal{I}^{\,}_{\Theta}\:=
\begin{pmatrix}
I^{\,}_{\Theta}
\\
\vdots
\\
I^{\,}_{\Theta}
\end{pmatrix},
\qquad
\mathcal{I}^{\,}_{\Pi}\:=
\begin{pmatrix}
I^{\,}_{\Pi}
\\
\vdots
\\
I^{\,}_{\Pi}
\end{pmatrix},
\label{eq: def reps widehat Theta and widehat Pi b}
\end{equation}
and the $MN\times MN$ diagonal matrices
\begin{equation}
\mathcal{D}^{\,}_{\Theta}\:=
\mathrm{diag}\,(\mathcal{I}^{\,}_{\Theta}),
\qquad
\mathcal{D}^{\,}_{\Pi}\:=
\mathrm{diag}\,(\mathcal{I}^{\,}_{\Pi}),
\label{eq: def reps widehat Theta and widehat Pi b bis}
\end{equation}
with the components of the vectors
$\mathcal{I}^{\,}_{\Theta}$
and
$\mathcal{I}^{\,}_{\Pi}$
as diagonal matrix elements,
are introduced.
The vectors $I^{\,}_{\Theta}$ and $I^{\,}_{\Pi}$
are not chosen arbitrarily. Demand that
the vectors 
$(1+\mathcal{P}^{\,}_{\Theta})\,\mathcal{I}^{\,}_{\Theta}$ 
and
$(1+\mathcal{P}^{\,}_{\Pi})\,\mathcal{I}^{\,}_{\Pi}$ 
are made of even 
[for the $+1$ in Eq.\ (\ref{eq: widehat Theta and widehat Phi square to pm1})]
and
odd 
[for the $-1$ in Eq.\ (\ref{eq: widehat Theta and widehat Phi square to pm1})]
integer entries only, while
\begin{equation}
e^{+\mathrm{i}\pi\,\mathcal{D}^{\,}_{\Theta}}\,
\mathcal{P}^{\,}_{\Theta}=
\pm
\mathcal{P}^{\,}_{\Theta}\,
e^{+\mathrm{i}\pi\,\mathcal{D}^{\,}_{\Theta}}
\label{eq: TRS on P and I }
\end{equation}
and
\begin{equation}
e^{+\mathrm{i}\pi\,\mathcal{D}^{\,}_{\Pi}}\,
\mathcal{P}^{\,}_{\Pi}=
\pm
\mathcal{P}^{\,}_{\Pi}\,
e^{+\mathrm{i}\pi\,\mathcal{D}^{\,}_{\Pi}},
\label{eq: PHS on P and I }
\end{equation} 
in order to meet 
$\Theta^{2}=\pm1$ and $\Pi^{2}=\pm1$, respectively.
The operations of reversal of time and interchanges of
particles and holes are then represented by
\begin{align}
\widehat{\Theta}\,
\widehat{\Psi}\,
\widehat{\Theta}^{-1}
=&\,
e^{+\mathrm{i}\,\pi\,\mathcal{D}^{\,}_{\Theta}}\,
\mathcal{P}^{\,}_{\Theta}\,\widehat{\Psi},
\label{eq: def reps widehat Theta and widehat Pi c}
\\
\widehat{\Pi}\,
\widehat{\Psi}\,
\widehat{\Pi}^{-1}
=&\,
e^{+\mathrm{i}\pi\,\mathcal{D}^{\,}_{\Pi}}\,
\mathcal{P}^{\,}_{\Pi}\,
\widehat{\Psi},
\label{eq: def reps widehat Theta and widehat Pi d}
\end{align}
for the fermions and
\begin{align}
\widehat{\Theta}\,
\widehat{\Phi}\,
\widehat{\Theta}^{-1}
=&\,
\mathcal{P}^{\,}_{\Theta}\,
\widehat{\Phi}
+
\pi\,
\mathcal{K}^{-1}\,
\mathcal{I}^{\,}_{\Theta},
\label{eq: def reps widehat Theta and widehat Pi e}
\\
\widehat{\Pi}\,
\widehat{\Phi}\,
\widehat{\Pi}^{-1}
=&
\mathcal{P}^{\,}_{\Pi}\,
\widehat{\Phi}
+
\pi\,
\mathcal{K}^{-1}\,
\mathcal{I}^{\,}_{\Pi},
\label{eq: def reps widehat Theta and widehat Pi f}
\end{align}
\end{subequations}
for the bosons. One verifies that Eq.
(\ref{eq: widehat Theta and widehat Phi square to pm1})
is fulfilled.

Hamiltonian~(\ref{eq: hat H bosonized}) is TRS if
\begin{subequations}
\label{eq: TRS H}
\begin{equation}
\widehat{\Theta}\,
\widehat{H}\,
\widehat{\Theta}^{-1}=
+\widehat{H}.
\label{eq: TRS H a}
\end{equation}
This condition is met if
\begin{align}
&
P^{\,}_{\Theta}\,
V\,
P^{-1}_{\Theta}
=
+V,
\label{eq: TRS H b}
\\
&
P^{\,}_{\Theta}\,
K\,
P^{-1}_{\Theta}
=
-
K,
\label{eq: TRS H c}
\\
&
h^{\,}_{\mathcal{T}}(x)=
h^{\,}_{-\mathcal{P}^{\,}_{\Theta}\mathcal{T}}(x),
\label{eq: TRS H d}
\\
&
\alpha^{\,}_{\mathcal{T}}(x)=
\alpha^{\,}_{-\mathcal{P}^{\,}_{\Theta}\mathcal{T}}(x)
-
\pi\,
\mathcal{T}^{\mathsf{T}}\,
\mathcal{P}^{\,}_{\Theta}\,
\mathcal{I}^{\,}_{\Theta}.
\label{eq: TRS H e}
\end{align}
\end{subequations}
The 
Hamiltonian~(\ref{eq: hat H bosonized}) is PHS if
\begin{subequations}
\label{eq: PHS H}
\begin{equation}
\widehat{\Pi}\,
\widehat{H}\,
\widehat{\Pi}^{-1}=
+\widehat{H}.
\label{eq: PHS H a}
\end{equation}
This condition is met if 
\begin{align}
&
P^{\,}_{\Pi}\,
V\,
P^{-1}_{\Pi}
=
+V,
\label{eq: PHS H b}
\\
&
P^{\,}_{\Pi}\,
K\,
P^{-1}_{\Pi}
=
+
K,
\label{eq: PHS H c}
\\
&
h^{\,}_{\mathcal{T}}(x)=
h^{\,}_{+\mathcal{P}^{\,}_{\Pi}\mathcal{T}}(x),
\label{eq: PHS H d}
\\
&
\alpha^{\,}_{\mathcal{T}}(x)=
\alpha^{\,}_{\mathcal{P}^{\,}_{\Pi}\mathcal{T}}(x)
+
\pi\,
\mathcal{T}^{\mathsf{T}}\,
\mathcal{P}^{\,}_{\Pi}\,
\mathcal{I}^{\,}_{\Pi}.
\label{eq: PHS H e}
\end{align}
\end{subequations}

\medskip\noindent
\textbf{\textit{7.5.3.2\hskip 10 true pt Particle-hole symmetry in interacting superconductors}}
\label{sec: PHS in superconductors}

\medskip\noindent
The total number of fermions is a good quantum number
in any metallic or insulating phase of fermionic matter.
This is not true anymore in the mean-field treatment of superconductivity.
In a superconductor, within a mean-field approximation, charge is
conserved modulo two as Cooper pairs can be created and annihilated.
The existence of superconductors and 
the phenomenological success of the mean-field approximation
suggest that the conservation of
the total fermion number operator should be relaxed
down to its parity in a superconducting phase of matter.
If one only demands that the parity of the total fermion number is conserved,  
one may then decompose any fermionic creation operator in the position basis
into its real and imaginary parts, thereby obtaining two Hermitean operators
called Majorana operators. Any Hermitean Hamiltonian that is build out of
even powers of Majorana operators necessarily 
preserves the parity of the total fermion number operator,
but it might break the conservation of the total fermion number.
By definition, any such Hamiltonian belongs to the symmetry class D. 

The tool of Abelian bosonization allows to represent a fermion
operator as a single exponential of a Bose field. In Abelian bosonization,
a Majorana operator is the sum of two exponentials, and this fact makes it 
cumbersome to apply Abelian bosonization for Majorana operators.
It is possible to circumvent this difficulty by representing any
Hamiltonian from the symmetry class D in terms of the components
of Nambu-Gorkov spinors obeying a reality condition. 
Indeed, one may double the dimensionality of the 
single-particle Hilbert space by introducing Nambu-Gorkov spinors
with the understanding that (i) a reality condition on the Nambu-Gorkov spinors
must hold within the physical subspace of the enlarged 
single-particle Hilbert space and (ii) the dynamics dictated by the
many-body Hamiltonian must be compatible with this reality condition.
The reality condition keeps track of the fact that there are many 
ways to express an even polynomial of Majorana operators in terms of
the components of a Nambu-Gorkov spinor. The complication brought about by 
this redundancy is compensated by the fact that it is straightforward 
to implement Abelian bosonization in the Nambu-Gorkov representation.

To implement this particle-hole doubling, assign to every
pair of fermionic operators $\hat{\psi}$ and $\hat{\psi}^{\dag}$
(whose indices have been omitted for simplicity)
related to each other by the reality condition
\begin{subequations}
\begin{equation}
\widehat{\Pi}\,
\hat{\psi}\,
\widehat{\Pi}^{\dag}=
\hat{\psi}^{\dag},
\end{equation}
the pair of bosonic field operators 
$\hat{\phi}$ 
and
$\hat{\phi}'$ 
related by the reality condition
\begin{equation}
\widehat{\Pi}\,
\hat{\phi}\,
\widehat{\Pi}^{\dag}=
-\hat{\phi}'.
\end{equation}
\end{subequations}
Invariance under this transformation has to be imposed on the 
(interacting) Hamiltonian in the doubled (Nambu-Gorkov) representation.
In addition to the PHS, it is also demanded, when describing the
superconducting symmetry classes, that the parity of the total fermion
number is conserved. This discrete global symmetry, 
the symmetry of the Hamiltonian under the reversal of sign of 
all fermion operators, becomes a continuous
$U(1)$ global symmetry that is responsible for the conservation 
of the electric charge in all non-superconducting symmetry classes.
In this way, all nine symmetry classes from the tenfold way 
descend from the symmetry class D by imposing a composition of TRS, 
$U(1)$ charge conservation, and the chiral (sublattice) symmetry.

The combined effects of disorder and interactions in superconductors was
studied in \cite{Jeng01,Foster06,Foster08,Foster14}
starting from the Nambu-Gorkov formalism to derive a non-linear-sigma model
for the Goldstone modes relevant to the interplay between the
physics of Anderson localization and that of interactions.
The stability of Majorana zero modes to interactions preserving
the particle-hole symmetry was studied in \cite{Goldstein12}. 

\medskip\noindent
\textbf{\textit{7.5.3.3\hskip 10 true pt Conditions for a spectral gap}}
\label{sec: Conditions for a spectral gap}

\medskip\noindent
Hamiltonian $\widehat{H}^{\,}_{\mathcal{V}}$ 
in the decomposition~(\ref{eq: hat H bosonized})
has $MN$ gapless modes. 
However, $\widehat{H}^{\,}_{\mathcal{V}}$  does not
commute with $\widehat{H}^{\,}_{\{\mathcal{T}\}}$
and the competition between 
$\widehat{H}^{\,}_{\mathcal{V}}$
and  
$\widehat{H}^{\,}_{\{\mathcal{T}\}}$
can gap some, if not all, 
the gapless modes of
$\widehat{H}^{\,}_{\mathcal{V}}$.
For example,
a tunneling amplitude that scatters the right mover 
into the left mover of each flavor in each wire 
will gap out the spectrum of
$\widehat{H}^{\,}_{\mathcal{V}}$.

A term in $\widehat{H}^{\,}_{\mathsf{\{\mathcal{T}\}}}$ has the
potential to gap out a gapless mode of $\widehat{H}^{\,}_{\mathcal{V}}$ 
if the condition (in the Heisenberg representation)
\cite{Neupert11,Haldane88}
\begin{equation}
\partial^{\,}_{x}
\left[
\mathcal{T}^{\mathsf{T}}\,
\mathcal{K}\,
\widehat{\Phi}(t,x)
+
\alpha^{\,}_{\mathcal{T}}(x)
\right]=
C^{\,}_{\mathcal{T}}(x)
\label{eq: locking condition bis}
\end{equation}
holds for some time-independent real-valued functions $C^{\,}_{\mathcal{T}}(x)$
on the canonical momentum 
\begin{equation}
(4\pi)^{-1}\,\mathcal{K}\,
(\partial^{\,}_{x}\widehat{\Phi})(t,x)
\end{equation}
that is conjugate to $\widehat{\Phi}(t,x)$,
when applied to the ground state. 
The locking condition~(\ref{eq: locking condition bis})
removes a pair of chiral bosonic modes with opposite chiralities
from the gapless degrees of freedom of the theory.
However, not all scattering vectors $\mathcal{T}$ 
can simultaneously lead to such a locking due to quantum fluctuations. 
The set of linear combinations
$\{\mathcal{T}^{\mathsf{T}}\,\mathcal{K}\,\widehat{\Phi}(t,x)\}$
that can satisfy the locking condition
(\ref{eq: locking condition bis})
simultaneously is labelled by the subset
$\left\{\mathcal{T}\right\}^{\,}_{\mathrm{locking}}$ 
of all tunneling matrices
$\{\mathcal{T}\}$ defined by 
Eqs.\ (\ref{eq: def mathcal T a})
and~(\ref{eq: def mathcal T b})
obeying the Haldane criterion
(\ref{eq: Haldane conditions})
\cite{Neupert11,Haldane88}
\begin{subequations}
\label{eq: Haldane conditions}
\begin{equation}
\mathcal{T}^{\mathsf{T}}\,\mathcal{K}\,\mathcal{T}=0
\end{equation}
for any 
$\mathcal{T}\in\{\mathcal{T}\}^{\,}_{\mathrm{locking}}$ 
and
\begin{equation}
\mathcal{T}^{\mathsf{T}}\,\mathcal{K}\,\mathcal{T}'=0
\end{equation}
\end{subequations}
pairwise for any
$\mathcal{T}\neq\mathcal{T}'\in\{\mathcal{T}\}^{\,}_{\mathrm{locking}}$.

\subsection{Reproducing the tenfold way}
\label{sec: Reproducing the tenfold way}

The first goal is to apply the wire construction in order to reproduce
the classification of non-interacting topological insulators 
(symmetry classes A, AIII, AI, AII, BDI, and CII in Table 
\ref{table: main table})
and superconductors
(symmetry classes D, DIII, C,and CI in Table \ref{table: main table})
in ($2+1$) dimensions
(see Table~\ref{table: main table})
\cite{Schnyder08,Schnyder09,Ryu10,Kitaev09}.
In this section, the classification scheme is carried out 
within the bosonized description of quantum wires. Here, 
the classification is restricted to one-body tunneling terms, i.e., $q=1$
in Eq.\ (\ref{eq:q-def}), for the non-superconducting symmetry classes,
and to two-body tunneling terms, i.e., $q=2$
in Eq.\ (\ref{eq:q-def}), for the superconducting symmetry classes.
In Section~\ref{sec: Fractionalized phases}, 
this construction is generalized to the cases 
$q>1$ and $q>2$ of multi-particle tunnelings in the non-superconducting
and superconducting symmetry classes, respectively.
The topological stability of edge modes will be an
immediate consequence of the observation that no symmetry-respecting
local terms can be added to the models to be constructed below.

Within the classification of non-interacting Hamiltonians,
superconductors are nothing but fermionic bilinears with a
particle-hole symmetry. The physical interpretation of the degrees of
freedom as Bogoliubov quasiparticles is of no consequence to the
analysis. In particular, they still carry an effective 
conserved $U(1)$ charge in the non-interacting description.

\medskip\noindent
\textbf{\textit{7.5.4.1\hskip 10 true pt Symmetry class A}}
\label{subsec: Symmetry class A}

\medskip\noindent
\textbf{SRE phases in the tenfold way.}
Topological insulators in symmetry class A can be realized without any
symmetry aside from the $U(1)$ charge conservation. The wire
construction starts from wires supporting spinless fermions, so that the
minimal choice $M=2$ only counts left- and right-moving degrees of
freedom. The $K$-matrix reads
\begin{subequations}
\label{eq: def H for class A}
\begin{equation}
K\:=\mathrm{diag}\,(+1,-1).
\label{eq: def H for class A a}
\end{equation}
The entry $+1$ of the $K$-matrix corresponds to a right mover.
It is depicted by the symbol $\otimes$ in
the first line of Table~\ref{table: main table}.
The entry $-1$ of the $K$-matrix corresponds to a left mover.
It is depicted by the symbol $\odot$ in
the first line of Table~\ref{table: main table}.
The operation for reversal of time in any one of the $N$ wires
is represented by 
[one verifies that Eq.\ (\ref{eq: TRS on P and I }) holds]
\begin{equation}
P^{\,}_{\Theta}\:=
\begin{pmatrix}
0&1
\\
1&0
\end{pmatrix},
\qquad
I^{\,}_{\Theta}\:=
\begin{pmatrix}
0\\0
\end{pmatrix}.
\label{eq: def H for class A b}
\end{equation}
\end{subequations}
Let us define $\widehat{H}^{\,}_{\{\mathcal{T}\}}$ 
by choosing $(N-1)$ scattering vectors,
whereby, for any $j=1,\cdots,(N-1)$,
\begin{subequations}
\label{eq: def H of T for class A}
\begin{equation}
\mathcal{T}^{(j)}_{(i,\gamma)}\:=
\delta^{\,}_{i,j}\,
\delta^{\,}_{\gamma,2}
-
\delta^{\,}_{i-1,j}\,
\delta^{\,}_{\gamma,1}
\label{eq: def H of T for class A d}
\end{equation}
with
$i=1,\cdots,N$
and $\gamma=1,2$.
In other words,
\begin{equation}
\mathcal{T}^{(j)}\:=
(0,0|\cdots|0,+1|-1,0|\cdots|0,0)^{\mathsf{T}}
\label{eq: def H of T for class A e}
\end{equation}
for $j=1,\cdots, N-1$.
Intent on helping with the interpretation of the tunneling vectors, 
the $|$'s in Eq.\ (\ref{eq: def H of T for class A e}) is used to
compartmentalize the elements within a given wire. Henceforth,
there are $M=2$ vector components within each pair of $|$'s that
encode the $M=2$ degrees of freedom within a given wire. The $j$th
scattering vector~(\ref{eq: def H of T for class A e}) labels a one-body
interaction in the fermion representation that fulfills 
Eq.\ (\ref{eq: def mathcal T b}) and breaks TRS, since the scattering vector
$(0,+1)^{\mathsf{T}}$ is mapped into the scattering vector
$(+1,0)^{\mathsf{T}}$ by the permutation $P^{\,}_{\Theta}$ that
represents reversal of time in a wire by exchanging right- with
left-movers. For any $j=1,\cdots,(N-1)$, introduce the
amplitude
\begin{equation}
h^{\,}_{\mathcal{T}^{(j)}}(x)\geq0
\label{eq: def H of T for class A f}
\end{equation}
and the phase
\begin{equation}
\alpha^{\,}_{\mathcal{T}^{(j)}}(x)\in\mathbb{R}
\label{eq: def H of T for class A g}
\end{equation}
\end{subequations}
according to  
Eqs.\ (\ref{eq: TRS H d}) and~(\ref{eq: TRS H e}), respectively.
The choices for the amplitude~(\ref{eq: def H of T for class A f})
and the phase~(\ref{eq: def H of T for class A g})
are arbitrary. In particular
the amplitude~(\ref{eq: def H of T for class A f})
can be chosen to be sufficiently large so that it is
$\widehat{H}^{\,}_{\mathcal{V}}$
that may be thought of as a perturbation of 
$\widehat{H}^{\,}_{\{\mathcal{T}\}}$ and not the converse.

One verifies that all $(N-1)$
scattering vectors~(\ref{eq: def H of T for class A d})
satisfy the Haldane criterion~(\ref{eq: Haldane conditions}), i.e.,
\begin{equation}
\mathcal{T}^{(i)\mathsf{T}}\,
\mathcal{K}\,
\mathcal{T}^{(j)}=0,
\qquad
i,j=1,\cdots,N-1.
\end{equation}
Correspondingly, the term $\widehat{H}^{\,}_{\{\mathcal{T}\}}$
gaps out $2(N-1)$ of the $2N$ gapless modes of
$\widehat{H}^{\,}_{\mathcal{V}}$. Two modes
of opposite chirality that propagate along the
first and last wire, respectively,
remain in the low energy sector of the theory. These edge
states are localized on wire $i=1$ and $i=N$, respectively,
for their overlaps with the gapped states from the bulk decay
exponentially fast as a function of the distance away from
the first and end wires. 
The energy splitting between the edge state localized on wire $i=1$
and the one localized on wire $i=N$ 
that is brought about by the bulk states vanishes
exponentially fast with increasing $N$. 
Two gapless edge states
with opposite chiralities emerge in the two-dimensional limit $N\to\infty$.

At energies much lower than the bulk gap, the effective
$\mathcal{K}$-matrix for the edge modes is
\begin{equation}
\begin{split}
\mathcal{K}^{\,}_{\mathrm{eff}}\:=&\,
\mathrm{diag}(+1,0|0,0|
\cdots|0,0|0,-1).
\end{split}
\end{equation} 
Here, $\mathcal{K}^{\,}_{\mathrm{eff}}$ follows from
replacing the entries in the 
$2N\times2N$ $\mathcal{K}$ matrix
for all gapped modes by 0.
The pictorial representation of the topological phase 
in the symmetry class A with one chiral edge state per end wire 
through the wire construction is shown in
Fig.\ \ref{Figures for Table wire construction}(a).
The generalization to an arbitrary number $n$ of gapless edge states
sharing a given chirality on the first wire 
that is opposite to that of the last
wire is the following. Enlarge $M=2$ to $M=2n$ by
making $n$ identical copies of the model depicted in 
Fig.\ \ref{Figures for Table wire construction}(a).
The stability of the $n$
chiral gapless edge states in wire $1$ and wire $N$
is guaranteed because back-scattering 
among these gapless edges state
is not allowed kinematically within wire $1$ or within wire $N$,
while back-scattering across the bulk is exponentially 
suppressed for $N$ large by locality and the gap in the bulk. 
The number of robust gapless edge states of a given chirality
is thus integer. This is why the entry 
in the third column on the first line
of Table~\ref{table: main table} {is $\mathbb{Z}$}.

\medskip\noindent
\textbf{SRE phases beyond the tenfold way.}
It is imperative to ask whether the phases constructed so far 
exhaust all possible SRE phases in the symmetry class A. 
By demanding that one-body interactions are dominant over
many-body interactions, all phases from the (exhaustive) classification 
for non-interacting fermions in class A and only those were constructed. 
In these phases, the same topological invariant controls the Hall and the thermal
conductivities.
However, it was observed that interacting fermion systems 
can host additional SRE phases in the symmetry class A 
where this connection is lost \cite{Lu12}. 
These phases are characterized by an edge that includes charge-neutral 
chiral modes. While such modes contribute to the quantized energy transport 
(i.e., the thermal Hall conductivity), 
they do not contribute to the quantized charge transport 
(i.e., the charge Hall conductivity). 
By considering the thermal and 
charge Hall conductivity as two independent quantized topological responses, 
this enlarges the classification of SPT phases in the symmetry class A to 
$\mathbb{Z}\times\mathbb{Z}$.

Starting from identical fermions of charge $e$, 
an explicit model for an array of wires will be constructed
that stabilizes a SRE phase of matter in the symmetry class A 
carrying a non-vanishing Hall conductivity 
but a vanishing thermal Hall conductivity.
In order to build a wire-construction of such a strongly interacting 
SRE phase in the symmetry class A, three spinless electronic wires are grouped
into one unit cell, i.e.,
\begin{subequations}
\begin{equation}
K\:=
\mathrm{diag}(+1,-1,+1,-1,+1,-1).
\end{equation} 
It will be useful to arrange the charges $Q^{\,}_{\gamma}=1$ 
measured in units of the electron charge $e$ 
for each of the modes $\hat{\phi}^{\,}_{\gamma}$,
$\gamma=1,\cdots,M$, into the vector
\begin{equation}
Q=(1,1,1,1,1,1)^{\mathsf{T}}.
\end{equation}
\end{subequations}

The physical meaning of the tunneling vectors (interactions)
to be defined below is most transparent when
employing the following linear transformation on the bosonic field variables
\begin{subequations}
\label{eq: T tilde if one wants to get bosons out of fermions}
\begin{align}
&
\widehat{\Phi}(x)\=:
\mathcal{W}\,
\widetilde{\Phi}(x),
\label{eq: T tilde if one wants to get bosons out of fermions a}
\\
&
\widetilde{\mathcal{K}}\:=
\mathcal{W}^{\mathsf{T}}\,
\mathcal{K}\,
\mathcal{W},
\label{eq: T tilde if one wants to get bosons out of fermions c}
\\
&
\widetilde{\mathcal{T}}\:=
\mathcal{W}^{-1}\,
\mathcal{T},
\label{eq: T tilde if one wants to get bosons out of fermions d}
\\
&
\widetilde{\mathcal{Q}}\:=
\mathcal{W}^{\mathsf{T}}\,
\mathcal{Q},
\label{eq: T tilde if one wants to get bosons out of fermions e}
\end{align}
\end{subequations}
where $\mathcal{W}$ is a $MN\times MN$ block-diagonal matrix 
with the non-orthogonal block $W$ having integer entries and 
the determinant one in magnitude.
The transformation $W$ and its inverse $W^{-1}$ are given by
\begin{equation}
\begin{split}
&
W\:=
\begin{pmatrix}
 0&-1&-1& 0& 0&0\\
+1&-1&-1& 0& 0& 0\\
+1& 0&-1& 0& 0& 0\\
 0& 0& 0&-1& 0&+1\\
 0& 0& 0&-1&-1&+1\\
 0& 0& 0&-1&-1& 0
\end{pmatrix},
\\
&
W^{-1}\:= 
\begin{pmatrix}
-1&+1& 0& 0& 0&0\\
 0&-1&+1& 0& 0& 0\\
-1&+1&-1& 0& 0& 0\\
 0& 0& 0&-1&+1&-1\\
 0& 0& 0&+1&-1& 0\\
 0& 0& 0& 0&+1&-1
\end{pmatrix},
\end{split}
\label{eq: trafo for bosonic chain}
\end{equation}
respectively.
It brings $K$ to the form
\begin{equation}
\widetilde{K}\:=
\left(
\begin{array}{cccccc}
 0 & +1 & 0 & 0 & 0 & 0 \\
 +1 & 0 & 0 & 0 & 0 & 0 \\
 0 & 0 & +1 & 0 & 0 & 0 \\
 0 & 0 & 0 & -1 & 0 & 0 \\
 0 & 0 & 0 & 0 & 0 & -1 \\
 0 & 0 & 0 & 0 & -1 & 0 \\
\end{array}
\right).
\end{equation}

As can be read off from Eq.~\eqref{eq: def hat phi's b}, the parity of
$K^{\,}_{\gamma\gamma}$ determines the self-statistics of particles of type
$\gamma=1,\cdots,N$. As Eq.~\eqref{eq: def hat phi's b}
is form invariant under the transformation
(\ref{eq: T tilde if one wants to get bosons out of fermions}),
one concludes that, with the choice~\eqref{eq: trafo for bosonic chain}, 
the transformed modes $\gamma=1,2$ as well as the modes $\gamma=5,6$ 
are pairs of bosonic degrees of freedom, 
while the third and fourth mode remain fermionic.
Furthermore, the charges transported by
the transformed modes $\widetilde{\phi}^{\,}_{\gamma}$ are given by
\begin{equation}
\widetilde{Q}=
W^{\mathsf{T}}\,Q=
(+2,-2,-3,-3,-2,+2)^{\mathsf{T}}.
\end{equation}

Let us define the charge-conserving tunneling vectors
($j=1,\cdots, N-1$)
\begin{equation}
\begin{split}
\widetilde{\mathcal{T}}^{(j)}_{1}\:=&
(0,0,0,0,0,0|\cdots|0,0,+1,-1,0,0|\cdots|0,0,0,0,0,0)^{\mathsf{T}},
\\
\widetilde{\mathcal{T}}^{(j)}_{2}\:=&
(0,0,0,0,0,0|\cdots|0,0,0,0,+1,0|0,-1,0,0,0,0|\cdots|0,0,0,0,0,0)^{\mathsf{T}}, 
\\
\widetilde{\mathcal{T}}^{(j)}_{3}\:=&
(0,0,0,0,0,0|\cdots|0,0,0,0,0,+1|-1,0,0,0,0,0|\cdots|0,0,0,0,0,0)^{\mathsf{T}}.
\end{split}
\label{eq: definition of interactions that produce bosons out of fermions}
\end{equation}
In the original basis, these three families of tunneling vectors are of
order 3, 2, and 2, respectively. They are explicitly given by
\begin{equation}
\begin{split}
\mathcal{T}^{(j)}_{1}\:=&
(0,0,0,0,0,0|\cdots|-1,-1,-1,+1,+1,+1|\cdots|0,0,0,0,0,0)^{\mathsf{T}},
\\
\mathcal{T}^{(j)}_{2}\:=&
(0,0,0,0,0,0|\cdots|0,0,0,0,-1,-1|+1,+1,0,0,0,0|\cdots|0,0,0,0,0,0)^{\mathsf{T}}, 
\\
\mathcal{T}^{(j)}_{3}\:=&
(0,0,0,0,0,0|\cdots|0,0,0,+1,+1,0|0,-1,-1,0,0,0|\cdots|0,0,0,0,0,0)^{\mathsf{T}}.
\end{split}
\label{eq: definition of interactions that produce bosons out of fermions bis}
\end{equation}

The tunneling vectors 
(\ref{eq: definition of interactions that produce bosons out of fermions})
gap all modes in the bulk and the 
remaining gapless edge modes on the left edge are
\begin{equation}
\widetilde{K}^{\,}_{\mathrm{eff,left}}=
\begin{pmatrix}0&1\\1&0\end{pmatrix},
\qquad
\widetilde{Q}^{\,}_{\mathrm{eff,left}}=
\begin{pmatrix}
+2
\\
-2
\end{pmatrix}.
\label{eq: edge theory of interacting A}
\end{equation}
The only charge-conserving tunneling vector that could gap out this
effective edge theory, $\widetilde{T}=(1,1)^{\mathsf{T}}$, is not
compatible with Haldane's criterion~\eqref{eq: Haldane conditions}. Thus,
the edge theory~\eqref{eq: edge theory of interacting A}
is stable against charge conserving perturbations. The Hall
conductivity supported by this edge theory is given by
\begin{equation}
\widetilde{Q}_{\mathrm{eff,left}}^{\mathsf{T}}\,
\widetilde{K}_{\mathrm{eff,left}}^{-1}\,
\widetilde{Q}_{\mathrm{eff,left}}
=-8
\end{equation}
in units of $e^2/h$. This is the minimal Hall conductivity of a SRE
phase of bosons, if each boson is interpreted as 
a pair of electrons carrying the electronic charge $2e$ \cite{Lu12}. 
On the other hand, the edge theory
\eqref{eq: edge theory of interacting A} 
supports two modes with opposite 
chiralities, for the symmetric matrix $\widetilde{K}^{\,}_{\mathrm{eff,left}}$
has the pair of eigenvalues $\pm1$. 
Thus, the net energy transported  along the left edge,
and with it the thermal Hall conductivity, 
vanishes.

\medskip\noindent
\textbf{\textit{7.5.4.2\hskip 10 true pt Symmetry class AII}} 
\label{subsec: symmetry class AII}

\medskip\noindent
Topological insulators in symmetry class AII can be realized by
demanding that $U(1)$ charge conservation holds and that
TRS with $\Theta^{2}=-1$ holds. The wire
construction starts from wires supporting spin-$1/2$ fermions
because $\Theta^{2}=-1$, so that the
minimal choice $M=4$ counts two pairs of Kramers degenerate
left- and right-moving degrees of freedom carrying opposite spin
projections on the spin quantization axis, i.e., 
two pairs of Kramers degenerate helical modes. 
The $K$-matrix reads
\begin{subequations}
\label{eq: def H of V for class AII}
\begin{equation}
K\:=\mathrm{diag}\,(+1,-1,-1,+1).
\label{eq: def H of V for class AII a}
\end{equation}
The entries in the $K$-matrix represent, from left to right, 
a right-moving particle with spin up, 
a left-moving particle with spin down, 
a left-moving particle with spin up,  
and a right-moving particle with spin down.
The operation for reversal of time in any one of the $N$ wires
is represented by
[one verifies that Eq.\ (\ref{eq: TRS on P and I }) holds]
\begin{equation}
P^{\,}_{\Theta}\:=
\begin{pmatrix}
0&1&0&0
\\
1&0&0&0
\\
0&0&0&1\\
0&0&1&0
\end{pmatrix},
\qquad
I^{\,}_{\Theta}\:=
\begin{pmatrix}
0\\1\\0\\1
\end{pmatrix}.
\label{eq: def H of V for class AII b}
\end{equation}
We define $\widehat{H}^{\,}_{\mathcal{V}}$ 
by choosing any symmetric $4\times 4$ matrix $V$
that obeys
\begin{equation}
V=
P^{\,}_{\Theta}\,
V\,
P^{-1}_{\Theta}.
\label{eq: def H of T for class AII c}
\end{equation}
\end{subequations}
We define $\widehat{H}^{\,}_{\{\mathcal{T}^{\,}_{\mathrm{SO}}\}}$ 
by choosing $2(N-1)$ scattering vectors
as follows. For any $j=1,\cdots,(N-1)$,
introduce the pair of scattering vectors
\begin{subequations}
\begin{equation}
\mathcal{T}^{(j)}_{\mathrm{SO}}\:=
(0,0,0,0|\cdots|0,0,+1,0|-1,0,0,0|\cdots|0,0,0,0)^{\mathsf{T}}
\label{eq: def H of T for class AII d}
\end{equation}
and
\begin{equation}
\overline{\mathcal{T}}^{(j)}_{\mathrm{SO}}\:=
-\mathcal{P}^{\,}_{\Theta}\,
\mathcal{T}^{(j)}_{\mathrm{SO}}.
\label{eq: def H of T for class AII e}
\end{equation}
The scattering vector~(\ref{eq: def H of T for class AII d})
labels a one-body interaction in the fermion representation
that fulfills Eq.\ (\ref{eq: def mathcal T b}).
It scatters a left mover with spin up from wire $j$ 
into a right mover with spin up in wire $j+1$.
For any $j=1,\cdots,(N-1)$,
introduce the pair of amplitudes
\begin{equation}
h^{\,}_{\mathcal{T}^{(j)}_{\mathrm{SO}}}(x)=
h^{\,}_{\overline{\mathcal{T}}^{(j)}_{\mathrm{SO}}}(x)\geq0
\label{eq: def H of T for class AII f}
\end{equation}
and the pair of phases
\begin{equation}
\alpha^{\,}_{\mathcal{T}^{(j)}_{\mathrm{SO}}}(x)=
\alpha^{\,}_{\overline{\mathcal{T}}^{(j)}_{\mathrm{SO}}}(x)\in\mathbb{R}
\label{eq: def H of T for class AII g}
\end{equation}
\end{subequations}
according to  
Eqs.\ (\ref{eq: TRS H d}) and~(\ref{eq: TRS H e}), respectively.
The choices for the amplitude~(\ref{eq: def H of T for class AII f})
and the phase~(\ref{eq: def H of T for class AII g})
are arbitrary. 
The subscript SO refers to the intrinsic spin-orbit coupling.  
The rational for using it shall be shortly explained.  

One verifies that all $2(N-1)$
scattering vectors~(\ref{eq: def H of T for class AII c})
and~(\ref{eq: def H of T for class AII d})
satisfy the Haldane criterion
(\ref{eq: Haldane conditions}), i.e.,
\begin{equation}
\mathcal{T}^{(i)\mathsf{T}}_{\mathrm{SO}}\,
\mathcal{K}\,
\mathcal{T}^{(j)}_{\mathrm{SO}}=
\overline{\mathcal{T}}^{(i)\mathsf{T}}_{\mathrm{SO}}\,
\mathcal{K}\,
\overline{\mathcal{T}}^{(j)}_{\mathrm{SO}}=
\mathcal{T}^{(i)\mathsf{T}}_{\mathrm{SO}}\,
\mathcal{K}\,
\overline{\mathcal{T}}^{(j)}_{\mathrm{SO}}=0,
\end{equation}
for $i,j=1,\cdots,N-1$.
Correspondingly, the term $\widehat{H}^{\,}_{\{\mathcal{T}^{\,}_{\mathrm{SO}}\}}$
gaps out $4(N-1)$ of the $4N$ gapless modes of
$\widehat{H}^{\,}_{\mathcal{V}}$. 
Two pairs of Kramers degenerate helical edge states 
that propagate along the first and last wire,
respectively, remain in the low energy sector of the theory.
These edge states are localized on wire $i=1$ and $i=N$, respectively,
for their overlaps with the gapped states from the bulk decay
exponentially fast as a function of the distance away from
the first and end wires. 
The energy splitting between the edge state localized on wire $i=1$
and wire $i=N$ brought about by the bulk states vanishes
exponentially fast with increasing $N$. Two pairs of
gapless Kramers degenerate helical edge states
emerge in the two-dimensional limit $N\to\infty$.

At energies much lower than the bulk gap,
the effective $\mathcal{K}$-matrix for 
the two pairs of helical edge modes is
\begin{equation}
\begin{split}
\mathcal{K}^{\,}_{\mathrm{eff}}\:=&\,
\mathrm{diag}(+1,-1,0,0|0,0,0,0|
\cdots|0,0,0,0|0,0,-1,+1).
\end{split}
\end{equation} 
Here, $\mathcal{K}^{\,}_{\mathrm{eff}}$ follows from
replacing the entries in the 
$4N\times4N$ $\mathcal{K}$ matrix
for all gapped modes by 0.
It will be shown that
the effective scattering vector 
\begin{equation}
\mathcal{T}^{\,}_{\mathrm{eff}}\:=
(+1,-1,0,0|0,0,0,0|\cdots)^{\mathsf{T}},
\label{eq: def T eff AII}
\end{equation}
with the potential to gap out the pair of 
Kramers degenerate helical edge modes on wire $i=1$
since it fulfills the Haldane criterion
(\ref{eq: Haldane conditions}),
is not allowed by TRS.
\footnote{%
Even integer multiples of
$\mathcal{T}^{\,}_{\mathrm{eff}}$ would gap the edge states, but they must
also be discarded as explained in \protect\cite{Neupert11}.
}
On the one hand,
$\mathcal{T}^{\,}_{\mathrm{eff}}$
maps to itself under reversal of time,
\begin{equation}
\mathcal{T}^{\,}_{\mathrm{eff}}=
-\mathcal{P}^{\,}_{\Theta}\,
\mathcal{T}^{\,}_{\mathrm{eff}}.
\end{equation}
On the other hand,
\begin{equation}
\mathcal{T}^{\mathsf{T}}_{\mathrm{eff}}\,
\mathcal{P}^{\,}_{\Theta}\,
\mathcal{I}^{\,}_{\Theta}=
-1.
\end{equation}
Therefore, the condition~(\ref{eq: TRS H e})
for $\mathcal{T}^{\,}_{\mathrm{eff}}$ to be a TRS perturbation is not met,
for the phase $\alpha^{\,}_{\mathcal{T}^{\,}_{\mathrm{eff}}}(x)$ associated to
$\mathcal{T}^{\,}_{\mathrm{eff}}$ then obeys
\begin{equation}
\alpha^{\,}_{\mathcal{T}^{\,}_{\mathrm{eff}}}(x)=
\alpha^{\,}_{\mathcal{T}^{\,}_{\mathrm{eff}}}(x)
-
\pi,
\label{eq: TRS condition on alpha Teff class AII}
\end{equation}
a condition that cannot be satisfied.

Had a TRS with $\Theta=+1$ been imposed instead of
$\Theta=-1$ as is suited for the symmetry class AI
that describes spinless fermions with TRS, one would
only need to replace $I^{\,}_{\Theta}$ in Eq.
(\ref{eq: def H of V for class AII b})
by the null vector.
If so, the scattering vector~(\ref{eq: def T eff AII})
is compatible with TRS since 
the condition~(\ref{eq: TRS H e})
for TRS then becomes
\begin{equation}
\alpha^{\,}_{\mathcal{T}^{\,}_{\mathrm{eff}}}(x)=
\alpha^{\,}_{\mathcal{T}^{\,}_{\mathrm{eff}}}(x)
\end{equation}
instead of 
Eq.\ (\ref{eq: TRS condition on alpha Teff class AII}).
This is the reason why symmetry class AI is 
always topologically trivial in two-dimensional space
from the point of view of the wire construction.

Note also that had one not insisted on the condition of charge neutrality 
(\ref{eq: def mathcal T b}),
the tunneling vector
 \begin{equation}
\mathcal{T}^{\prime}_{\mathrm{eff}}\:=
(+1,+1,0,0|0,0,0,0|\cdots)^{\mathsf{T}},
\end{equation}
that satisfies the Haldane criterion and is compatible with TRS could gap 
out the Kramers degenerate pair of helical edge states.

To address the question of what happens if $M=4$ is changed to
$M=4n$ with $n$ any strictly positive integer
in each wire from the array, consider, without loss of generality, 
the case of $n=2$. To this end,
it suffices to repeat all the steps that
lead to Eq.\ (\ref{eq: def T eff AII}), except for the change
\begin{equation}
\begin{split}
\mathcal{K}^{\,}_{\mathrm{eff}}\:=
\mathrm{diag}\, &
(+1,-1,0,0;+1,-1,0,0|0,0,0,0;0,0,0,0|
\\
&\qquad
\cdots
|0,0,0,0;0,0,0,0|0,0,-1,+1;0,0,-1,+1). 
\end{split}
\end{equation}
One verifies that the scattering vectors
\begin{equation}
\mathcal{T}^{\prime}_{\mathrm{eff}}\:=
(+1,0,0,0;0,-1,0,0|0,0,0,0;0,0,0,0|\cdots)^{\mathsf{T}}
\label{eq: def mathcal T'eff AII}
\end{equation}
and
\begin{equation}
\mathcal{T}^{\prime\prime}_{\mathrm{eff}}\:=
(0,-1,0,0;+1,0,0,0|0,0,0,0;0,0,0,0|\cdots)^{\mathsf{T}}
\label{eq: def mathcal T''eff AII}
\end{equation}
are compatible with the condition that TRS holds in that the pair
is a closed set under reversal of time,
\begin{equation}
\mathcal{T}^{\prime}_{\mathrm{eff}}=
-
\mathcal{P}^{\,}_{\Theta}\,
\mathcal{T}^{\prime\prime}_{\mathrm{eff}}.
\end{equation} 
One verifies that these scattering vectors
fulfill the Haldane criterion~(\ref{eq: Haldane conditions}).
Consequently, inclusion in $\widehat{H}^{\,}_{\{\mathcal{T}^{\,}_{\mathrm{SO}}\}}$ 
of the two cosine potentials with 
$\mathcal{T}^{\prime}_{\mathrm{eff}}$ and $\mathcal{T}^{\prime\prime}_{\mathrm{eff}}$
entering in their arguments, respectively, gaps out the pair of 
Kramers degenerate helical modes on wire $i=1$. The same treatment
of the wire $i=N$ leads to the conclusion that TRS does not
protect the gapless pairs of Kramers degenerate
edge states from perturbations when $n=2$.
The generalization to 
$M=4n$ channels
is that it is only when $n$ is odd that a pair of Kramers degenerate
helical edge modes is robust to the most generic
$\widehat{H}^{\,}_{\{\mathcal{T}^{\,}_{\mathrm{SO}}\}}$
of the form depicted in
Fig.\ \ref{Figures for Table wire construction}(c).
Since it is the parity of $n$
in the number $M=4n$ of channels per wire that matters
for the stability of the Kramers degenerate helical edge states,
the group of two integers $\mathbb{Z}^{\,}_{2}$ under
addition modulo 2 appears in the third column of the third row of 
Table~\ref{table: main table}.

If conservation of the projection of the
spin-$1/2$ quantum number on the quantization axis was imposed,
then processes by which a spin is flipped 
must be precluded from all scattering vectors.
In particular, the scattering vectors
(\ref{eq: def mathcal T'eff AII})
and
(\ref{eq: def mathcal T''eff AII})
are not admissible anymore. By imposing the $U(1)$ residual symmetry
of the full $SU(2)$ symmetry group for a spin-$1/2$ degree of freedom,
the group of integers $\mathbb{Z}$ under the addition
that encodes the topological stability
in the quantum spin Hall effect (QSHE)
is recovered.

This discussion of the symmetry class AII is closed by justifying the
interpretation of the index SO as an abbreviation for the intrinsic
spin-orbit coupling. To this end, introduce a set of 
$(N-1)$ pairs of scattering vectors
\begin{subequations}
\label{eq: def mathcal T R AII}
\begin{equation}
\mathcal{T}^{(j)}_{\mathrm{R}}\:=
(0,0,0,0|\cdots|0,+1,0,0|-1,0,0,0|\cdots|0,0,0,0)^{\mathsf{T}}
\label{eq: def mathcal T R AII a}
\end{equation}
and
\begin{equation}
\overline{\mathcal{T}}^{(j)}_{\mathrm{R}}\:=
-
\mathcal{P}^{\,}_{\Theta}\,
\mathcal{T}^{(j)}_{\mathrm{R}}
\label{eq: def mathcal T R AII b}
\end{equation}
for $j=1,\cdots,N-1$.
The scattering vector~(\ref{eq: def mathcal T R AII a})
labels a one-body interaction in the fermion representation
that fulfills Eq.\ (\ref{eq: def mathcal T b}).
The index R is an acronym for Rashba as it describes a 
backward scattering process 
by which a left mover with spin down from wire $j$
is scattered into a right mover with spin up on wire $j+1$
and conversely.
For any $j=1,\cdots,(N-1)$,
introduce the pair of amplitudes
\begin{equation}
h^{\,}_{\mathcal{T}^{(j)}_{\mathrm{R}}}(x)=
h^{\,}_{\overline{\mathcal{T}}^{(j)}_{\mathrm{R}}}(x)\geq0
\label{eq: def mathcal T R AII c}
\end{equation}
and the pair of phases
\begin{equation}
\alpha^{\,}_{\mathcal{T}^{(j)}_{\mathrm{R}}}(x)=
\alpha^{\,}_{\overline{\mathcal{T}}^{(j)}_{\mathrm{R}}}(x)
+
\pi
\in\mathbb{R}
\label{eq: def mathcal T R AII d}
\end{equation}
\end{subequations}
according to  
Eqs.\ (\ref{eq: TRS H d}) and~(\ref{eq: TRS H e}), respectively.
In contrast to the intrinsic spin-orbit scattering vectors,
the Rashba scattering vectors~(\ref{eq: def mathcal T R AII a})
fail to meet the Haldane criterion~(\ref{eq: Haldane conditions})
as
\begin{equation}
\mathcal{T}^{(j)\mathsf{T}}_{\mathrm{R}}\,
\mathcal{K}\,
\overline{\mathcal{T}}^{(j+1)}_{\mathrm{R}}=
-1,
\qquad
j=1,\cdots, N-1.
\end{equation}
Hence, the Rashba scattering processes fail to open a gap in the bulk,
as is expected of a Rashba coupling in a two-dimensional electron gas.
On the other hand, the intrinsic spin-orbit coupling can
lead to a 
phase with a gap in the bulk that supports the spin quantum Hall effect
in a two-dimensional electron gas.

\medskip\noindent
\textbf{\textit{7.5.4.3\hskip 10 true pt Symmetry class D}}
\label{subsec: Symmetry class D}

\medskip\noindent
The simplest example among the topological superconductors 
can be found in the symmetry class D that is defined by the presence
of a PHS with $\Pi^{2}=+1$ and the absence of TRS.

With the understanding of PHS as discussed in Section
\ref{sec: PHS in superconductors}, 
a representative phase in class D is constructed from identical wires supporting
right- and left-moving spinless fermions, each of which carry a particle or 
a hole label, i.e., $M=4$. The $K$-matrix reads
\begin{subequations}
\label{eq: def H of V for class D}
\begin{equation}
K\:=
\mathrm{diag}(+1,-1,-1,+1).
\label{eq: def H of V for class D a}
\end{equation}
The entries in the $K$-matrix represent, from left to right, 
a right-moving particle, 
a left-moving particle, 
a left-moving hole, 
and a right-moving hole.
The operation for the exchange of particles and holes 
in any one of the $N$ wires is represented by 
[one verifies that Eq.\ (\ref{eq: PHS on P and I }) holds]
\begin{equation}
P^{\,}_{\Pi}\:=
\begin{pmatrix}
0&0&0&1
\\
0&0&1&0
\\
0&1&0&0
\\
1&0&0&0
\end{pmatrix},
\qquad
I^{\,}_{\Pi}\:=
\begin{pmatrix}
0
\\
0
\\
0
\\
0
\end{pmatrix}.
\label{eq: def H of V for class D b}
\end{equation}
We define $\widehat{H}^{\,}_{\mathcal{V}}$ 
by choosing any symmetric $4\times4$ matrix $V$ that obeys
\begin{equation}
V=
+
P^{\,}_{\Pi}\,
V\,
P^{-1}_{\Pi}.
\label{eq: def H of V for class D c}
\end{equation}
\end{subequations}
We define $\widehat{H}^{\,}_{\{\mathcal{T}\}}$ 
by choosing $2N-1$ scattering vectors
as follows. For any wire $j=1,\cdots,N$,
introduce the scattering vector
\begin{subequations}
\label{eq: def H of T for class D}
\begin{equation}
\mathcal{T}^{(j)}\:=
(0,0,0,0|\cdots|+1,-1,-1,+1|\cdots|0,0,0,0)^{\mathsf{T}}.
\label{eq: def H of T for class D d}
\end{equation}
Between any pair of neighboring wires 
introduce the scattering vector
\begin{equation}
\begin{split}
&
\overline{\mathcal{T}}^{(j)}\:=
\\
&
(0,0,0,0|\cdots|0,+1,-1,0|-1,0,0,+1|\cdots|0,0,0,0)^{\mathsf{T}},
\end{split}
\label{eq: def H of T for class D e}
\end{equation}
for $j=1,\cdots,(N-1)$.
Observe that both $\mathcal{T}^{(j)}$ and $\overline{\mathcal{T}}^{(j)}$
are eigenvectors of the particle-hole transformation in that
\begin{equation}
\mathcal{P}^{\,}_{\Pi}\,
\mathcal{T}^{(j)}=
+\mathcal{T}^{(j)},
\qquad
\mathcal{P}^{\,}_{\Pi}\,
 \overline{\mathcal{T}}^{(j)}=
-\overline{\mathcal{T}}^{(j)}.
\end{equation}
Thus, to comply with PHS, demand that the phases
\begin{equation}
\alpha^{\,}_{\overline{\mathcal{T}}^{(j)}}(x)
=0,
\label{eq: def H of T for class D g}
\end{equation}
\end{subequations}
while
$\alpha^{\,}_{\mathcal{T}^{(j)}}(x)$ are unrestricted.
Similarly, the amplitudes $h^{\,}_{\mathcal{T}^{(j)}}(x)$
and $h^{\,}_{\overline{\mathcal{T}}^{(j)}}(x)$ can take arbitrary real values.

One verifies that the set of scattering vectors defined by
Eqs.\ (\ref{eq: def H of T for class D d}) and~
(\ref{eq: def H of T for class D e}) 
satisfies the Haldane criterion.  
Correspondingly, the term
$\widehat{H}^{\,}_{\{\mathcal{T}\}}$ gaps out $(4N-2)$ of the $4N$ gapless
modes of $\widehat{H}^{\,}_{\mathcal{V}}$. Furthermore, one identifies with
\begin{equation}
\overline{\mathcal{T}}^{(0)}
=(-1,0,0,+1|0,0,0,0|\cdots|0,0,0,0|0,+1,-1,0)^{\mathsf{T}}
\end{equation} 
a unique (up to an integer multiplicative factor)
scattering vector that satisfies the Haldane criterion with all
existing scattering vectors Eqs.\ (\ref{eq: def H of T for class D d})
and~(\ref{eq: def H of T for class D e}) and could thus potentially
gap out the remaining pair of modes. However,
the tunneling $\overline{\mathcal{T}}^{(0)}$ is non-local 
for it connects the two edges of the system
when open boundary conditions are chosen.
One thus concludes that the two remaining modes are exponentially
localized near wire $i=1$ and wire $i=N$, respectively, and propagate
with opposite chirality.

To give a physical interpretation of the resulting topological (edge)
theory in this wire construction, one has to keep in mind that the
degrees of freedom were artificially doubled.  One finds, in this
doubled theory, a single chiral boson (with chiral central charge
$c=1$). To interpret it as the edge of a chiral 
$(p^{\,}_{x}+\mathrm{i}p^{\,}_{y})$
superconductor, the reality condition is imposed 
to obtain a single chiral Majorana mode with chiral central charge $c=1/2$.

The pictorial representation of the topological phase 
in the symmetry class D through the wire construction 
is shown in Fig.\ \ref{Figures for Table wire construction}(g).
The generalization to an arbitrary number $n$ of 
gapless chiral edge modes
is analogous to the case discussed in symmetry class A. 
The number of robust gapless chiral edge states of a given chirality is thus 
integer. 
This is the reason why the group of integers
$\mathbb{Z}$ is found in the third column of the fifth row
of Table~\ref{table: main table}.

\medskip\noindent
\textbf{\textit{7.5.4.4\hskip 10 true pt Symmetry classes DIII and C}}

\medskip\noindent
The remaining two topological nontrivial superconducting classes DIII
(TRS with $\Theta^{2}=-1$ and PHS with $\Pi^{2}=+1$) 
and C (PHS with $\Pi^{2}=-1$)
involve spin-$1/2$ fermions. Each wire thus features no less than
$M=8$ internal degrees of freedom corresponding to the spin-$1/2$,
chirality, and particle/hole indices. The construction is very similar
to the cases already presented. Details are relegated to 
\cite{Neupert14}.

The scattering vectors that are needed to gap out the bulk for each class of 
class DIII and C are represented pictorially 
in Fig.\ \ref{Figures for Table wire construction}(e,i).

\medskip\noindent
\textbf{\textit{7.5.4.5\hskip 10 true pt Summary}}

\medskip\noindent
An explicit construction was provided by way of an array of wires
supporting fermions that realizes all five insulating and
superconducting topological phases of matter with a nondegenerate
ground state in two-dimensional space according to the tenfold
classification of band insulators and superconductors. The topological
protection of edge modes in the bosonic formulation follows from
imposing the Haldane criterion~(\ref{eq: Haldane conditions}) along
with the appropriate symmetry constraints. 
It remains to extend the wire construction
to allow many-body tunneling processes
that delivers fractionalized phases with degenerate ground states.

\subsection{Fractionalized phases}
\label{sec: Fractionalized phases}

The power of the wire construction goes much beyond what was used 
in Section~\ref{sec: Reproducing the tenfold way} 
to reproduce the classification of the SRE phases.
It is possible to construct models for interacting
phases of matter with intrinsic topological order and fractionalized
excitations by relaxing the condition on the tunnelings 
between wires that they be of the one-body type.
While these phases are more complex, the principles for
constructing the models and proving the stability of edge modes
remain the same: All allowed tunneling vectors have to obey the Haldane
criterion~(\ref{eq: Haldane conditions}) and the respective
symmetries. 

\medskip\noindent
\textbf{\textit{7.5.5.1\hskip 10 true pt 
Symmetry class A: Fractional quantum Hall states}}
\label{subsec: Symmetry class A: Fractional quantum Hall states}

\medskip\noindent
First, the models of quantum wires
that are topologically equivalent to the
Laughlin state in the FQHE are reviewed \cite{Laughlin83},
following the construction in \cite{Kane02} 
for Abelian fractional quantum Hall states.
Here, the choice of scattering vectors is determined by
the Haldane criterion (\ref{eq: Haldane conditions})
and at the same time prepare the grounds for the
construction of fractional topological insulators with TRS in
Section 7.5.5.2.

Needed are the fermionic Laughlin states indexed by 
the positive odd integer $m$ \cite{Laughlin83}.
(By the same method, other fractional quantum Hall phases
from the Abelian hierarchy could be constructed
\cite{Kane02}.)
The elementary degrees of freedom in each wire are spinless 
right- and left-moving fermions with the $K$-matrix
\begin{subequations}
\label{eq: def array quantum wires TSE A} 
\begin{equation}
K=
\mathrm{diag}\,(+1,-1),
\label{eq: def array quantum wires TSE A a} 
\end{equation}
as is done in 
Eq.\ (\ref{eq: def H for class A a}). 
Reversal of time is defined through
$P^{\,}_{\Theta}$ and $I^{\,}_{\Theta}$ given in
Eq~(\ref{eq: def H for class A b}). Instead of
Eq~(\ref{eq: def H of T for class A}), 
the scattering vectors that describe the interactions
between the wires are now defined by
\begin{equation}
\mathcal{T}^{(j)}\:=
\left(
0, 0
\left|
\cdots
\left|
m^{\,}_{+},
-m^{\,}_{-}
\left|
m^{\,}_{-},
-m^{\,}_{+}
\right.
\right|
\cdots
\right|
0,0
\right)^{\mathsf{T}},
\label{eq: def array quantum wires TSE A b} 
\end{equation}
\end{subequations}
for any $j=1,\cdots,N-1$,
where $m^{\,}_{\pm}=(m\pm 1)/2$
[see Table~\ref{table: main table} for an
illustration of the scattering process]. 

For any $j=1,\cdots,N-1$,
the scattering (tunneling) vectors
(\ref{eq: def array quantum wires TSE A b})
preserve the conservation of the total fermion number 
in that they obey Eq.~\eqref{eq: def mathcal T b},
and they encode a tunneling interaction of order $q=m$, 
with $q$ defined in Eq.~\eqref{eq:q-def}.
As a set, all tunneling interactions satisfy
the Haldane criterion~(\ref{eq: Haldane conditions}), since
\begin{equation}
\begin{split}
\mathcal{T}^{(i)\mathsf{T}}\,
\mathcal{K}\,
\mathcal{T}^{(j)}
=&\,
0,
\quad
i,j=1,\cdots,N-1.
\end{split}
\label{eq: def Tj for A}
\end{equation} 
Note that the choice of tunneling vector in 
Eq.~\eqref{eq: def array quantum wires TSE A b} 
is unique (up to an integer multiplicative factor)
if one insists on charge conservation,
compliance with the Haldane
criterion~(\ref{eq: Haldane conditions}), and only includes scattering
between neighboring wires. 

The bare counting of tunneling vectors shows that the wire model gaps
out all but two modes. However, one still needs to show that
the remaining two modes (i) live on the edge, (ii) cannot be gapped
out by other (local) scattering vectors and (iii) are made out of
fractionalized quasiparticles.

To address (i) and (ii), note that the remaining two modes can be
gapped out by a unique (up to an integer multiplicative factor)
charge-conserving scattering vector that
satisfies the Haldane criterion~(\ref{eq: Haldane conditions}) 
with all existing scatterings, namely
\begin{equation}
\mathcal{T}^{(0)}\:=
\left(\left.\left.
m^{\,}_{-},
-m^{\,}_{+}
\right|
0,
0
\right|
\cdots
\left|
0,
0
\left|
m^{\,}_{+},
-m^{\,}_{-}
\right.\right.
\right)^{\mathsf{T}}.
\label{eq: def T0 for A}
\end{equation}
Connecting the opposite ends of the array of wires 
through the tunneling
$\mathcal{T}^{(0)}$ is not an admissible perturbation,
for it violates locality in the 
two-dimensional thermodynamic limit $N\to\infty$.
Had periodic boundary conditions 
corresponding to a cylinder geometry (i.e., a tube as in
Fig.~\ref{fig: BoundaryConditions}) by which 
the first and last wire are nearest neighbors been chosen,
$\mathcal{T}^{(0)}$ would be admissible. 
Hence, the gapless nature of the remaining modes 
when open boundary conditions are chosen depends
on the boundary conditions. These gapless modes have support 
near the boundary only and are topologically protected.

Applying the transformation~\eqref{eq: general linear trsf}
with
\begin{subequations}
\begin{equation}
W\:=
\begin{pmatrix}
-m^{\,}_{-}&m^{\,}_{+}
\\
m^{\,}_{+}&-m^{\,}_{-}
\end{pmatrix},
\qquad
\mathrm{det}\,W=-m,
\qquad
W^{-1}=
\frac{1}{m}
\begin{pmatrix}
m^{\,}_{-}&m^{\,}_{+}
\\
m^{\,}_{+}&m^{\,}_{-}
\end{pmatrix},
\label{eq: trasf for fields Phi A b R}
\end{equation}
transforms the matrix $K$ into
\begin{equation}
\begin{split}
\widetilde{K}
=&\,
W^{\mathsf{T}}\,K\,W
=
\begin{pmatrix}
-m^{\,}_{-}& m^{\,}_{+}
\\
 m^{\,}_{+}&-m^{\,}_{-}
\end{pmatrix}
\begin{pmatrix}
+1&0
\\
0&-1
\end{pmatrix}
\begin{pmatrix}
-m^{\,}_{-}& m^{\,}_{+}
\\
 m^{\,}_{+}&-m^{\,}_{-}
\end{pmatrix}
=
\begin{pmatrix}
-m&0
\\
0&+m
\end{pmatrix}.
\end{split}
\end{equation}
As its determinant is not unity, 
the linear transformation~(\ref{eq: trasf for fields Phi A b R})
changes the compactification radius of the new field $\widetilde{\Phi}(x)$ 
relative to the compactification radius of the old field $\widehat{\Phi}(x)$ 
accordingly. Finally, the transformed tunneling and charge vectors 
are given by
\begin{align}
&
\widetilde{\mathcal{T}}^{(j)}=
\mathcal{W}^{-1}\,\mathcal{T}^{(j)}=
(0,0|\cdots|0,0|0,+1|-1,0|0,0|\cdots|0,0)^{\mathsf{T}}\neq
\mathcal{T}^{(j)},
\label{eq: T tilde FQHE}
\\
&
\widetilde{\mathcal{Q}}=
\mathcal{W}^{\mathsf{T}}\,\mathcal{Q}=
(1,1|\cdots|1,1|1,1|1,1|1,1|\cdots|1,1)^{\mathsf{T}}=
\mathcal{Q},
\label{eq: Q tilde FQHE}
\end{align}
\end{subequations}
respectively,
where $\mathcal{W}\:=\myopenone^{\,}_{N}\otimes W$ and
$j=1,\cdots,N-1$. Contrary to the tunneling vectors,
the charge vector is invariant under
the non-orthogonal linear transformation
(\ref{eq: trasf for fields Phi A b R}).

In view of Eq.~\eqref{eq: T tilde FQHE}, 
the remaining effective edge theory is
described by
\begin{equation}
\widetilde{\mathcal{K}}_{\mathrm{eff}}=
\mathrm{diag}\,
(-m,0|0,0|\cdots|0,0|0,+m).
\label{eq: effective K FQHE}
\end{equation} 
This is a chiral theory at each edge that cannot be gapped by local
perturbations. In combination with Eq.\ (\ref{eq: Q tilde FQHE}),
Eq.~\eqref{eq: effective K FQHE} is precisely the
edge theory for anyons with statistical angle $1/m$ and charge $e/m$
\cite{Wen91b},
where $e$ is the charge of the original fermions.

\medskip\noindent
\textbf{\textit{7.5.5.2\hskip 10 true pt 
Symmetry Class AII: Fractional topological insulators}}
\label{subsec: Symmetry Class AII: Fractional topological insulators}

\medskip\noindent
Having understood how fractionalized quasiparticles emerge out of a
wire construction, it is imperative to ask what other phases can be
obtained when symmetries are imposed on the topologically ordered
phase. Such symmetry enriched topological phases have been classified
by methods of group cohomology \cite{Chen13}.
Here, the case of TRS with $\Theta^{2}=-1$ will provide an example for
how the wire construction can be used to build up an intuition 
for these phases and to study the stability of their edge theory.

The elementary degrees of freedom in each wire are spin-$1/2$ 
right- and left-moving fermions with the $K$-matrix
\begin{subequations} 
\label{eq: def array quantum wires TSE AII} 
\begin{equation}  
K\:=
\mathrm{diag}\,(+1,-1,-1,+1),
\label{eq: def array quantum wires TSE AII a}  
\end{equation}
as is done in Eq.\ (\ref{eq: def H of V for class AII a}).
Reversal of time is defined through
$P^{\,}_{\Theta}$ and $I^{\,}_{\Theta}$ given in
Eq~(\ref{eq: def H of V for class AII b}). Instead of
Eq~(\ref{eq: def H of T for class AII d}), 
the scattering vectors that describe the interactions between
the wires are now defined by
\begin{equation}
\mathcal{T}^{(j)}\:=
\left(
0,0,0,0
\left|
\cdots
\left|
-m^{\,}_{-},
0,
+m^{\,}_{+},
0
\left|
-m^{\,}_{+},
\right.
0,
+m^{\,}_{-},
0
\right|
\cdots
\right|
0,0,0,0
\right)^{\mathsf{T}}
\label{eq: def array quantum wires TSE AII b} 
\end{equation}
and
\begin{equation}
\overline{\mathcal{T}}^{(j)}\:=
-
\mathcal{P}^{\,}_{\Theta}\,\mathcal{T}^{(j)},
\label{eq: def array quantum wires TSE AII c} 
\end{equation}
\end{subequations}
for any $j=1,\cdots,N-1$, $m$ a positive odd integer, and $m^{\,}_{\pm}=(m\pm1)/2$.

For any $j=1,\cdots,N-1$,
the scattering (tunneling) vectors
(\ref{eq: def array quantum wires TSE AII b} )
preserve conservation of the total fermion number in that
they obey Eq.~\eqref{eq: def mathcal T b},
and they encode a tunneling interaction of order $q=m$ 
with $q$ defined in Eq.~\eqref{eq:q-def}.
They also satisfy the Haldane criterion~(\ref{eq: Haldane conditions})
as a set  
(see Fig.\ \ref{Figures for Table wire construction} 
for an illustration of the scattering process).

Applying the transformation~\eqref{eq: general linear trsf}
with 
\begin{equation}
W\:=
\begin{pmatrix}
-m^{\,}_{-}&0&m^{\,}_{+}&0\\
0&-m^{\,}_{-}&0&m^{\,}_{+}\\
m^{\,}_{+}&0&-m^{\,}_{-}&0\\
0&m^{\,}_{+}&0&-m^{\,}_{-}
\end{pmatrix}
\label{eq: linear trafo for FTI}
\end{equation}
to the bosonic fields,
leaves the representation of time-reversal invariant 
\begin{equation}
W^{-1}\,P^{\,}_{\Theta}\,W=P^{\,}_{\Theta},
\label{eq: W commutes with P SEP class AII}
\end{equation}
while casting the theory in a new form with the transformed
matrix $\widetilde{K}$ given by
\begin{equation}
\widetilde{K}=
\mathrm{diag}\,(-m,+m,+m,-m),
\end{equation}
and, for any $j=1,\cdots,N-1$,
with the transformed pair of scattering vectors 
$(\widetilde{\mathcal{T}}^{j},\widetilde{\overline{\mathcal{T}}}^{j})$
given by
\begin{equation}
\widetilde{\mathcal{T}}^{(j)}=
(
0,0,0,0
|
\cdots
|
+1,0,0,0
|
0,0,-1,0
|
\cdots
|
0,0,0,0
)^{\mathsf{T}}
\end{equation}
and
\begin{equation}
\widetilde{\overline{\mathcal{T}}}^{(j)}=
(
0,0,0,0
|
\cdots
|
0,-1,0,0
|
0,0,0,+1
|
\cdots
|
0,0,0,0
)^{\mathsf{T}}.
\end{equation}
When these scattering vectors have gapped out all modes in the bulk,
the effective edge theory is described by
\begin{equation}
\widetilde{\mathcal{K}}_{\mathrm{eff}}=
\mathrm{diag}\,
(0,0,+m,-m|
0,0,0,0|
\cdots
|0,0,0,0|
-m,+m,0,0).
\label{eq: effective K}
\end{equation}
This effective $K$-matrix describes 
a single Kramers degenerate pair of $1/m$ anyons propagating along
the first wire and another 
single Kramers degenerate pair of $1/m$ anyons propagating along
the last wire. Their robustness to local perturbations
is guaranteed by TRS.

In contrast to the tenfold way, the correspondence between the bulk
topological phase and the edge theories of LRE phases is not
one-to-one. For example, while a bulk topological LRE phase supports
fractionalized topological excitations in the bulk, its edge modes
may be gapped out by symmetry-allowed perturbations. For the phases
discussed in this section, namely the Abelian and TRS fractional
topological insulators, it was shown in \cite{Neupert11}
and \cite{Levin09} that the edge, consisting of Kramers
degenerate pairs of edge modes, supports at most one stable Kramers
degenerate pair of delocalized quasiparticles that are stable
against disorder.  (Note that this does not preclude the richer edge
physics of non-Abelian TRS fractional topological
insulators \cite{Scharfenberger11}.)

It turns out that the wire constructions with edge modes
given by Eq.~\eqref{eq: effective K} exhaust all stable edge
theories of Abelian topological phases which are protected by TRS
with $\Theta^{2}=-1$ alone.

We suppose that the single protected Kramers degenerate pair 
{is} characterized 
by the linear combination of bosonic fields
\begin{equation}
\hat{\varphi}(x)\:=
\mathcal{T}^{\mathsf{T}}\,
\mathcal{K}'\,
\widehat{\Phi}(x)
\end{equation}
and its time-reversed partner
\begin{equation}
\hat{\bar{\varphi}}(x)\:=
\overline{\mathcal{T}}^{\mathsf{T}}\,
\mathcal{K}'\,
\widehat{\Phi}(x),
\end{equation}
where the tunneling vector $\mathcal{T}$ was constructed from the
microscopic information from the theory in \cite{Neupert11}
and $\mathcal{K}'$ is the $K$-matrix of a TRS bulk Chern-Simons
theory from the theory in \cite{Neupert11}. 
[In other words, the theory encoded by $\mathcal{K}'$
has nothing to do a priory with the array of quantum wires defined
by Eq.\ (\ref{eq: def array quantum wires TSE AII}).]  
The Kramers degenerate pair of modes 
$(\hat{\varphi},\hat{\bar{\varphi}})$ 
is stable against TRS perturbations supported on a single edge 
if and only if
\begin{equation}
\frac12|\mathcal{T}^{\mathsf{T}}\,\mathcal{Q}|
\hbox{ is an odd number.}
\end{equation}
Here, $\mathcal{Q}$ is the charge vector with integer entries that
determines the coupling of the different modes to the electromagnetic
field. Provided $(\hat{\varphi},\hat{\bar{\varphi}})$ 
is stable, its equal-time
commutation relations follow from 
Eq.\ (\ref{eq: def hat phi's b})
as  
\begin{subequations}
\begin{align}
\left[\hat{\varphi}(x),\hat{\varphi}(x')\right]
=&\,
-\mathrm{i}\pi\,
\left(
\mathcal{T}^{\mathsf{T}}\,
\mathcal{K}'\,
\mathcal{T}\,
\mathrm{sgn}(x-x')
+
\mathcal{T}^{\mathsf{T}}\,
\mathcal{L}\,
\mathcal{T}
\right),
\\
\left[\hat{\bar{\varphi}}(x),\hat{\bar{\varphi}}(x')\right]
=&\,
-\mathrm{i}\pi\,
\left(
-\mathcal{T}^{\mathsf{T}}\,
\mathcal{K}'\,
\mathcal{T}\,
\mathrm{sgn}(x-x')
+
\overline{\mathcal{T}}^{\mathsf{T}}\,
\mathcal{L}\,
\overline{\mathcal{T}}
\right),
\end{align}
\end{subequations} 
where the fact that $\mathcal{K}'$ anticommutes with 
$\mathcal{P}^{\,}_{\Theta}$ 
according to Eq.\ (\ref{eq: TRS H c}) was used. 
By the same token, 
one can show that the fields 
$\hat{\varphi}$ and 
$\hat{\bar{\varphi}}$ commute, 
since
\begin{equation}
\mathcal{T}^{\mathsf{T}}\,
\mathcal{K}'\,
\overline{\mathcal{T}}=
\mathcal{T}^{\mathsf{T}}\,
\mathcal{P}^{\,}_{\Theta}\,
\mathcal{K}'\,\mathcal{T}=
-
\overline{\mathcal{T}}^{\mathsf{T}}\,
\mathcal{K}'\,
\mathcal{T}=0.
\end{equation}
We conclude that the effective edge theory for \textit{any} Abelian TRS
fractional topological insulator build from fermions has the effective
form of one Kramers degenerate pairs
\begin{equation}
\mathcal{K}_{\mathrm{eff}}=
\begin{pmatrix}
\mathcal{T}^{\mathsf{T}}\mathcal{K}'\mathcal{T}&0\\
0&-\mathcal{T}^{\mathsf{T}}\mathcal{K}'\mathcal{T}
\end{pmatrix},
\end{equation}
and is thus entirely defined by the single integer
\begin{equation}
m\:=\mathcal{T}^{\mathsf{T}}\mathcal{K}'\mathcal{T}.
\end{equation}
With the scattering vectors
(\ref{eq: def array quantum wires TSE AII c}) 
An explicit wire construction for each of these cases was given, 
thus exhausting all possible stable edge theories for Abelian
fractional topological insulators.

For each positive odd integer $m$, 
the fractionalized mode has a $\mathbb{Z}^{\,}_{2}$ character. 
It can have either one or none stable Kramers degenerate pair of $m$ 
quasiparticles. 

\medskip\noindent
\textbf{\textit{7.5.5.3\hskip 10 true pt 
Symmetry Class D: Fractional superconductors}}
\label{subsec: Symmetry Class D: Fractional superconductors}

\medskip\noindent
In Section 7.5.5.2,
TRS was imposed on the wire construction of fractional quantum Hall states 
from which the fractional topological insulators in symmetry class AII
followed.  In complete analogy, one may impose PHS with
$\Pi^{2}=+1$ on the wire construction of a fractional quantum
Hall state, thereby promoting it to symmetry class D.  Physically,
there follows a model for a superconductor with ``fractionalized" 
Majorana fermions or Bogoliubov quasiparticles. 

Lately, interest in this direction has been revived by the investigation 
of exotic quantum dimensions of twist defects embedded 
in an Abelian fractional quantum Hall liquid
\cite{Maissam12,Maissam13a,Maissam13b},
along with heterostructures of superconductors combined 
with fractional quantum Hall effect
\cite{Netanel12,Vaezi13,Clarke13}, 
or fractional topological insulators
\cite{Cheng12}. 
Furthermore, the Kitaev quantum wire has been generalized to 
$\mathbb{Z}^{\,}_{n}$ clock models hosting parafermionic edge modes
\cite{Kitaev01,Fendley12},
along with efforts to transcend the
Read-Rezayi quantum Hall state \cite{Read99} 
to spin liquids \cite{Greiter09,Greiter14} 
and superconductors \cite{gangof11}, 
all of which exhibit parafermionic quasiparticles.

As in the classification of non-interacting insulators, 
the Bogoliubov quasiparticles are treated with Abelian bosonization
as if they were Dirac fermions. The fractional 
phase is driven by interactions among the Bogoliubov quasiparticles. 

The elementary degrees of freedom in each wire are spinless right- and
left-moving fermions and holes as was defined for symmetry class D in
Eqs.~\eqref{eq: def H of V for class D a}%
-\eqref{eq: def H of V for class D c}.  
Construct the fractional
topological insulator using the set of PHS scattering vectors 
$\mathcal{T}^{(j)}$ , for $j=1,\cdots,N$
with $\mathcal{T}^{(j)}$ as defined in 
Eq.~\eqref{eq: def H of T for class D d} in each wire
and the PHS as defined in Eq.~\eqref{eq: def H of V for class D b}.
Complement them with the set of PHS scattering vectors 
$\overline{\mathcal{T}}^{(j)}$, for $j=1,\cdots,N-1$ defined by
($m^{\,}_{\pm}=(m\pm1)/2$)
\begin{equation}
\overline{\mathcal{T}}^{(j)}
=\left(
0,0,0,0\left|
\cdots
\left|\left.
-m^{\,}_{-},m^{\,}_{+},-m^{\,}_{+}, m^{\,}_{-}
\right|
-m^{\,}_{+},m^{\,}_{-},-m^{\,}_{-}, m^{\,}_{+}
\right|
\cdots
\right|0,0,0,0
\right)^{\mathsf{T}},
\label{eq: def overline T for class D j=1 ... N-1}
\end{equation}
with $m$ an odd positive integer.  
Notice that 
$\overline{\mathcal{T}}^{(j)}\:=
-\mathcal{P}_{\Pi}\,\overline{\mathcal{T}}^{(j)}$ so that one has to
demand that $\alpha^{\,}_{\overline{\mathcal{T}}^{(j)}}=0$ has to comply with PHS. 
Thus, together the $\mathcal{T}^{(j)}$ and $\overline{\mathcal{T}}^{(j)}$
gap out $(4N-2)$ of the $4N$ chiral modes in the wire. 
Identify the unique (up to an integer multiplicative factor) 
scattering vector ($m^{\,}_{\pm}=(m\pm1)/2$)
\begin{equation}
\overline{\mathcal{T}}^{(0)}=
\left(\left.\left.
-m^{\,}_{+},m^{\,}_{-},-m^{\,}_{-}, m^{\,}_{+}
\right|0,0,0,0\right|
\cdots
\left|0,0,0,0\left|
-m^{\,}_{-},m^{\,}_{+},-m^{\,}_{+}, m^{\,}_{-}
\right.\right.
\right)^{\mathsf{T}},
\label{eq: T0 for SET D}
\end{equation}
with $m$ the same odd positive integer as in Eq.\ 
(\ref{eq: def overline T for class D j=1 ... N-1}) 
that satisfies the Haldane
criterion with all $\mathcal{T}^{(j)}$ and
$\overline{\mathcal{T}}^{(j)}$ and thus can potentially gap out the 2
remaining modes. However, it is physically forbidden for it represents
a non-local scattering from one edge to the other. It is concluded that
each boundary supports a single remaining chiral mode that is an
eigenstate of PHS.

To understand the nature of the single remaining chiral mode on each boundary, 
the local linear transformation $W$ of the bosonic fields
\begin{equation}
W=
\begin{pmatrix}
-m^{\,}_{-}&+m^{\,}_{+}&0&0\\
+m^{\,}_{+}&-m^{\,}_{-}&0&0\\
0&0&-m^{\,}_{-}&+m^{\,}_{+}\\
0&0&+m^{\,}_{+}&-m^{\,}_{-}
\end{pmatrix},
\quad
m^{\,}_{\pm}=\frac{m\pm1}{2},
\end{equation}
with determinant $\mathrm{det}\,W=m^{4}$ is used.
When applied to the non-local scattering vector $\overline{\mathcal{T}}^{(0)}
$ that connects the two remaining chiral edge modes, this gives
\begin{equation}
\begin{split}
\widetilde{\overline{\mathcal{T}}}^{(0)}&=\mathcal{W}^{-1}\,\overline{\mathcal{T}}^{(0)}
\\&
=(0,-1,+1,0|0,0,0,0|\cdots|0,0,0,0|+1,0,0,-1),
\end{split}
\end{equation}
while the matrix $K$ changes under this transformation to
\begin{equation}
\widetilde{K}=
\mathrm{diag}\,(-m,m,m,-m).
\end{equation}
Noting that the representation of PHS is unchanged \begin{equation}
W^{-1}\,P^{\,}_{\Pi}\,W=P^{\,}_{\Pi},
\label{eq: W commutes with P SEP class D}
\end{equation}
we may interpret the remaining chiral edge mode as a PHS superposition
of a Laughlin quasiparticle and a Laughlin quasihole.  It thus 
describes a fractional chiral edge mode on either side of the
two-dimensional array of quantum wires.
The definite chirality is
an important difference to the case of the fractional
$\mathbb{Z}^{\,}_{2}$ topological insulator discussed in
Section
\ref{subsec: Symmetry Class AII: Fractional topological insulators}.  
It guarantees that any integer number $n\in\mathbb{Z}$
layers of this theory is stable, for no tunneling vector that acts
locally on one edge can satisfy the Haldane criterion
(\ref{eq: Haldane conditions}). For each $m$, one may thus say that the
parafermion mode has a $\mathbb{Z}$ character, as does the SRE phase
in symmetry class D.

\medskip\noindent
\textbf{\textit{7.5.5.4\hskip 10 true pt 
Symmetry classes DIII and C: More fractional superconductors}}

\medskip\noindent
The construction is very similar for classes DIII and C.
to the cases already presented. 
Details are relegated to \cite{Neupert14}.
For class DIII, the edge excitations (and bulk quasiparticles) 
of the phase are TRS fractionalized Bogoliubov quasiparticles
that have also been discussed in one-dimensional realizations.
(In the latter context, these TRS 
fractionalized Bogoliubov quasiparticles are
rather susceptible to perturbations
\cite{Klinovaja13c,Oreg14}).

\subsection{Summary}
\label{sec: Discussion}

In this section, a wire construction was developed to build models of 
short-range entangled and long-range entangled topological phases 
of two-dimensional quantum matter,
so as to yield immediate information about 
the topological stability of their edge modes. As such, 
the periodic table for integer topological phases 
was promoted to its fractional counterpart.
The following paradigms were applied.\\
(1) Each Luttinger liquid wire describes (spinfull or spinless) electrons. 
Abelian bosonization was used.\\
(2) Back-scattering and short-range interactions within and between wires 
are added. Modes are gapped out if these terms acquire 
a finite expectation value.\\
(3) A mutual compatibility condition, 
the Haldane criterion, 
is imposed among the terms that acquire an expectation value. 
It is an incarnation of the statement that the operators have to commute
if they are to be replaced simultaneously by their expectation values.\\
(4) A set of discrete and local symmetries are imposed on all terms 
in the Hamiltonian. When modes become massive, 
they may not break these symmetries.\\
(5) The model was analyzed in a strong-coupling limit,
instead of the weak coupling limit in which one derives
the renormalization group flows for the interactions.

It has become fashionable to write 
papers in condensed matter physics that take Majorana fermions as 
the building blocks of lattice models. Elegant mathematical
results have been obtained in this way, some of which
having the added merit for bringing conceptual clarity. 
However, the elementary building blocks of condensed matter are ions and
electrons whose interactions are governed by quantum electrodynamics.
Majorana fermions in condensed matter physics
can only emerge in a non-perturbative way through 
(i) the interactions between the electrons from the valence bands of a material,
or (ii) as the low-energy excitations of exotic quantum magnets. 
For Majorana fermions to be observable in condensed matter physics, a
deconfining transition must have taken place, a notoriously non-perturbative
phenomenon. One of the challenges that was undertaken in this section
is to find interacting models for itinerant electrons with local
interactions that support Majorana fermions at low energies and long
wave lengths. This goal was achieved, starting from non-interacting itinerant
electrons, by constructing local many-body interactions that conserve 
the electron charge and that stabilize two-dimensional bulk superconductors 
supporting gapless Majorana fermions along their two-dimensional boundaries.
This is why strictly many-body interactions are needed in the symmetry classes
D, DIII, and C to realize SRE topological phases
in the fourth column of Table~\ref{table: main table}.

\section*{Acknowledgments}
I am grateful to Claudio Chamon, Titus Neupert, Luiz Santos,
Shinsei Ryu, and Ronnie Thomale with whom our collaborations
have shaped Sections
\ref{sec: Stability analysis for the edge theory in the symmetry class AII} 
and 
\ref{sec: Construction of two-dimensional topological phases from coupled wires}.
I am also grateful to Maurizio Storni 
for his help with Section
\ref{sec: Fractionalization from Abelian bosonization}.
{Finally, I would like to thank my student Jyong-Hao Chen for
observing that Eq.\ (\ref{eq: Levin-Stern relation for AII}) holds.}

\bibliographystyle{OUPnum}
\bibliography{bibliooup}

\end{document}